\documentclass[10pt,reqno,twoside]{article}

\usepackage[margin=0.8in]{geometry}

\title{False Discovery and Its Control in Low-Rank Estimation} % CORRECT TITLE
\author{Armeen Taeb $^{a}$, Parikshit Shah $^{b}$, and Venkat Chandrasekaran $^a{^,}^c$
\thanks{Correspondence email: ataeb@caltech.edu} \vspace{.25in} \\ $^a$ Department of Electrical Engineering, California Institute of Technology \\  $^b$ Yahoo Research and Wisconsin Institutes for Discovery at the University of Wisconsin \\ $^c$ Department of Computing and Mathematical Sciences, California Institute of Technology}
\date{}
\makeatletter\@addtoreset{section}{part}\makeatother%

\newcommand{\upperRomannumeral}[1]{\uppercase\expandafter{\romannumeral#1}}

\renewcommand{\hat}{\widehat}
\usepackage{cite} 
\usepackage{natbib}
\usepackage{siunitx}
\usepackage{parskip}
\usepackage{graphicx}
\usepackage{hyperref}

\newcommand{\R}{\mathbb{R}} 
\newcommand{\ProjM}{\mathcal{P}_{\mathrm{span}(M_i)}}

\newcommand{\Proj}{\mathcal{P}}

\usepackage{algorithm,algorithmic}
\usepackage{amssymb,amsmath,amsthm}

\DeclareMathOperator*{\argmax}{argmax} % no space, limits underneath in displays
\DeclareMathOperator*{\argmin}{argmin} % no space, limits underneath in displays
\newtheorem{theorem}{Theorem}
\newtheorem{defn}[theorem]{Definition}
\newtheorem{proposition}[theorem]{Proposition}

\usepackage{enumitem}

\usepackage{verbatim}

\usepackage{titlesec}

\numberwithin{equation}{section}

\titlespacing{\section}{2pt}{\parskip}{-\parskip}
\titlespacing{\subsection}{6pt}{\parskip}{-\parskip}
\titlespacing{\subsubsection}{6pt}{\parskip}{-\parskip}
\usepackage{titlesec}
\titlespacing*{\section}{6pt}{0.4\baselineskip}{\baselineskip}
\usepackage{placeins}
\usepackage{subfigure}
\newtheorem{lemma}{Lemma}
\date{October 19, 2018; revised June 14, 2020}

\begin{document}
\setlength{\abovedisplayskip}{15pt}
\setlength{\belowdisplayskip}{15pt}
\maketitle

\begin{abstract}
Models specified by low-rank matrices are ubiquitous in contemporary applications.  In many of these problem domains, the row/column space structure of a low-rank matrix carries information about some underlying phenomenon, and it is of interest in inferential settings to evaluate the extent to which the row/column spaces of an estimated low-rank matrix signify discoveries about the phenomenon.  However, in contrast to variable selection, we lack a formal framework to assess true/false discoveries in low-rank estimation; in particular, the key source of difficulty is that the standard notion of a discovery is a discrete one that is ill-suited to the smooth structure underlying low-rank matrices.  We address this challenge via a \emph{geometric} reformulation of the concept of a discovery, which then enables a natural definition in the low-rank case.  We describe and analyze a generalization of the Stability Selection method of Meinshausen and B\"uhlmann to control for false discoveries in low-rank estimation, and we demonstrate its utility compared to previous approaches via numerical experiments.\\[0.2in]
%\vspace{-0.1in}
% Please include a  of smaximumeven keywords
{\bf keywords:} algebraic geometry, determinantal varieties, testing, model selection, regularization, stability selection
\end{abstract}

\section{Introduction}
\label{section:introduction}
Models described by low-rank matrices are ubiquitous in many contemporary problem domains.  The reason for their widespread use is that low-rank matrices offer a flexible approach to specify various types of low-dimensional structure in high-dimensional data.  For example, low-rank matrices are used to describe user preferences in collaborative filtering \citep{collab}, small collections of end-member signatures in hyperspectral imaging \citep{Manolakis}, directions of moving targets in radar measurements \citep{Fa}, low-order systems in control theory \citep{hankel}, coherent imaging systems in optics \citep{optics}, and latent-variable models in factor analysis \citep{Shapiro}.  In many of these settings, the row/column space structure of a low-rank matrix carries information about some underlying phenomenon of interest; for instance, in hyperspectral imaging for mineralogy problems, the column space represents the combined signatures of relevant minerals in a mixture.  Similarly, the row/column spaces of matrices obtained from radar measurements signify the directions of moving targets. Therefore, in inferential contexts in which low-rank matrices are estimated from data, it is of interest to evaluate the extent to which the row/column spaces of the estimated matrices signify true/false \emph{discoveries} about the relevant phenomenon.

In seeking an appropriate framework to assess discoveries in low-rank estimation, it is instructive to consider the case of variable selection, which may be viewed conceptually as low-rank estimation with diagonal matrices.  Stated in terms of subspaces, the set of discoveries in variable selection is naturally represented by a subspace that is spanned by the standard basis vectors corresponding to the subset of variables that are declared as significant.  The number of true discoveries then corresponds to the dimension of the intersection between this `discovery subspace' and the `population subspace' (i.e., the subspace spanned by standard basis vectors corresponding to significant variables in the population), and the number of false discoveries is the dimension of the `discovery subspace' minus the number of true discoveries.  Generalizing this perspective to low-rank estimation, it is perhaps appealing to declare that the number of true discoveries is the dimension of the intersection of the estimated row/column spaces and the population row/column spaces, and the number of false discoveries is the dimension of the remaining components of the estimated row/column spaces.  The difficulty with this approach is that we cannot expect any inference procedure to perfectly estimate with positive probability even a one-dimensional subspace of the population row/column spaces as the collection of these spaces is not discrete; in particular, the set of all subspaces of a given dimension is the Grassmannian manifold, whose underlying smooth structure is unlike that of the finite collection of coordinate subspaces that correspond to discoveries in variable selection.  Therefore, the number of true discoveries would generically be zero.  One method to improve upon this idea is to define the number of true discoveries as the dimension of the largest subspaces of the estimated row/column spaces that are within a specified angle of the population row/column spaces, and to treat the dimension of the remaining components of the estimated row/column spaces as the number of false discoveries.  An unappealing feature of this second approach is that it depends on an extrinsic parameter, and minor perturbations of this parameter could result in potentially large changes in the number of true/false discoveries.  In some sense, these preceding attempts fail as they are based on a sharp binary choice that declares components of the estimated row/column spaces exclusively as true or false discoveries, which is ill-suited to the smooth structure underlying low-rank matrices.

As our first contribution, we develop in Section~\ref{section:false_discovery} a \emph{geometric} framework for evaluating false discoveries in low-rank estimation.  We begin by expressing the number of true/false discoveries in variable selection in terms of functionals of the projection matrices associated to the discovery/population subspaces described above; this expression varies smoothly with respect to the underlying subspaces, unlike dimensions of intersections of subspaces.  Next, we interpret the discovery/population subspaces in variable selection as tangent spaces to algebraic varieties of sparse vectors.  Finally, we note that tangent spaces with respect to varieties of low-rank matrices encode the row/column space structure of a matrix, and therefore offer an appropriate generalization of the subspaces discussed in the context of variable selection.  Putting these observations together, we substitute tangent spaces with respect to varieties of low-rank matrices into our reformulation of discoveries in variable selection in terms of projection matrices, which leads to a natural formalism of the number of true/false discoveries that is suitable for low-rank estimation.  We emphasize that although our definition respects the smooth geometric structure underlying low-rank matrices, one of its appealing properties is that it specializes transparently to the usual discrete notion of true/false discoveries in the setting of variable selection if the underlying low-rank matrices are diagonal.

Our next contribution concerns the development of a procedure for low-rank estimation that provides false discovery control.  In Section~\ref{section:stability}, we generalize the `stability selection' procedure of \cite{stability} for controlling false discoveries in variable selection.  Their method operates by employing variable selection methods in conjunction with subsampling; in particular, one applies a variable selection algorithm to subsamples of a dataset, and then declares as discoveries those variables that are selected most frequently.  In analogy to their approach, our algorithm -- which we call `subspace stability selection' -- operates by combining existing low-rank estimation methods in conjunction with subsampling.  Our framework employs row/column space selection procedures (based on standard low-rank estimation algorithms) on subsamples of a dataset, and then outputs as discoveries a set of row/column spaces that are `close to' most of the estimated row/column spaces; the specific notion of distance here is based on our tangent space formalism.  Building on the results in \citep{stability,Samworth}, we provide a theoretical analysis of the performance of our algorithm. {A key quantity in our results is the commutator between projection matrices associated to estimated tangent spaces and to the population tangent space, which highlights the distinction between the discrete nature of variable selection and the smooth geometry underlying low-rank estimation.}

Finally, in Section~\ref{section:experimental} we contrast subspace stability selection with previous methods in a range of low-rank estimation problems involving simulated as well as real data.  The tasks involving real data are on estimating user-preference matrices for recommender systems and identifying signatures of relevant minerals in hyperspectral images.  The estimates provided by subspace stability selection offer improvements in multiple respects.  First, the row/column spaces of the subspace stability selection estimates are far closer to their population counterparts in comparison to other standard approaches; in other words, our experiments demonstrate that subspace stability selection provides estimates with far fewer false discoveries, without a significant loss in power (both false discovery and power are based on the definitions introduced in this paper).  Second, in settings in which regularized formulations are employed, subspace stability selection estimates are much less sensitive to the specific choice of the regularization parameter.  Finally, a common challenge with approaches based on cross-validation for low-rank estimation is that they overestimate the complexity of a model, i.e., they produce higher rank estimates (indeed, a similar issue arises in variable selection, which was one of the motivations for the development of stability selection in \cite{stability}).  We observe that the estimates produced by subspace stability selection have substantially lower rank than those produced by cross-validation, with a similar or improved prediction performance.

The outline of this paper is as follows.  In Section~\ref{section:false_discovery}, we briefly review the relevant concepts from algebraic geometry and then formulate a false discovery framework for low-rank estimation.  Our subspace stability selection algorithm is described in Section~\ref{section:stability}, with theoretical support presented in Section~\ref{section:theoretical}. In Section~\ref{section:experimental}, we demonstrate the utility of our approach in experiments with synthetic and real data.  We conclude with a discussion of further research directions in Section~\ref{section:conclusions}.

\paragraph{Related work} We are aware of prior work for low-rank estimation based on testing the significance level of the singular values of an observed matrix (see, for example, \cite{Choi}, \cite{Liu}, \cite{Song}).  However, in contrast to our framework, these methods do not directly control deviations of row/column spaces, which carry significant information about various phenomena of interest in applications.  Further, these previous approaches have limited applicability as they rely on having observations of all the entries of a matrix; this is not the case, for example, in low-rank matrix completion problems which arise commonly in many domains.  In comparison, our methodology is general-purpose and is applicable to a broad range of low-rank estimation problems.  On the computational front, our algorithm and its analysis are a generalization of some of the ideas in \citep{stability,Samworth}.  However, the geometry underlying the collection of tangent spaces to low-rank matrices leads to a number of new challenges in our context.

%, in which the authors propose a `complementary bagging' approach for subsampling

\paragraph{Notation}For a subspace $\mathbb{V}$, we denote projection onto $\mathbb{V}$ by $\Proj_{\mathbb{V}}$.  Given a self-adjoint linear map $M: \bar{\mathbb{V}} \rightarrow \bar{\mathbb{V}}$ on a vector space $\bar{\mathbb{V}}$ and a subspace $\mathbb{V} \subset \bar{\mathbb{V}}$, the minimum singular value of $M$ restricted to $\mathbb{V}$ is given by $\sigma_{\min}(\Proj_{\mathbb{V}} M \Proj_{\mathbb{V}}) = \inf_{x \in \mathbb{V} \backslash \{0\} } ~ \frac{\|Mx\|_{\ell_2}}{\|x\|_{\ell_2}}$. We denote Kronecker product between two matrices $A$ and $B$ by $A \otimes B$. Finally, the nuclear norm (sum of singular values) is denoted by $\|\cdot\|_\star$, and the frobenious norm is denoted by $\|\cdot\|_F$.

\section{A Geometric False Discovery Framework}
\label{section:false_discovery}

We describe a geometric framework for assessing discoveries in low-rank estimation.  Our discussion proceeds by first reformulating true/false discoveries in variable selection in geometric terms, which then enables a transparent generalization to the low-rank case.  We appeal to elementary ideas from algebraic geometry on varieties and tangent spaces \citep{Harris}.  We also describe a procedure to obtain an estimate of a low-rank matrix given an estimate of a tangent space.

\subsection{False Discovery in Low-Rank Estimation}
\vspace{0.1in}
The performance of a variable selection procedure $\hat{\mathcal{S}} \subset \{1,\dots,p\}$, which estimates a subset of a collection of $p$ variables as being significant, is evaluated by comparing the number of elements of $\hat{\mathcal{S}}$ that are also present in the `true' subset of significant variables $\mathcal{S}^\star \subset \{1,\dots,p\}$ -- the number of true discoveries is $|\hat{\mathcal{S}} \cap \mathcal{S}^\star|$, while the number of false discoveries is $|\hat{\mathcal{S}} \cap {\mathcal{S}^\star}^{c}|$.  We give next a geometric perspective on this combinatorial notion.  As described in the introduction, one can associate to each subset $\mathcal{S} \subset \{1,\dots,p\}$ the coordinate aligned \emph{subspace} $T(\mathcal{S}) = \{x \in \R^p ~|~ \mathrm{support}(x) \subseteq \mathcal{S}\}$, where $\mathrm{support}(x)$ denotes the locations of the nonzero entries of $x$.  With this notation, the number of false discoveries in an estimate $\hat{\mathcal{S}}$ is given by:
\begin{equation*}
\# \texttt{false-discoveries} = |\hat{\mathcal{S}} \cap {\mathcal{S}^\star}^{c}| = \mathrm{dim}(T(\hat{\mathcal{S}}) \cap T(\mathcal{S}^\star)^\perp) = \mathrm{trace}\left(\Proj_{T(\hat{\mathcal{S}})} \Proj_{T(\mathcal{S}^\star)^\perp} \right).
\end{equation*}
 Similarly, the number of true discoveries is given by $\mathrm{trace}\left(\Proj_{T(\hat{\mathcal{S}})} \Proj_{T({\mathcal{S}^\star})}\right)$. These latter reformulations in terms of projection operators have no obvious `discrete' attribute to them.  In particular, for any subspaces $\mathcal{W}, \tilde{\mathcal{W}}$, the expression $\mathrm{trace}(\Proj_{\mathcal{W}} \Proj_{\tilde{\mathcal{W}}})$ is equal to the sum of the squares of the cosines of the principal angles between $\mathcal{W}$ and $\tilde{\mathcal{W}}$ \citep{Golub}; as a result,  the quantity $\mathrm{trace}(\Proj_{\mathcal{W}} \Proj_{\tilde{\mathcal{W}}})$ varies smoothly with respect to perturbations of $\mathcal{W}, \tilde{\mathcal{W}}$.  The discrete nature of a discovery is embedded inside the encoding of the subsets $\hat{\mathcal{S}}, \mathcal{S}^\star$ using the subspaces $T(\hat{\mathcal{S}}), T(\mathcal{S}^\star)$.  Consequently, to make progress towards a suitable definition of true/false discoveries in the low-rank case, we require an appropriate encoding of row/column space structure via subspaces in the spirit of the mapping $\mathcal{S} \mapsto T(\mathcal{S})$.  Towards this goal, we interpret next the subspace $T(\mathcal{S})$ associated to a subset $\mathcal{S} \subset \{1,\dots,p\}$ as a tangent space to an algebraic variety.

Formally, for any integer $k \in \{1,\dots,p\}$ let $\mathcal{V}_{\mathrm{sparse}}(k) \subset \R^p$ denote the algebraic variety of elements of $\R^p$ with at most $k$ nonzero entries.  Then for any point in $\mathcal{V}_{\mathrm{sparse}}(k)$ consisting of exactly $k$ nonzero entries at locations given by the subset $\mathcal{S} \subset \{1,\dots,p\}$ (here $|\mathcal{S}| = k$), the tangent space at that point with respect to $\mathcal{V}_{\mathrm{sparse}}(k)$ is given by $T(\mathcal{S})$.  In other words, the tangent space at a smooth point of $\mathcal{V}_{\mathrm{sparse}}(k)$ is completely determined by the locations of the nonzero entries of that point.  This geometric perspective extends naturally to the low-rank case.

Consider the \emph{determinantal variety} $\mathcal{V}_{\text{low-rank}}(r) \subset \R^{p_1 \times p_2}$ of matrices of size $p_1 \times p_2$ with rank at most $r$ (here $r \in \{1,\dots,\min(p_1,p_2)\}$).  Then for any matrix in $\mathcal{V}_{\text{low-rank}}(r)$ with rank equal to $r$ and with row and column spaces given by $\mathcal{R} \subset \R^{p_2}$ and $\mathcal{C} \subset \R^{p_1}$, respectively, the tangent space at that matrix with respect to $\mathcal{V}_{\text{low-rank}}(r)$ is given by \citep{Harris}:
\begin{equation}
T(\mathcal{C}, \mathcal{R}) \triangleq \{M_R + M_C ~|~ M_R,M_C \in \R^{p_1 \times p_2}, \text{row-space}(M_{R}) \subseteq \mathcal{R}, \text{column-space}(M_{C}) \subseteq \mathcal{C}\}. \label{eq:tspace}
\end{equation}
The dimension of $T(\mathcal{C}, \mathcal{R})$ equals $r(p_1+p_2) - r^2$ and the dimension of its orthogonal complement $T(\mathcal{C}, \mathcal{R})^\perp$ equals $(p_1-r)(p_2-r)$.  Further, the projection operators onto $T(\mathcal{C}, \mathcal{R})$ and onto $T(\mathcal{C}, \mathcal{R})^\perp$ can be expressed in terms of the projection maps onto $\mathcal{C}$ and $\mathcal{R}$ as follows:
\begin{equation}
\begin{aligned}
\Proj_{T(\mathcal{C}, \mathcal{R})} &= \Proj_{\mathcal{C}} \otimes I + I \otimes \Proj_{\mathcal{R}} - \Proj_{\mathcal{C}} \otimes \Proj_{\mathcal{R}} \\ \Proj_{T(\mathcal{C}, \mathcal{R})^\perp} &= (I - \Proj_{\mathcal{C}}) \otimes (I - \Proj_{\mathcal{R}}) = \Proj_{\mathcal{C}^\perp} \otimes \Proj_{\mathcal{R}^\perp}.
\end{aligned}
\label{eqn:tang_def}
\end{equation}
where $\otimes$ denotes a Kronecker product. Consequently, the action of projection operators $\Proj_{T(\mathcal{C}, \mathcal{R})}$ and $\Proj_{T(\mathcal{C}, \mathcal{R})^\perp}$ on a matrix $M \in \mathbb{R}^{p_1 \times p_2}$ yields
$\Proj_{T(\mathcal{C}, \mathcal{R})}(M) =\Proj_{\mathcal{C}}M + M\Proj_{\mathcal{R}}-\Proj_{\mathcal{C}}M\Proj_{\mathcal{R}}$ and $  \Proj_{T(\mathcal{C}, \mathcal{R})^\perp}(M) = \Proj_{\mathcal{C}^\perp}M\Proj_{\mathcal{R}^\perp}$. In analogy to the previous case with variable selection, the tangent space at a rank-$r$ matrix with respect to $\mathcal{V}_{\text{low-rank}}(r)$ encodes -- and is in one-to-one correspondence with -- the row/column space structure at that point.  Indeed, estimating the row/column spaces of a low-rank matrix can be viewed equivalently as estimating the tangent space at that matrix with respect to a determinantal variety.  With this notion in hand, we give our definition of true/false discoveries in low-rank estimation:
\vspace{0.1in}
\begin{defn}
Let $\mathcal{C}^\star \subset \R^{p_1}$ and $\mathcal{R}^\star \subset \R^{p_2}$ denote the column and row spaces of a low-rank matrix in $\R^{p_1 \times p_2}$; in particular, $\mathrm{dim}(\mathcal{C}^\star) = \mathrm{dim}(\mathcal{R}^\star)$.  Given observations from a model parametrized by this matrix, let $(\hat{\mathcal{C}}, \hat{\mathcal{R}}) \subset \R^{p_1} \times \R^{p_2}$ be an estimator of the pair of subspaces $(\mathcal{C}^\star, \mathcal{R}^\star)$ with $\mathrm{dim}(\hat{\mathcal{C}}) = \mathrm{dim}(\hat{\mathcal{R}})$.  Then the \emph{expected false discovery} of the estimator is defined as:\vspace{0.1in}
\begin{equation}
\mathrm{FD} = \mathbb{E}\left[\mathrm{trace}\left(\Proj_{T(\hat{\mathcal{C}},\hat{\mathcal{R}})} \Proj_{T(\mathcal{C}^\star,\mathcal{R}^\star)^\perp}\right)\right],
\label{eqn:gfd}
\end{equation}
and the \emph{power} of the estimator is defined as:\vspace{0.1in}
\begin{equation}
\mathrm{PW} = \mathbb{E}\left[\mathrm{trace}\left(\Proj_{T(\hat{\mathcal{C}},\hat{\mathcal{R}})} \Proj_{T(\mathcal{C}^\star,\mathcal{R}^\star)}\right)\right].
\label{eqn:gpw}
\end{equation}
The expectations in both cases are with respect to randomness in the data employed by the estimator, and the tangent spaces $T(\hat{\mathcal{C}},\hat{\mathcal{R}}), T(\mathcal{C}^\star,\mathcal{R}^\star)$ are as defined in \eqref{eq:tspace}.
\end{defn}   
{\par}With respect to our objective of identifying a suitable notion of discovery for low-rank estimation, the definitions of $\mathrm{FD}$ and of $\mathrm{PW}$ possess a number of favorable attributes. These definitions do not depend on a choice of basis for the tangent space $T(\mathcal{C}^\star,\mathcal{R}^\star)$. Further, for the reasons described above small changes in row/column space estimates lead to small changes in the performance of an estimator, as evaluated by $\mathrm{FD}$ and $\mathrm{PW}$. Despite these definitions respecting the smooth structure underlying low-rank matrices, they specialize transparently to the usual discrete notion of true/false discoveries in the setting of variable selection if the underlying low-rank matrices are
diagonal. We also have that the expected false discovery is bounded as  $0 \leq \mathrm{FD} \leq \mathrm{dim}(T(\mathcal{C}^\star,\mathcal{R}^\star)^\perp)$ and the power is bounded as $0 \leq \mathrm{PW} \leq \mathrm{dim}(T(\mathcal{C}^\star,\mathcal{R}^\star))$, which is in agreement with the intuition that the spaces $T(\mathcal{C}^\star,\mathcal{R}^\star)$ and $T(\mathcal{C}^\star,\mathcal{R}^\star)^\perp$ represent the total true and false discoveries, respectively, that can be made by any estimator.  Similarly, we observe that $\mathrm{FD} + \mathrm{PW} = \mathbb{E}[\mathrm{dim}(T(\hat{\mathcal{C}},\hat{\mathcal{R}}))]$, which is akin to the expected total discovery made by the estimator $(\hat{\mathcal{C}},\hat{\mathcal{R}})$. 

{One can also arrive at the definitions \eqref{eqn:gfd} and \eqref{eqn:gpw} in an ``axiomatic'' manner as follows.  Suppose we wish to identify a suitable notion of alignment between the estimate $T(\hat{\mathcal{C}},\hat{\mathcal{R}})$ and the population $T({\mathcal{C}}^\star,{\mathcal{R}}^\star)$ via a real-valued function $f(\cdot, \cdot)$ whose arguments consist of a pair of tangent spaces.  First, $f$ should remain invariant to simultaneous isometric linear transformations of the row/column spaces of the population and of the estimate; as a parallel, the appropriate invariance in variable selection is simultaneous relabeling of the variables in the estimate and the population.  We conclude from this that $f$ must be a function purely of the \emph{principal angles} between its arguments, which correspond to the spectrum of the product of the associated projection matrices.  Second, our definition of $f$ should satisfy the condition that the sum $f(T(\hat{\mathcal{C}},\hat{\mathcal{R}}), T(\mathcal{C}^\star, \mathcal{R}^\star)^\perp) + f(T(\hat{\mathcal{C}}, \hat{\mathcal{R}}), T(\mathcal{C}^\star,\mathcal{R}^\star))$ equals $\mathrm{dim}(T(\hat{\mathcal{C}},\hat{\mathcal{R}}))$ -- that is, the sum of the false discovery and the true discovery must equal the total amount of discovery.  Based on this requirement as well as the deduction from the first argument, one can arrive at the definitions \eqref{eqn:gfd} and \eqref{eqn:gpw} after taking expectations.}

Additionally, we note that the definition of $\mathrm{FD}$ may be modified to obtain an analog of the \emph{false discovery rate} \citep{B-H}, which is of interest in contemporary multiple testing as well as in high-dimensional estimation:\vspace{0.05in}

\begin{equation*}
\mathrm{FDR} = \mathbb{E}\left[\frac{\mathrm{trace}\left(\Proj_{T(\hat{\mathcal{C}},\hat{\mathcal{R}})} \Proj_{T(\mathcal{C}^\star,\mathcal{R}^\star)^\perp}\right)} {\mathrm{dim}(T(\hat{\mathcal{C}},\hat{\mathcal{R}}))}\right].
\end{equation*}
We focus in the present paper on controlling the quantity $\mathrm{FD}$, and we discuss in Section~\ref{section:conclusions} some challenges associated with controlling FDR in low-rank estimation.

{Finally, while the main focus of this paper is on a false discovery framework for low-rank estimation in which we seek reliable estimates of both the row and column spaces, the geometric perspective outlined here can be adapted to settings in which one only seeks an estimate of the column-space of a low-rank matrix.  (Such a problem arises in hyperspectral imaging, as illustrated in Section~\ref{section:experimental}.)  In such situations, the ideas described previously can be extended as follows:
%\begin{eqnarray}
\begin{gather}
\widetilde{\mathrm{FD}} = \mathbb{E}\left[\mathrm{trace}\left(\Proj_{\hat{\mathcal{C}}} \Proj_{{\mathcal{C}^\star}^\perp}\right)\right]~~;~~
 \widetilde{\mathrm{PW}} = \mathbb{E}\left[\mathrm{trace}\left(\Proj_{\hat{\mathcal{C}}} \Proj_{{\mathcal{C}^\star}}\right)\right]~~;~~
{\widetilde{\mathrm{FDR}}  = \mathbb{E}\left[\frac{\mathrm{trace}\left(\Proj_{\hat{\mathcal{C}}} \Proj_{{\mathcal{C}^\star}^\perp}\right)}{\text{dim}(\hat{\mathcal{C}})}\right]}.
\label{eqn:fd_colspace}
\end{gather}
Here $\mathcal{C}^\star \subset \mathbb{R}^p$ represents the population column space and $\hat{\mathcal{C}} \subset \mathbb{R}^p$ is an estimator.  These expressions can be derived by considering tangent spaces with respect to  quotients of the determinantal variety under certain equivalence relations; supplementary material Section~A.9 provides the details.} 
{
{\subsection{From Tangent Space to Parameter Estimation}}
\vspace{0.1in}
Although the primary emphasis of this paper is on a framework to evaluate and control the expected false discovery of tangent spaces estimated from data, in many practical settings (e.g. some of the prediction tasks with real datasets in Section 4), the ultimate object of interest is an estimate of a low-rank matrix.  One can obtain such an estimate by solving a subsequent matrix estimation problem in which the tangent space of the matrix is constrained to lie within the tangent space identified from our framework.  Concretely, let $T(\mathcal{C},\mathcal{R}) \subset \R^{p_1 \times p_2}$ be a tangent space that corresponds to column and row spaces $\mathcal{C} \subset \R^{p_1}, \mathcal{R} \subset \mathbb{R}^{p_2}$, and given a collection of observations $\mathcal{D}$, we wish to solve the following optimization problem:
\begin{equation}
\label{eqn:constrainedLowRankr}
\hat{L} = \argmin_{L \in \mathbb{R}^{p_1 \times p_2}} \text{Loss}~(L~;~\mathcal{D}) ~~~ \text{subject to} ~~~ T(\text{column-space}(L),\text{row-space}(L))  \subseteq {T}(\mathcal{C},\mathcal{R}),
\end{equation}
in which the decision variable $L$ is constrained to have a tangent space that lies within the prescribed tangent space ${T}(\mathcal{C},\mathcal{R})$. Furthermore, this constraint may be simplified as follows.  Suppose that the subspaces $\mathcal{R}, \mathcal{C}$ are of dimension $k$.  Let $U_{C} \in \R^{p_1 \times k}$ and $U_R \in \R^{p_2 \times k}$ be any matrices with columns spanning the spaces $\mathcal{C}$ and $\mathcal{R}$, respectively.  Then one can check that the set $\{U_C M U_R' ~|~ M \in \R^{k \times k}\}$ is precisely the collection of matrices whose tangent spaces are contained in $T(\mathcal{C},\mathcal{R})$.  Consequently \eqref{eqn:constrainedLowRankr} may be reformulated as:
\begin{equation}
\label{eqn:tractable}
\hat{L} = \argmin_{L \in \mathbb{R}^{p_1 \times p_2}, ~M \in \R^{k \times k}} \text{Loss}~\left(L~;~\mathcal{D}\right) ~~~ \text{subject to} ~~~ L = U_C M U_R'.
\end{equation}
Note that the constraint here is linear in the decision variables $L, M$.  Consequently, an appealing property of \eqref{eqn:tractable} is that if the loss function $\text{Loss}(\cdot~;~\mathcal{D})$ is convex, then \eqref{eqn:tractable} is a convex optimization problem. For example, when $\text{Loss}(\cdot~;~\mathcal{D})$ is the squared loss, an optimal solution can be obtained in closed form.

In a similar fashion, in situations in which one is only concerned with estimating low-rank matrices with an accurate column space, one can solve an analog of \eqref{eqn:tractable} in which the decision variable satisfies the linear constraint that its column space lies inside a prescribed column space.}

\section{False Discovery Control via Subspace Stability Selection}
\label{section:stability}
Building on the discussion in the preceding section, our objective is the accurate estimation of the tangent space associated to a low-rank matrix, as this is in one-to-one correspondence with the row/column spaces of the matrix.  In this section, we formulate an approach based on the stability selection procedure of \cite{stability} to estimate such a tangent space. We will also describe how this method can be specialized for problems involving subspace estimation. 

Stability selection is a general technique to control false discoveries in variable selection.  The procedure can be paired with any variable selection procedure as follows: instead of applying a selection procedure (e.g. the Lasso) to a collection of observations, one instead applies the procedure to many subsamples of the data and then chooses those variables that are most consistently selected in the subsamples.  The virtue of the subsampling and averaging framework is that it provides control over the expected number of falsely selected variables (see Theorem 1 in \cite{stability} and Theorem 1 in \cite{Samworth}).  We develop a generalization of this framework in which existing row/column space selection procedures (based on any low-rank estimation procedure) are employed on subsamples of the data, and then these spaces are aggregated to produce a tangent space that provides false discovery control.

\emph{Subsampling procedure}: Although our framework is applicable with general subsamples of the data, we adopt the subsampling method outlined in \cite{Samworth} in our experimental demonstrations and our theoretical analysis; in particular, given a dataset $\mathcal{D}$ and a positive (even) integer $B$, we consider $B$ subsamples or bags obtained from $B/2$ \emph{complementary partitions} of $\mathcal{D}$ of the form $\{(\mathcal{D}_{2j-1},\mathcal{D}_{2j}): j = 1,2,3, \dots, B/2\}$, where $|\mathcal{D}_{2j-1}| = |\mathcal{D}|/2$ and $\mathcal{D}_{2j} = \mathcal{D} \backslash \mathcal{D}_{2j-1}$.

\emph{Setup for numerical demonstrations}: For our numerical illustrations in this section, we consider the following stylized low-rank matrix completion problem.  The population parameter $L^\star \in \mathbb{R}^{70\times{70}}$ is a rank-$10$ matrix with singular values (and associated multiplicities) given by $1 (x3), 0.5 (x5),$ and $0.1 (x2)$, and with row/column spaces sampled uniformly at random according to the Haar measure.  We are given noisy observations $Y_{i,j} = L^\star_ {i,j}+ \epsilon_{i,j}$ with $\epsilon_{i,j} {\sim} \mathcal{N}(0,\sigma^2)$ and $(i,j) \in \Omega$, where $\Omega \subset \{1,\dots,70\}^2$ is chosen uniformly at random with $|\Omega| = 3186$.  The variance $\sigma^2$ is chosen to set the signal-to-noise (SNR) ratio (defined as $\mathbb{E}[{\|L^\star\|_F}/{\|\epsilon\|_F}]$) at a desired level, and this is specified later.  As our subsamples, we consider a collection of $B=100$ subsets each consisting of $|\Omega|/2 = 1593$ entries obtained from $50$ random complementary partitions of the data.  On each subsample -- corresponding to a subset $S \subset \Omega$ of observations with $|S| = 1593$ -- we employ the following convex program
\citep{srebro,matrix_completion} 
\begin{eqnarray}
\hat{L} = \underset{L \in \mathbb{R}^{70 \times{70}}}{\mathrm{argmin}}~~\sum_{\{i,j\} \in S} \|(L-Y)_{i,j}\|_{F}^2 + \lambda\|L\|_{\star},
\label{eqn:nulcear_norm}
\end{eqnarray}
and we report the tangent space $T(\texttt{column-space}(\hat{L}),\texttt{row-space}(\hat{L}))$ as the estimate associated to the subsample.  Here $\lambda > 0$ is a regularization parameter (to be specified later) and $\|\cdot\|_\star$ is the nuclear norm (the sum of the singular values), which is commonly employed to promote low-rank structure in a matrix \citep{Fazel}.  We emphasize that our development is relevant for general low-rank estimation problems, and this problem is merely for illustrative purposes in the present section;  for a more comprehensive set of experiments in more general settings, we refer the reader to Section~\ref{section:experimental}.

\subsection{Stable Tangent Spaces}
\vspace{0.1in}
\label{section:subspace_stability}
The first step in stability selection is to combine estimates of significant variables obtained from different subsamples.  This is accomplished by computing for each variable the frequency with which it is selected across the subsamples.  We generalize this idea to our context via projection operators onto tangent spaces as follows:\vspace{0.1in}
\begin{defn}[Average Projection Operator] Suppose $\hat{T}$ is an estimator of a tangent space of a low-rank matrix, and suppose further that we are given a set of observations $\mathcal{D}$ and a corresponding collection of subsamples $\{\mathcal{D}_\ell\}_{\ell=1}^B$ with each $\mathcal{D}_\ell \subset \mathcal{D}$.  Then the \emph{average projection operator} of the estimator $\hat{T}$ with respect to the subsamples $\{\mathcal{D}_\ell\}_{\ell=1}^B$ is defined as:
\begin{equation}
\mathcal{P}_\text{avg}  \triangleq \frac{1}{B} \sum_{\ell=1}^{B} \Proj_{\hat{T}(\mathcal{D}_\ell)},
\label{eqn:avg_proj}
\end{equation}
where $\hat{T}(\mathcal{D}_\ell)$ is the tangent space estimate based on the subsample $\mathcal{D}_\ell$.
\label{defn:avg_proj_low}
\end{defn}

Here $\mathcal{P}_{\texttt{avg}} : \mathbb{R}^{p_1 \times p_2} \rightarrow \mathbb{R}^{p_1 \times p_2}$ is self-adjoint, and its eigenvalues lie in the interval $[0,1]$ as each $\mathcal{P}_{\hat{T}(\mathcal{D}_\ell)}$ is self-adjoint with eigenvalues equal to $0$ or $1$.  To draw a comparison with variable selection, the tangent spaces in that case correspond to subspaces spanned by coordinate vectors in $\R^p$ (with $p$ being the total number of variables of interest) and the average projection operator is a diagonal matrix of size $p \times p$, with each entry on the diagonal specifying the fraction of subsamples in which a particular variable is selected.  The virtue of averaging over tangent spaces estimated across a large number of subsamples is that most of the `energy' of the average projection operator $\mathcal{P}_{\texttt{avg}}$ tends to be better aligned with the underlying population tangent space.  We illustrate this point next with an example.

\begin{figure}[thbp]
\centering
\subfigure[$SNR = 0.8$]{
\includegraphics[scale = 0.4]{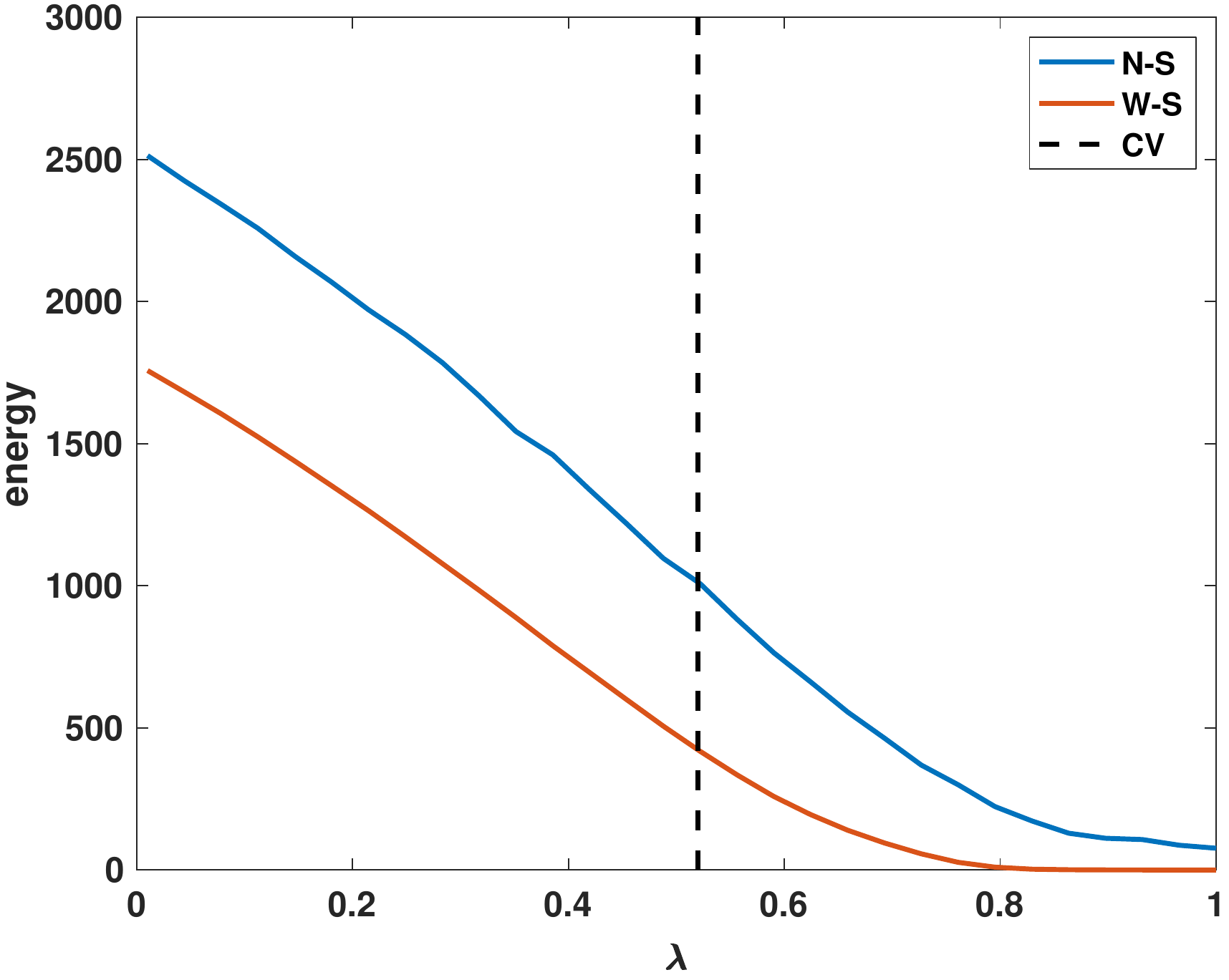}}
\subfigure[$SNR  = 1.6 $]{
\includegraphics[scale = 0.4]{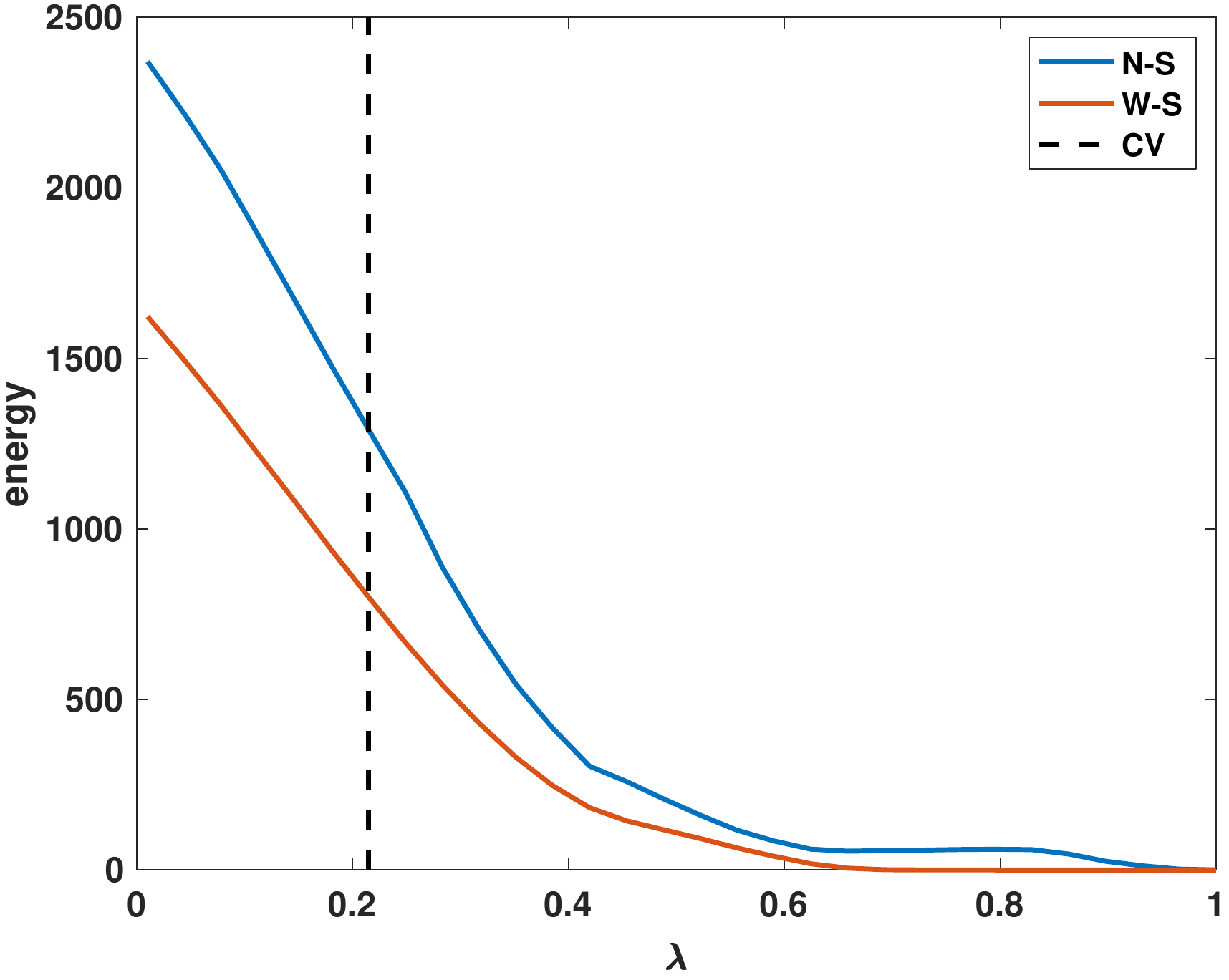}}
\caption{The quantities $\mathbb{E}\left[\mathrm{trace}\left(\Proj_{\hat{T}({\mathcal{D}})} \Proj_{{T^\star}^\perp}\right)\right]$ (in blue) and $\mathbb{E}\left[\mathrm{trace}\left(\Proj_{\texttt{avg}}\Proj_{{T^\star}^\perp}\right)\right]$ (in red) as a function of $\lambda$ for SNR = 1.6 and SNR = 0.8 in the synthetic matrix completion setup. The cross-validated choice of $\lambda$ is shown as the dotted black line. Here `N-S' denotes no subsampling and `W-S ' denotes with subsampling.}
\label{fig:kappa}
\end{figure}

\emph{Illustration: the value of averaging projection maps} -- Consider the stylized low-rank matrix completion problem described at the beginning of Section~\ref{section:stability}.  To support the intuition that the average projection matrix $\Proj_{\texttt{avg}}$ has reduced in energy in directions corresponding to ${T^\star}^\perp$ (i.e., the orthogonal complement of the population tangent space), we compare the quantities $\mathbb{E}\left[\mathrm{trace}\left(\Proj_{\texttt{avg}}\Proj_{{T^\star}^\perp}\right)\right]$ and $\mathbb{E}\left[\mathrm{trace}\left(\Proj_{\hat{T}({\mathcal{D}})}\Proj_{{T^\star}^\perp}\right)\right]$, where the expectation is computed over $100$ instances.  {Generically speaking, the operator $\mathcal{P}_{\texttt{avg}}$ is not a projection operator onto a tangent space and thus the quantity $\mathbb{E}\left[\mathrm{trace}\left(\Proj_{\texttt{avg}}\Proj_{{T^\star}^\perp}\right)\right]$ is not a valid false discovery, rather it evaluates the average false discovery over the subsampled models}. The second quantity, $\mathbb{E}\left[\mathrm{trace}\left(\Proj_{\hat{T}({\mathcal{D}})}\Proj_{{T^\star}^\perp}\right)\right]$, is based on employing the nuclear norm regularization procedure on the full set of observations.  The variance $\sigma$ is selected so that $\text{SNR} = \{0.8,1.6\}$. As is evident from Figure~\ref{fig:kappa}, $\mathbb{E}\left[\mathrm{trace}\left(\Proj_{\texttt{avg}}\Proj_{{T^\star}^\perp}\right)\right]$ is smaller than $\mathbb{E}\left[\mathrm{trace}\left(\Proj_{\hat{T}({\mathcal{D}})} \Proj_{{T^\star}^\perp}\right)\right]$ for the entire range of $\lambda$, with the gap being larger in the low SNR regime.  In other words,  averaging the subsampled tangent spaces reduces energy in the directions spanned by ${T^\star}^\perp$. 
% * <ataeb@caltech.edu> 2018-03-23T17:47:55.083Z:
%
% ^.

While the average projection aggregated over many subsamples appears to have less energy in ${T^\star}^\perp$, this operator is not a proper projection. Thus it still remains for us to identify a single tangent space as our estimate from $\mathcal{P}_{\texttt{avg}}$.  We formulate the following criterion to establish a measure of closeness between a single tangent space and the aggregate over subsamples:
\vspace{0.1in}
\begin{defn}[Stable Tangent Spaces] Suppose $\hat{T}$ is an estimator of a tangent space of a low-rank matrix, and suppose further that we are given a set of observations $\mathcal{D}$ and a corresponding collection of subsamples $\{\mathcal{D}_\ell\}_{\ell=1}^B$ with each $\mathcal{D}_\ell \subset \mathcal{D}$.  For a parameter $\alpha \in (0,1)$, the set of \emph{stable tangent spaces} is defined as
\begin{equation}
  \begin{aligned}
  \mathcal{T}_{\alpha} &\triangleq \Big\{ T ~ | ~  \sigma_{\text{min}}\left(\mathcal{P}_{T} \mathcal{P}_{\texttt{avg}}\mathcal{P}_{T}\right) \geq \alpha ~\mathrm{and}~ T \text{is a tangent space to a determinantal variety} \Big\}
  \end{aligned}
  \label{eqn:setT}
\end{equation}
where $\mathcal{P}_{\texttt{avg}}$ is computed based on Definition \ref{defn:avg_proj_low}.
\label{defn:low_rank}
\end{defn}

As the spectrum of $\mathcal{P}_{\texttt{avg}}$ lies in the range $[0,1]$, this is also the only meaningful range of values for $\alpha$.  The set $\mathcal{T}_\alpha$ consists of all those tangent spaces $T$ to a determinantal variety such that the Rayleigh quotient of every nonzero element of $T$ with respect to $\mathcal{P}_{\texttt{avg}}$ is at least $\alpha$.  To contrast again with variable selection, we note that both $\mathcal{P}_T$ and $\mathcal{P}_{\texttt{avg}}$ are diagonal matrices in that case (and thus simultaneously diagonalizable).  As a consequence, the set $\mathcal{T}_\alpha$ has a straightforward characterization for variable selection problems; it consists of subspaces spanned by any subset of standard basis vectors corresponding to variables that are selected as significant in at least an $\alpha$ fraction of the subsamples.

As averaging the tangent spaces obtained from the subsampled data reduces energy in the directions contained in ${T^\star}^\perp$, each element of $\mathcal{T}_\alpha$ is also far from being closely aligned with ${T^\star}^\perp$ (for large values of $\alpha$).  We build on this intuition by proving next that a tangent space estimator that selects any element of $\mathcal{T}_\alpha$ provides false discovery control at a level that is a function of $\alpha$.  In Section~\ref{section:algorithm} we describe efficient methods to choose an element of $\mathcal{T}_{\alpha}$. 

As a final remark, the ideas described here can be readily applied to subspace estimation problems. Specifically, we define the average projection operator $\Proj_\texttt{avg}^{\mathcal{C}}$ (analogous to \eqref{eqn:avg_proj}) as the average of projection matrices onto column-space estimates obtained from $n/2$ subsamples.  Then, the stable subspace set \eqref{eqn:setT} is modified to be the collection of subspaces $\mathcal{C} \in \mathbb{R}^{p}$ that satisfy the criterion $\sigma_\texttt{min}(\Proj_\mathcal{C}\Proj_{\texttt{avg}}^{\mathcal{C}}\Proj_\mathcal{C}) \geq \alpha$.

\subsection{{False Discovery Control of Stable Tangent Spaces: Theoretical Analysis}}
\label{section:theoretical}
\vspace{0.1in}
{\par}\emph{Setup}: Consider a low-rank matrix $L^\star \in \R^{p_1 \times p_2}$ with associated tangent space $T^\star$, and suppose we are given i.i.d. observations from a model parametrized by $L^\star$.  The objective is to obtain an accurate estimate of $T^\star$.  We intentionally keep our discussion broad so our results are relevant for a wide range of low-rank estimation problems, e.g., low-rank matrix completion, factor analysis, etc.  Let $\hat{T}$ denote a tangent space estimator that operates on samples drawn from the model parametrized by $L^\star$.  Let $\mathcal{D}(n)$ denote a dataset consisting of $n$ i.i.d observations from this model; we assume that $n$ is even and that we are given $B$ subsamples $\{\mathcal{D}_\ell\}_{\ell=1}^B$ via complementary partitions of $\mathcal{D}(n)$.

We present a general result that bounds the expected false discovery of stable tangent spaces under the sole assumption that the dataset provided consists of i.i.d. observations.  Under additional assumptions that take the form of ``better than random guessing'' and a geometric analog of exchangeability, we specialize our result to obtain a more refined bound that is similar in spirit to the bound of \cite{stability}. Finally, inspired by Theorem 1 of \cite{Samworth}, we also specialize our result to produce a bag-independent false discovery bound that is valid for any $B \geq 2$.  The results in this section extend naturally to settings in which one only seeks accurate estimates of the column-space of a matrix; for precise statements in that setting, see supplementary material Section~A.10.

{

Our theoretical findings are centered on the following intuition: for subspace stability selection to be effective, the tangent space estimates across subsamples should contain many directions around $T^\star$ (i.e., the signal component) and the remaining components (i.e., the noise) should be evenly spread over all the other directions.  Due to the smooth structure underlying low-rank matrices, there are ``many'' directions in which deviations about $T^\star$ can occur in a low-rank estimation procedure (a significant contrast to variable selection where the collection of tangent spaces is a discrete set); thus, the requirement on the noise portion of the estimates from the subsamples is a stringent one.  This situation is alleviated if the noise components in the subsamples are concentrated around ${T^\star}^\perp$, i.e., the tangent space estimates across subsamples contain directions that mostly lie close to $T^\star$ or ${T^\star}^\perp$.  Mathematically, this intuition can be quantified via \emph{commutators}.  The commutator between self-adjoint operators $A,B$ is denoted $\left[A,B\right] = AB-BA$, and this map evaluates how far away $A,B$ are from commuting with each other.  For projection operators $\Proj_{T_1},\Proj_{T_2}$ associated to subspaces $T_1$ and $T_2$, the singular values of $[\Proj_{T_1},\Proj_{T_2}]$ are $\pm\frac{1}{2}\sin(2\theta_i)$ where $\{\theta_i\}$ are the principal angles between $T_1$ and $T_2$ \citep{commutator}. Consequently, $\left\|\left[\Proj_{T_1},\Proj_{T_2}\right]\right\|_F^2 = \frac{1}{2}\sum_{i} \sin(2\theta_i)^2$ and $\left\|\left[\Proj_{T_1},\Proj_{T_2}\right]\right\|_2^2 = \frac{1}{4}\max_i\sin(2\theta_i)^2$. A small commutator between the tangent space estimates from subsamples and ${T^\star}^\perp$ ensures that the tangent space estimates consist of components that are closely aligned with $T^\star$ or with ${T^\star}^\perp$. (As a contrast, in variable selection the associated projection operators commute; in particular, $\theta_i \in \{0,\frac{\pi}{2}\}$ in variable selection.)\\[0.1in]

} 

\begin{theorem}
[False Discovery Control of Subspace Stability Selection]~Consider the setup described above.  Let $\hat{T}(\mathcal{D}_\ell)$ denote the tangent space estimates obtained from each of the subsamples, and let $\mathcal{P}_{\texttt{avg}}$ denote the associated average projection operator computed via \eqref{eqn:avg_proj} over $B$ complementary bags.  Fix any $\alpha \in (1/2,1)$ and let $T$ denote any selection of an element of the associated set $\mathcal{T}_\alpha$ of stable tangent spaces.  Then for any fixed orthonormal basis $\{M_i\}_{i = 1}^{\mathrm{dim}({T^\star}^\perp)}$ for ${T^\star}^\perp$, we have that:
\begin{eqnarray}
\mathbb{E}\left[\mathrm{trace}\left(\Proj_{T}\Proj_{{T^\star}^\perp}\right)\right] &\leq& F+\kappa_{\text{bag}}(\alpha)+{2({1-\alpha})}\mathbb{E}[\mathrm{dim}(T)], \label{eqn:main2}
\end{eqnarray}
\allowbreak
where for a basis-dependent bound take $F =\sum_{i = 1}^{\mathrm{dim}({T^\star}^\perp)} \mathbb{E}[\|\Proj_{\hat{T}(\mathcal{D}(n/2))}(M_i)\|_F]^2$ and $\kappa_{\text{bag}}(\alpha) =\sum_{i = 1}^{\text{dim}({T^\star}^\perp)}\tfrac{2}{B} \sum_{j=1}^{B/2}\mathbb{E}\allowbreak \max_{k\in\{0,1\}}\allowbreak\mathrm{trace}([\Proj_T, \Proj_{\hat{T}{(\mathcal{D}_{2j-k})}^\perp}] \times [\Proj_{\mathrm{span}(M_i)}, \Proj_{\hat{T}(\mathcal{D}_{2j-k})}])]$, and for a basis-independent bound take 
$F = \mathbb{E}[\mathrm{trace}(\allowbreak\Proj_{\hat{T}(\mathcal{D}(n/2))}\Proj_{{T^\star}^\perp})^{1/2}]^2$ and  $\kappa_{\text{bag}}(\alpha) =\tfrac{2}{B} \sum_{j=1}^{B/2}\mathbb{E}[ \max_{k\in\{0,1\}}\mathrm{trace}([\Proj_T, \Proj_{\hat{T}{(\mathcal{D}_{2j-k})}^\perp}] \times [\Proj_{{T^\star}^\perp}, \Proj_{\hat{T}(\mathcal{D}_{2j-k})}])]$. The expectations are with respect to randomness in the data and the set $\mathcal{D}(n/2)$ denotes $n/2$ i.i.d. observations drawn from the model parametrized by $L^\star$.    
\label{thm:main}
\end{theorem}
The proof of Theorem~\ref{thm:main} is presented in supplementary material Section A.1. The result states that the expected false discovery of a stable tangent space is bounded by a sum of three quantities. The first term $F$ characterizes the quality of the estimator employed on subsamples consisting of n/2 observations. The terms $\kappa_{\text{bag}}(\alpha)$ and ${2({1-\alpha})}\mathbb{E}[\mathrm{dim}(T)]$ are functions of the user-specified parameter $\alpha$, the number of bags $B$, and product of commutators. In Proposition 5, we show that $\alpha$ close to one leads to a small $\kappa_{\text{bag}}(\alpha)$, and thus, as expected, a smaller expected false discovery.  Further, one must select $\alpha > 1/2$ for \eqref{eqn:main2} to be non-vacuous as we always have that $\mathbb{E}\left[\mathrm{trace}(\Proj_T\Proj_{{T^\star}^\perp})\right] \leq \mathbb{E}[\mathrm{dim}(T)]$.\\[0.1in]
{\indent}\emph{Remark 1}: The quantities $\sum_{i = 1}^{\mathrm{dim}({T^\star}^\perp)} \mathbb{E}[\|\Proj_{\hat{T}(\mathcal{D}(n/2))}(M_i)\|_F]^2$ and $\mathbb{E}[\mathrm{trace}(\Proj_{\hat{T}(\mathcal{D}(n/2))}\Proj_{{T^\star}^\perp})^{1/2}]^2$ for $F$ highlight the role of bagging in reducing variance. For ease of exposition, we define $\beta \in \mathbb{R}^{\mathrm{dim}({T^\star}^\perp)}$ as $\beta_i \triangleq \|\Proj_{\hat{T}(\mathcal{D}(n/2))}(M_i)\|_F$, so that $\sum_{i = 1}^{\mathrm{dim}({T^\star}^\perp)} \mathbb{E}[\|\Proj_{\hat{T}(\mathcal{D}(n/2))}(M_i)\|_F]^2 = \mathrm{trace}(\mathbb{E}[\beta]\mathbb{E}[\beta]')$ and $\mathrm{trace}(\Proj_{\hat{T}(\mathcal{D}(n/2))}\Proj_{{T^\star}^\perp}) = \mathrm{trace}(\beta \beta')$.  Jensen's inequality yields $\mathbb{E}[\mathrm{trace}(\beta \beta')^{1/2}]^2 \leq \mathbb{E}[\mathrm{trace}(\beta \beta')]$, so that the improvement of bagging over just using a subsample $\mathcal{D}(n/2)$ once is given by $\text{var}\left(\mathrm{trace}(\beta \beta')^{1/2} \right)$.  Next, by appealing to the positive-definiteness of a covariance matrix, we have that $\mathrm{trace}(\mathbb{E}[\beta]\mathbb{E}[\beta]') \leq \mathbb{E}[\mathrm{trace}(\beta\beta')]$; in this case, the variance reduction is given by $\mathrm{trace}(\mathrm{cov}(\beta))$.  In both these cases, the variance is maximally reduced under conditions that follow from the Bhatia-Davis inequality.  Specifically, given a fixed $\mathbb{E}[\mathrm{trace}(\beta \beta')^{1/2}]$, the Bhatia–Davis inequality states that $\text{var}(\mathrm{trace}(\beta \beta')^{1/2})$ is enhanced when the distribution of $\mathrm{trace}(\beta \beta')^{1/2}$ concentrates around $0$ and $\sqrt{\text{dim}({T^\star}^\perp)}$ (i.e., most discoveries are either true or false). Similarly, given a fixed $\mathbb{E}[\beta]$, $\mathrm{trace}(\text{cov}(\beta))$ is enhanced when the distribution of each $\beta_i$ concentrates around $0$ or $1$ (i.e., the estimate $\hat{T}(\mathcal{D}(n/2))$ is mostly aligned with or orthogonal to each $M_i \in {T^\star}^\perp$). Such concentration of $\beta_i$ can be precisely translated to the commutators $\|\mathbb{E}[[\Proj_{\hat{T}(\mathcal{D}(n/2)},\Proj_{\text{span}(M_i)}]]\|_F$ being small, which is exploited in Proposition 6 to bound $F$. In Section~\ref{section:experimental}, we use this intuition to describe synthetic experiments that illustrate the improvement (in terms of expected false discovery) of a stable tangent space over using the original estimator without subsampling. \\[0.1in] 
{\indent}\emph{Remark 2}: The terms $\sum_{i = 1}^{\mathrm{dim}({T^\star}^\perp)} \mathbb{E}[\|\Proj_{\hat{T}(\mathcal{D}(n/2))}(M_i)\|_F]^2$ and $\mathbb{E}[\mathrm{trace}(\Proj_{\hat{T}(\mathcal{D}(n/2))}\Proj_{{T^\star}^\perp})^{1/2}]^2$ for $F$ are incomparable in general.  The term $\mathbb{E}[\|\Proj_{\hat{T}(\mathcal{D}(n/2))}(M_i)\|_F]^2$ depends on the specific choice of basis, and it is useful in scenarios in which a particular choice of $\{M_i\}_{i = 1}^{\mathrm{dim}({T^\star}^\perp)}$ is natural, such as in variable selection problems in which the standard basis has a clear interpretation. On the other hand, $\mathbb{E}[\mathrm{trace}(\Proj_{\hat{T}(\mathcal{D}(n/2))}\Proj_{{T^\star}^\perp})^{1/2}]^2$ is basis-independent and is more useful in problem settings in which no particular choice of a basis is natural.\\[0.1in]
{\indent}\emph{Remark 3}: The quantity $\kappa_\text{bag}(\alpha)$ depends on commutators of projection operators associated to various tangent spaces.  As such, this quantity is closer to zero if the principal angles between $T$ and $\hat{T}(\mathcal{D}(n/2))^\perp$ and between ${T^\star}^\perp$ and $\hat{T}(\mathcal{D}(n/2))$ are close to $0$ or $\frac{\pi}{2}$.  Notice that in variable selection problems all the underlying projection matrices commute, and as a result we have that $\kappa_{\text{bag}}(\alpha) = 0$. In this sense, $\kappa_\text{bag}(\alpha)$ highlights the distinction between low-rank estimation and variable selection. \\[0.1in]
\indent\emph{Remark 4}: Building on the previous remark, the commutativity property in the variable selection setting enables additional simplifications of our bounds.  Although the bound \eqref{eqn:main2} is valid for variable selection, exploiting the fact that the projection matrices commute in that case and with the choice of the standard basis for $\{M_i\}_{i = 1}^{\mathrm{dim}({T^\star}^\perp)}$, we obtain additional simplifications. Specifically, letting $\{M_i\}_{i = 1}^{\mathrm{dim}({T^\star}^\perp)}$ be the subset of the standard basis that lies in ${T^\star}^\perp$ and noting that $\kappa_\text{bag}$ vanishes, one can modify the proof of Theorem~\ref{thm:main} to obtain the following bound:
\begin{eqnarray}
\mathbb{E}\left[\mathrm{trace}\left(\Proj_{T}\Proj_{{T^\star}^\perp}\right)\right] &\leq& \sum_{i=1}^{\mathrm{dim}({T^\star}^\perp)}\frac{\mathbb{E}\left[\left\|\Proj_{\hat{T}(\mathcal{D}({n/2}))}(M_i)\right\|_F\right]^2}{2\alpha-1}
=\sum_{i = 1}^{\mathrm{dim}({T^\star}^\perp)}\frac{\mathbb{P}[i'\text{th null selected by }\hat{T}(\mathcal{D}(n/2))]}{2\alpha-1}.
\label{eqn:alternate_bound}
\end{eqnarray}
This improved bound follows from a careful accounting of the first and third terms in \eqref{eqn:main2}; see the supplementary material Section~A.3. The equality here is a consequence of the observations that $\Proj_{\hat{T}(\mathcal{D}(n/2))}$ is a diagonal projection matrix and that each $M_i$ is an element of the standard basis.  Thus, we recover the interpretation that the overall expected false discovery for the special case of variable selection can be
bounded in terms of the probability that the procedure $\hat{T}$ selects null variables on subsamples. The final expression \eqref{eqn:alternate_bound} matches Theorem 1 of \cite{Samworth} (in particular, it holds for any $B \geq 2$). As a final comparison between the low-rank estimation and variable selection settings, the dependence on $\alpha$ in \eqref{eqn:alternate_bound} is multiplicative as opposed to additive as in \eqref{eqn:main2}.  In particular, in the low-rank case even if the estimator $\hat{T}$ performs exceedingly well on the subsamples, the expected false discovery may still be large depending on the choice of $\alpha$ and $\text{dim}({T^\star}^\perp)$; in contrast, for variable selection if the estimator $\hat{T}$ performs exceedingly well on the subsamples, the expected false discovery is small provided $\alpha$ is chosen to be close to one.  This distinction is fundamental to the geometry underlying the sparse and determinantal varieties.  Specifically, in the low-rank case even if $\mathcal{P}_{\texttt{avg}} \approx \mathcal{P}_{T^\star}$ the set of stable tangent spaces $\mathcal{T}_\alpha$ necessarily includes many tangent spaces that are near the population tangent space $T^\star$ but are not perfectly aligned with it.  This is due to the fact that the collection of row/column spaces forms a Grassmannian manifold rather than a finite/discrete set.  On the other hand, if $\mathcal{P}_{\texttt{avg}} \approx \mathcal{P}_{T^\star}$ in variable selection, the only elements of the set of stable tangent spaces (for large $\alpha$) are those corresponding to subsets of the true significant variables.\\[0.1in]
Next we provide a bound on both the basis-independent and basis-dependent versions of $\kappa_\text{bag}(\alpha)$, which leads to a bag-independent bound on the expected false discover by combining with Theorem~\ref{thm:main}:\\[0.1in]
\begin{proposition} [Bounding $\kappa_\text{bag}(\alpha)$ and a Bag Independent Result] Consider the setup of Theorem~\ref{thm:main}. Then the following bound holds for both the basis-independent and basis-dependent versions of $\kappa_\text{bag}(\alpha)$: $\kappa_\text{bag}(\alpha) \leq 2\sqrt{1-\alpha}\mathbb{E}[\text{dim}(T)]$.  Further, letting the average number of discoveries from $n/2$ observations be denoted by $q := \mathbb{E}[\mathrm{dim}(\hat{T}(\mathcal{D}(n/2)))]$, we also have that $\mathbb{E}[\text{dim}(T)] \leq \frac{q}{\alpha}$.  Thus, we obtain the following false discovery bound for any $B \geq 2$:
\begin{eqnarray}
\mathbb{E}\left[\mathrm{trace}\left(\Proj_{T}\Proj_{{T^\star}^\perp}\right)\right] ~~\leq~~ F+ 2({1-\alpha}+\sqrt{1-\alpha})\mathbb{E}[\text{dim}(T)] ~~\leq~~ F+\frac{2q}{\alpha}({1-\alpha}+\sqrt{1-\alpha})
\label{eqn:bag_dependent}
\end{eqnarray}
for either the basis-dependent or the basis-independent form of $F$ from Theorem~\ref{thm:main}. \label{prop:bagindep}
\end{proposition}
\vspace{-0.03in}
\noindent\emph{Remark 5}: The proof of this result is presented in supplementary material A.2. This bound highlights the role of $\alpha$, where $\kappa_\text{bag}(\alpha)$ becomes smaller as $\alpha$ is chosen close to one. The bag-independent bound \eqref{eqn:bag_dependent} on expected false discovery of a stable tangent space holds for any $B \geq 2$, and thus can be looser than \eqref{eqn:main2}. In particular, the bound \eqref{eqn:bag_dependent} is relevant for $\alpha \gtrapprox 0.9$ (as the bound otherwise exceeds $q$), which is more stringent than the condition $\alpha > \frac{1}{2}$ in Theorem~\ref{thm:main}. Despite the more restrictive range of values for $\alpha$, these bag independent results may nonetheless have utility in regimes in which the signal strength is high so that larger values of $\alpha$ may be considered.

Next we describe a more refined false discovery bound under additional assumptions on the estimator $\hat{T} = (\hat{C},\hat{R})$:
\begin{eqnarray}
\begin{gathered}
\text{Assumption 1:}~~  \frac{\mathbb{E}\left[\text{trace}\left(\Proj_{{T^\star}^\perp}\Proj_{\hat{T}(\mathcal{D}({n/2}))}\right)\right]}{\text{dim}({T^\star}^\perp)} \leq \frac{\mathbb{E}\left[\text{trace}\left(\Proj_{{T^\star}}\Proj_{\hat{T}(\mathcal{D}({n/2}))}\right)\right]}{\text{dim}({T^\star})} \\
\text{Assumption 2:} ~~\text{distribution of }  \|\Proj_{\hat{\mathcal{T}}(\mathcal{D}(n/2))}(M)\|_F \text{ is the same for all rank-1 }M\in {T^\star}^\perp, \|M\|_F = 1 
\end{gathered}
\label{eqn:exchangeable}
\end{eqnarray} 
In words, Assumption 1 states that the estimator's normalized power is greater than its normalized expected false discovery and Assumption 2 states that the energy of a
any normalized rank-1 element in ${T^\star}^\perp$ onto tangent spaces obtained from subsamples consisting of $n/2$ observations is identically distributed. In the case of variable selection, Assumption 1 reduces precisely to the ``better than random guessing" assumption employed by \cite{stability}, namely that the probability that the procedure $\hat{T}$ selects a null variable when employed on the subsamples is better than random guessing.  As a second condition, \cite{stability} require that the random variables $\{\mathbb{I}_{k \in \hat{T}(\mathcal{D}(n/2))}\}$ are exchangeable. This assumption implies that the distribution of $\mathbb{I}_{k \in \hat{T}(\mathcal{D}(n/2))}$ is the same for all null $k$.  Our Assumption 2 when specialized to variable selection reduces to the weaker requirement that each of the random variables $\mathbb{I}_{k \in \hat{T}(\mathcal{D}(n/2))}$ has the same distribution.  In supplementary material Section~A.4, we show that Assumptions 1 and 2 in \eqref{eqn:exchangeable} are satisfied by some natural ensembles and estimators in low-rank estimation problems.  We prove next a bound on the expected false discovery under these additional assumptions.\\[0.1in]

\begin{proposition} [Refined False Discovery Control] Consider the setup of Theorem~\ref{thm:main}.  Suppose additionally that Assumptions 1 and 2 in \eqref{eqn:exchangeable} are satisfied. For any $M \in {T^\star}^\perp$ with $\text{rank}(M)=1, \|M\|_F = 1$, the false discovery of a stable tangent space $T$ is bounded by:
\begin{eqnarray}
\mathbb{E}\left[\mathrm{trace}\left(\Proj_{T}\Proj_{{T^\star}^\perp}\right)\right] \leq \frac{q^2}{p_1p_2}+ f\left(\kappa_\text{indiv}\right) + \frac{2q}{\alpha}(1-\alpha+\sqrt{1-\alpha}),
\label{eqn:gammabound}
\end{eqnarray}
where $\kappa_\text{indiv} := \mathbb{E}\left[\|[\Proj_{\mathrm{span}(M)},\Proj_{\hat{T}(\mathcal{D}(n/2))}]\|_F\right]$ and $f(\kappa_\text{indiv}) = p_1p_2\kappa_\text{indiv}^2+2q\kappa_\text{indiv}$.
\label{prop:gamma}
\end{proposition}

\emph{Remark 6}: The proof of this proposition can be found in supplementary material Section~A.5. It proceeds by showing that in the bag-dependent setting (although the specific choice does not matter due to Assumption 2), $F \leq \frac{q^2}{p_1p_2}+f\left(\kappa_\text{indiv}\right)$ and employs the bounds $\kappa_{\text{bag}} \leq 2\sqrt{1-\alpha}\mathbb{E}[\text{dim}(T)]$ and  $\mathbb{E}[\text{dim}(T)] \leq \frac{q}{\alpha}$ (from Proposition~\ref{prop:bagindep}).  Consider the term $\frac{q^2}{p_1p_2}$ in this result, where $q$ can be approximated by $q \approx \mathrm{trace}(\Proj_{\texttt{avg}})$. Suppose typical outputs of the estimates obtained from subsamples $\{\mathcal{D}_\ell\}_{\ell = 1}^{B}$ have rank $k$ which is far smaller than the ambient dimensions, i.e. $k \ll \min\{p_1,p_2\}$, yielding $\mathrm{trace}(\Proj_{\texttt{avg}}) = \mathcal{O}(k(p_1+p_2))$; as a result, $\frac{q^2}{p_1p_2}$ is much smaller than $q$. The second term is an increasing function of the commutator-dependent quantity $\kappa_\mathrm{indiv}$.  To bound $\kappa_\mathrm{indiv}$ we note that it suffices to consider a single $M \in {T^\star}^\perp$ with $\mathrm{rank}(M)=1,\|M\|_F = 1$.  A natural data-driven heuristic to obtain such an $M$ is to consider a rank-one matrix that is ``least-aligned'' with $\Proj_{\mathrm{avg}}$, i.e., in some sense choosing the opposite of a stable tangent space.  Concretely, letting $u,v$ be the singular vectors corresponding to the smallest singular values of $\Proj_\texttt{avg}^\mathcal{C}, \Proj_\texttt{avg}^\mathcal{R}$, respectively, we propose setting $\tilde{M} = uv'$.  This choice can be justified theoretically provided the estimator $\hat{T}(\mathcal{D}(n/2))$ has good power; see supplementary material Section A.6. We then obtain the following data-driven approximation $\kappa_{\mathrm{indiv}} = \frac{1}{B}\sum_{\ell = 1}^B \|[\Proj_{\hat{T}(\mathcal{D}_\ell)}, \Proj_{\mathrm{span}(\tilde{M})}]\|_F$.  Finally, the third term can be controlled by choosing $\alpha$ sufficiently close to $1$.\\[0.1in]
\emph{Remark 7}: For the case of variable selection, $\kappa_\text{indiv} = 0$, so that $F \leq \frac{q^2}{\text{total variables}}$. Plugging this into \eqref{eqn:alternate_bound}, we obtain the bound on the expected false discovery of $\frac{\mathbb{E}[\# \text{discoveries in } \hat{T}(\mathcal{D}(n/2))]^2}{2(1-\alpha)(\# \text{total variables})}$. This bound was obtained by \cite{Samworth} as a consequence of their Theorem 1 and it holds for any $B \geq 2$ (an identical bound was also obtained by \cite{stability}, although that result requires averaging over all subsamples).

\subsection{Subspace Stability Selection Algorithm}
\label{section:algorithm}
\vspace{0.1in}
As described in the previous subsection, every tangent space in $\mathcal{T}_\alpha$ provides control on the expected false discovery.  The goal then is to select an element of $\mathcal{T}_\alpha$ to optimize power.  A natural approach to achieve this objective is to choose a tangent space of largest dimension from $\mathcal{T}_\alpha$ to maximize the total discovery.

Consider the following optimization problem for each $r = 1,\dots,\min\{p_1,p_2\}$:
\begin{eqnarray}
T_{\texttt{OPT}}(r) = \argmax_{T \text{ tangent space to a point in } \mathcal{V}_{\text{low-rank}}(r)}  ~~~ \sigma_{\text{min}}\left(\Proj_{T}\Proj_{\texttt{avg}}\Proj_T\right).\label{eqn:optimal}
\end{eqnarray}
A conceptually appealing approach to select an optimal tangent space is via the following optimization problem:
\begin{eqnarray}
T_{\texttt{OPT}} \in \argmax_{T \in T_{\texttt{OPT}}(r) \cap \mathcal{T}_\alpha} ~~~ r,
\label{eqn:optoverrank}
\end{eqnarray}
{where by construction, the set $T_{\texttt{OPT}}(r) \cap \mathcal{T}_\alpha$ is non-empty if $\mathcal{T}_\alpha$ is a non-empty set}. In the case of variable selection, this procedure would result in the selection of all those variables that are estimated as being significant in at least an $\alpha$ fraction of the bags, which is in agreement with the procedure of \cite{stability}.  In our setting of low-rank estimation, however, we are not aware of a computationally tractable approach to solve the problem \eqref{eqn:optimal}.  The main source of difficulty lies in the geometry underlying the collection of tangent spaces to determinantal varieties.  In particular, solving \eqref{eqn:optimal} in the case of variable selection is easy because the operators $\Proj_T, \Proj_{\texttt{avg}}$ are both diagonal (and hence trivially simultaneously diagonalizable) in that case; as a result, one can decompose \eqref{eqn:optimal} into a set of one-variable problems.  In contrast, the operators $\Proj_T, \Proj_{\texttt{avg}}$ are not simultaneously diagonalizable in the low-rank case, and consequently there doesn't appear to be any clean separability in \eqref{eqn:optimal} in general with determinantal varieties.

We describe next a heuristic to approximate \eqref{eqn:optimal}.  Our approximation entails computing optimal row-space and column-space approximations from the bags separately rather than in a combined fashion via tangent spaces.  Specifically, suppose $\{(\hat{\mathcal{C}}({\mathcal{D}_\ell}), \hat{\mathcal{R}}({\mathcal{D}_\ell}))\}_{\ell= 1}^{B}$ denote the row/column space estimates from $B$ subsamples $\{\mathcal{D}_\ell\}_{\ell= 1}^{B} \subset \mathcal{D}$ of the data.  We average the projection operators associated to these row/column spaces:
\begin{equation}
\Proj^{\mathcal{C}}_{\texttt{avg}} = \frac{1}{B} \sum_{\ell = 1}^{B} \Proj_{\hat{\mathcal{C}}({\mathcal{D}_\ell})}, ~~~~~ \Proj^{\mathcal{R}}_{\texttt{avg}} = \frac{1}{B} \sum_{\ell = 1}^{B} \Proj_{\hat{\mathcal{R}}({\mathcal{D}_\ell})} \label{eqn:rowcolavg}
\end{equation}
Note that the average operator $\Proj_{\texttt{avg}}$ based on estimates from subsamples of tangent spaces to determinantal varieties is a self-adjoint map on the space $\R^{p_1 \times p_2}$, while the averages $\Proj^{\mathcal{C}}_{\texttt{avg}}$ and $\Proj^{\mathcal{R}}_{\texttt{avg}}$ are self-adjoint maps on the spaces $\R^{p_1}$ and $\R^{p_2}$, respectively.  Based on these separate column-space and row-space averages, we approximate \eqref{eqn:optimal} as follows:
\begin{eqnarray}
T_{\texttt{approx}}(r) = T\left(\argmax_{\substack{\mathcal{C} \subset \R^{p_1} \text{ subspace of dimension } r}} \sigma_{\text{min}}\left(\Proj_{\mathcal{C}}\Proj^{\mathcal{C}}_{\texttt{avg}}\Proj_\mathcal{C}\right), \argmax_{\substack{\mathcal{R} \subset \R^{p_2} \text{ subspace of dimension } r}} \sigma_{\text{min}}\left(\Proj_{\mathcal{R}}\Proj^{\mathcal{R}}_{\texttt{avg}}\Proj_\mathcal{R}\right) \right). \label{eqn:optimalapprox}
\end{eqnarray}
The advantage of this latter formulation is that the inner-optimization problems of identifying the best row-space and column-space approximations of rank $r$ can be computed tractably.  In particular, the optimal column-space (resp. row-space) approximation of dimension $r$ is equal to the span of the eigenvectors corresponding to the $r$ largest eigenvalues of $\Proj^{\mathcal{C}}_{\texttt{avg}} $ (resp. $\Proj^{\mathcal{R}}_{\texttt{avg}}$).  We have that $\sigma_{\text{min}}\left(\Proj_{T_{\texttt{approx}}(r)} \Proj_{\texttt{avg}} \Proj_{T_{\texttt{approx}}(r)} \right) \leq \sigma_{\text{min}}\left(\Proj_{T_{\texttt{OPT}}(r)} \Proj_{\texttt{avg}} \Proj_{T_{\texttt{OPT}}(r)} \right)$ and we expect this inequality to be strict in general, even though tangent spaces to determinantal varieties are in one-to-one correspondence with the underlying row/column spaces.  To see why this is the case, consider a column-space and row-space pair $(\mathcal{C},\mathcal{R}) \subset \R^{p_1} \times \R^{p_2}$, with $\mathrm{dim}(\mathcal{C}) = \mathrm{dim}(\mathcal{R}) = r$.  The collection of matrices $\mathcal{M}_\mathcal{C} \subseteq \R^{p_1 \times p_2}$ with column-space contained in $\mathcal{C}$ has dimension $p_2 r$ and the collection of matrices $\mathcal{M}_\mathcal{R} \subseteq \R^{p_1 \times p_2}$ with row-space contained in $\mathcal{R}$ has dimension $p_1 r$.  However, the tangent space $T(\mathcal{C},\mathcal{R}) \subset \R^{p_1 \times p_2}$, which is the sum of $\mathcal{M}_\mathcal{C}$ and $\mathcal{M}_\mathcal{R}$ has dimension $p_1 r + p_2 r - r^2$.  In other words, the spaces $\mathcal{M}_\mathcal{C}, \mathcal{M}_\mathcal{R}$ do not have a transverse intersection (i.e. $\mathcal{M}_\mathcal{C} \cap  \mathcal{M}_\mathcal{R} \neq \{0\}$), and therefore optimal tangent-space estimation does not appear to be decoupled into (separate) optimal column-space estimation and optimal row-space estimation.  Although this heuristic is only an approximation, it does yield good performance in practice, as described in the illustrations in the next subsection as well as in the experiments with real data in the Section~\ref{section:experimental}.  Further, our final estimate of a tangent space still involves the solution of \eqref{eqn:optoverrank} using the approximation \eqref{eqn:optimalapprox} instead of \eqref{eqn:optimal}.  Consequently, we continue to retain our guarantees from Section~\ref{section:theoretical} on false discovery control.  The full procedure is presented in Algorithm~\ref{algo:1}.

{The tuning parameter $\alpha \in [0,1]$ in Algorithm~\ref{algo:1} plays an important role in how much signal is selected by subspace stability selection. In our experience, the output of subspace stability selection is rather robust to $\alpha$ in moderate to high SNR settings. As a result, in all our experiments we select $\alpha$ to equal $0.70$. For detailed analysis on the sensitivity to $\alpha$ see supplementary material Section A.7.}

{\par\bf \emph{Computational Cost of Algorithm 1}} -- We do not account for the cost of obtaining the row/column space estimates $\{(\hat{\mathcal{C}}({\mathcal{D}_\ell}), \hat{\mathcal{R}}({\mathcal{D}_\ell}))\}_{\ell= 1}^{B}$ on each subsample in Step $2$, and focus exclusively on the cost of combining these estimates via Steps $3-5$.  In Step $3$, the computational complexity of computing the averages $\mathcal{P}^{\mathcal{R}}_{\texttt{avg}}, \mathcal{P}^{\mathcal{C}}_{\texttt{avg}}$ requires $\mathcal{O}(B \max\{p_1,p_2\}^2)$ operations and computing the average $\mathcal{P}_{\texttt{avg}}$ requires $\mathcal{O}(B p_1^2p_2^2)$ operations.  Step $4$ entails the computation of two singular value decompositions of matrices of size $p_1 \times p_1$ and $p_2 \times p_2$, which leads to a cost of $\mathcal{O}(\max\{p_1,p_2\}^3)$ operations.  Finally, in Step $5$, to check membership in $\mathcal{T}_\alpha$ we multiply three maps of size $p_1 p_2 \times p_1 p_2$ and compute the singular value decomposition of the result, which requires a total of $\mathcal{O}(p_1^3p_2^3)$ operations.  Thus, the computational cost of Algorithm $1$ to aggregate estimates produced by $B$ bags is $\mathcal{O}(\max\{B p_1^2, B p_2^2, B p_1^2 p_2^2, p_1^3, p_2^3, p_1^3 p_2^3\})$.

\FloatBarrier
\begin{algorithm}
\caption{Subspace Stability Selection Algorithm}
\begin{algorithmic}[1]
\vspace{0.1in}
\STATE {\bf Input}: A set of observations $\mathcal{D}$; a collection  of subsamples $\{\mathcal{D}_\ell\}_{\ell= 1}^{B} \subset \mathcal{D}$; a row/column space (equivalently, tangent space) estimation procedure $(\hat{\mathcal{C}},\hat{\mathcal{R}})$; a parameter ${\alpha} \in (0,1)$. \\
\vspace{.04in}
\STATE{\bf Obtain Tangent Space Estimates}: For each bag $\{\mathcal{D}_\ell , \ell = 1,2,\dots,B\}$, obtain row/column space estimates $\{(\hat{\mathcal{C}}({\mathcal{D}_\ell}), \hat{\mathcal{R}}({\mathcal{D}_\ell}))\}_{\ell= 1}^{B}$ and set $\hat{T}(\mathcal{D}_\ell) = T(\hat{\mathcal{C}}({\mathcal{D}_\ell}), \hat{\mathcal{R}}({\mathcal{D}_\ell}))$.
\vspace{0.04in}
\STATE{\bf Compute Average Projection Operators}: Compute the average tangent space projection operator $\Proj_{\texttt{avg}}$ according to \eqref{eqn:avg_proj} and the average row/column space projection operators $\Proj_{\texttt{avg}}^{\mathcal{R}}, \Proj_{\texttt{avg}}^{\mathcal{C}}$ according to \eqref{eqn:rowcolavg}.
\vspace{.04in}
\STATE{\bf Compute Optimal Row/Column Space Approximations}: Compute ordered singular vectors $\{u_1,u_2,\dots,u_{p_1}\} \subset \R^{p_1}$ and $\{v_1,v_2,\dots,v_{p_2}\} \subset \R^{p_2}$ of $\Proj^{\mathcal{C}}_\texttt{avg}$ and $\Proj^{\mathcal{R}}_\texttt{avg}$, respectively.  For each $r = 1,\dots,\min\{p_1,p_2\}$, set $\mathcal{C}^\star(r) = \text{span}(u_1,\dots,u_r)$ and $\mathcal{R}^\star(r) = \text{span}(v_1,\dots,v_r)$.
\vspace{0.04in}
\STATE{\bf Tangent Space Selection via \eqref{eqn:optoverrank}}: Let $r_{\texttt{S3}}$ denote the largest $r$ such that $T(\mathcal{C}^\star(r),\mathcal{R}^\star(r)) \in \mathcal{T}_\alpha$.
\vspace{0.04in}
\STATE{\bf Output}: Tangent space $T_{\texttt{S3}} = T(\mathcal{C}^\star(r_{\texttt{S3}}),\mathcal{R}^\star(r_{\texttt{S3}}))$. 
\end{algorithmic} \label{algo:1}
\end{algorithm}
\FloatBarrier

Although the scaling of Algorithm $1$ is polynomial in the size of the inputs, when either $p_1$ or $p_2$ is large the overall cost due to terms such as $p_1^3 p_2^3$ may be prohibitive.  In particular, the reason for the expensive terms $B p_1^2 p_2^2$ and $p_1^3 p_2^3$ in the final expression is due to computations involving projection maps onto tangent spaces (which belong to $\R^{p_1p_2}$).  We describe next a modification of Algorithm $1$ so that the resulting procedure only consists of computations involving projection maps onto row and column spaces (which belong to $\R^{p_2}$ and $\R^{p_1}$ respectively).

{\bf \emph{Modification of Algorithm $1$ and Associated Cost}} -- The inputs to this modified procedure are the same as those of the original procedure.  We modify Step $3$ of Algorithm $1$ by only computing the average row/column space projection maps $\mathcal{P}^{\mathcal{R}}_{\texttt{avg}}, \mathcal{P}^{\mathcal{C}}_{\texttt{avg}}$.  Let $\mathcal{P}^{\mathcal{C}}_{\texttt{avg}} = U \Gamma U'$ and let $\mathcal{P}^{\mathcal{R}}_{\texttt{avg}} = V \Delta V'$ be the singular value decomposition computations of Step $4$.  We modify Step $5$ of Algorithm $1$ to choose the largest $r'_{\texttt{S3}}$ so that $\Gamma_{r'_{\texttt{S3}},r'_{\texttt{S3}}} \geq \alpha$ and $\Delta_{r'_{\texttt{S3}},r'_{\texttt{S3}}} \geq \alpha$.  One can check that the cost associated to this modified procedure is $\mathcal{O}(\max\{B p_1^2, B p_2^2, p_1^3, p_2^3\})$.

This modified method has the property that the row and column spaces are individually well-aligned with the corresponding averages from the subsamples; the following result shows that the resulting tangent space belongs to a set of stable tangent spaces:
\vspace{0.1in}
\begin{proposition}[Modified Algorithm $1$ Satisfies Subspace Stability Selection Criterion]
Let $T_{\texttt{S3-modified}}$ be the output of the modified Algorithm 1 with input parameter $\alpha$. Then, $T_{\texttt{S3-modified}} \in \mathcal{T}_{1-4({1-{\alpha}})}$.
\label{lemma:algo12}
\end{proposition}
Proposition~\ref{lemma:algo12} guarantees that our modification of Algorithm 1 continues to provide false discovery control.  We use this modified approach in some of our larger experiments in Section~\ref{section:experimental}. The proof of this proposition can be found in supplementary material Section~A.8.

Finally we remark that in subspace estimation problems (see Section 2.1), the subspace stability selection can be readily employed to find a stable tangent space. In particular, recall from Section~\ref{section:subspace_stability} that the stability selection criterion \eqref{eqn:setT} reduces to finding $\mathcal{C}$ such that $\sigma_\texttt{min}\left(\Proj_{\mathcal{C}}\Proj_\texttt{avg}^{\mathcal{C}} \Proj_{\mathcal{C}}\right) \geq \alpha$. Naturally, a projection operator $\Proj_{\mathcal{C}}$ that satisfies the criterion above can be obtained via singular-value thresholding. Furthermore, this subspace estimate is optimal according to \eqref{eqn:optoverrank}.

\subsection{Further Illustrations}
\vspace{0.1in}
In the remainder of this section, we explore various facets of Algorithm 1 via illustrations on the synthetic matrix completion problem setup described at the beginning of Section~\ref{section:stability}.   For further demonstrations of the utility of subspace stability selection with real data, we refer the reader to the experiments of Section 4. 

\begin{figure}[ht!]
\centering
\subfigure[$SNR = 50$]{
\includegraphics[scale = 0.2]{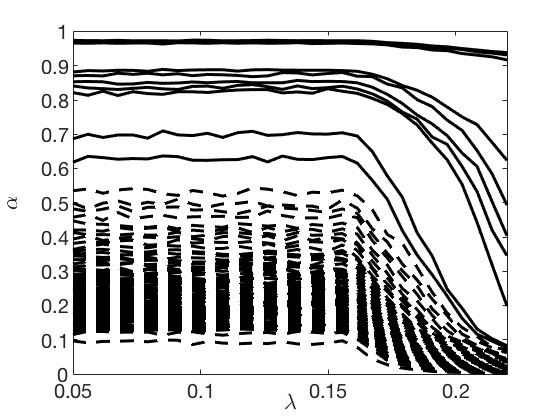}}
\subfigure[$SNR  = 1.2 $]{
\includegraphics[scale = 0.2]{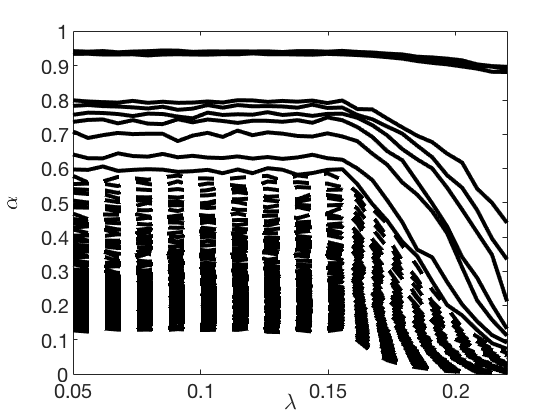}}
\subfigure[$SNR  = 0.8 $]{
\includegraphics[scale = 0.2]{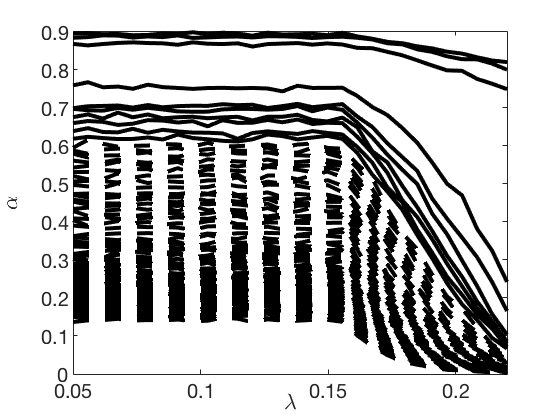}}
\subfigure[$SNR  = 0.4 $]{
\includegraphics[scale = 0.2]{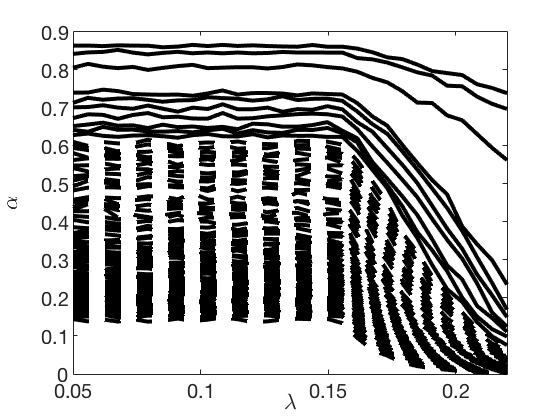}}
\caption{Relationship between $r_\texttt{s3}$ and $\alpha$ in Algorithm 1 for a large range of $\lambda$ and SNR $= \{0.4, 0.8, 1.2, 50\}$.}
\label{fig:alpha_curve}
\end{figure}

{\par \emph{Illustration : $\alpha$ vs. $r_{\texttt{S3}}$}} -- The threshold parameter $\alpha$ determines the eventual optimal rank $r_{\texttt{S3}}$, with larger values of $\alpha$ yielding a smaller $r_{\texttt{S3}}$.  To better understand this relationship, we  plot in Figure~\ref{fig:alpha_curve} $\sigma_{\texttt{min}}(\Proj_{{T}_{\texttt{S3}}}\Proj_{\texttt{avg}}\Proj_{{T}_{\texttt{S3}}})$ as a function of $r_{\texttt{S3}}$ for a large range of values of the regularization parameter $\lambda$ and SNR $= \{0.4,0.8,1.2,50\}$.  Each curve in the different plots corresponds to a particular value of $r_{\texttt{S3}}$, with the solid curves representing $r_{\texttt{S3}} = 1,\dots, 10$ and the dotted curves representing $r_{\texttt{S3}} = 11,\dots,70$.  As smaller values of $r_{\texttt{S3}}$ lead to larger values of $\sigma_{\texttt{min}}(\Proj_{{T}_{\texttt{S3}}}\Proj_{\texttt{avg}}\Proj_{{T}_{\texttt{S3}}})$, the curves are ordered such that the top curve corresponds to $r_{\texttt{S3}} = 1$ and the bottom curve corresponds to $r_{\texttt{S3}} = 70$.  We first observe that for a fixed $r_{\texttt{S3}}$, the associated curve is generally decreasing as a function of $\lambda$.  For large values of $\lambda$, both signal and noise are substantially reduced due to a significant amount of regularization.  Conversely, for small values of $\lambda$, both signal and noise are present to a greater degree in the estimates on each subsample; however, the averaging procedure reduces the effect of noise, which results in high-quality aggregated estimates for smaller values of $\lambda$.  Next, we observe that the curves indexed by $r_{\texttt{S3}}$ cluster in the high SNR regime, with the first three corresponding to $r_{\texttt{S3}} = 1,2,3$, the next five corresponding to $r_{\texttt{S3}} = 4,\dots,8$, the next two corresponding to $r_{\texttt{S3}} = 9,10$, and finally the remaining curves corresponding to $r_{\texttt{S3}} > 10$.  This phenomenon is due to the clustering of the singular values of the underlying population $L^\star$.  On the other hand, for low values of SNR, the clustering is less pronounced as the components of $L^\star$ with small singular values are overwhelmed by noise.

\begin{figure}[thbp]
\centering
\subfigure[SNR = 1.6]{
\includegraphics[scale = 0.35]{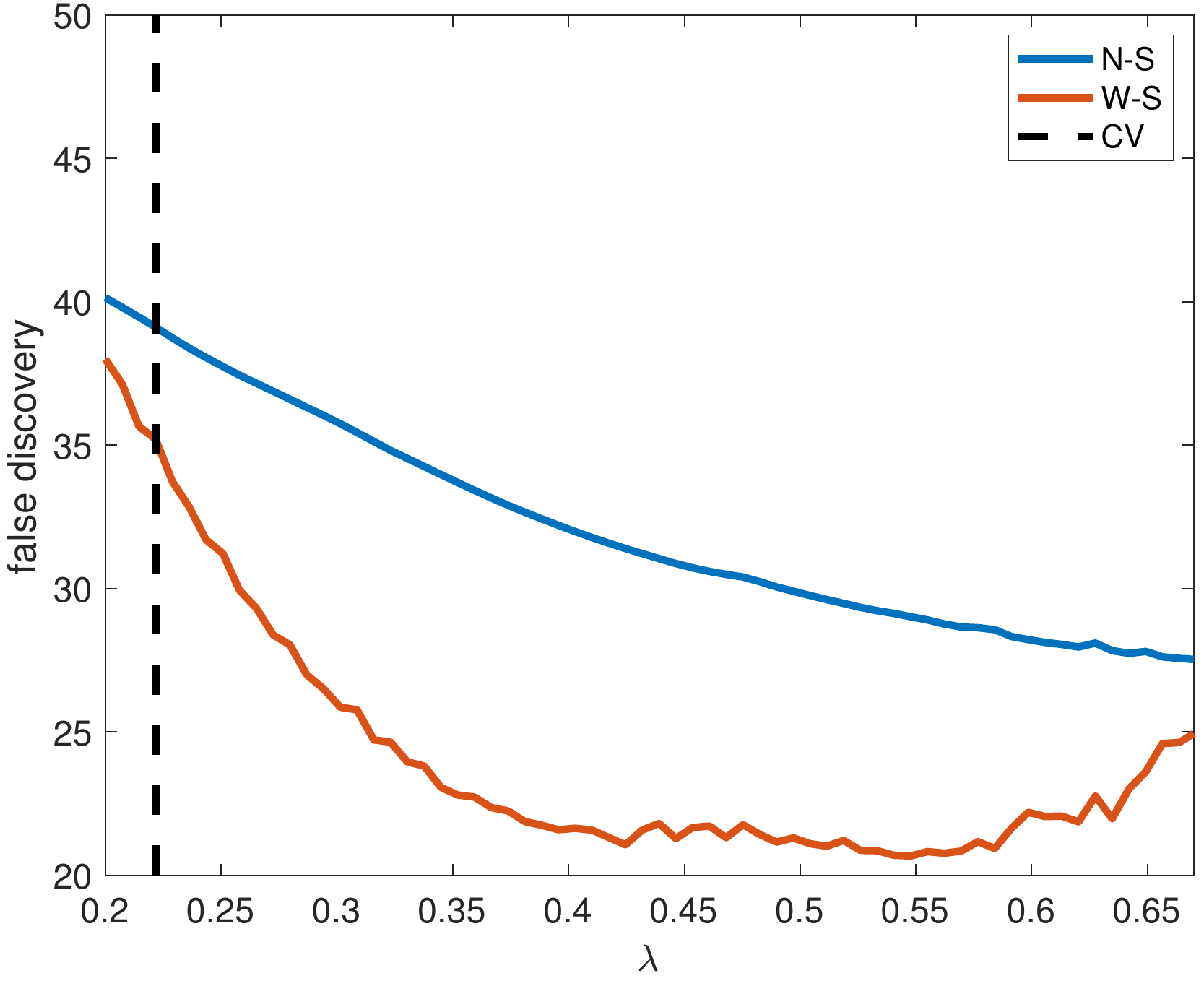}}
\subfigure[SNR = 0.8]{
\includegraphics[scale = 0.35]{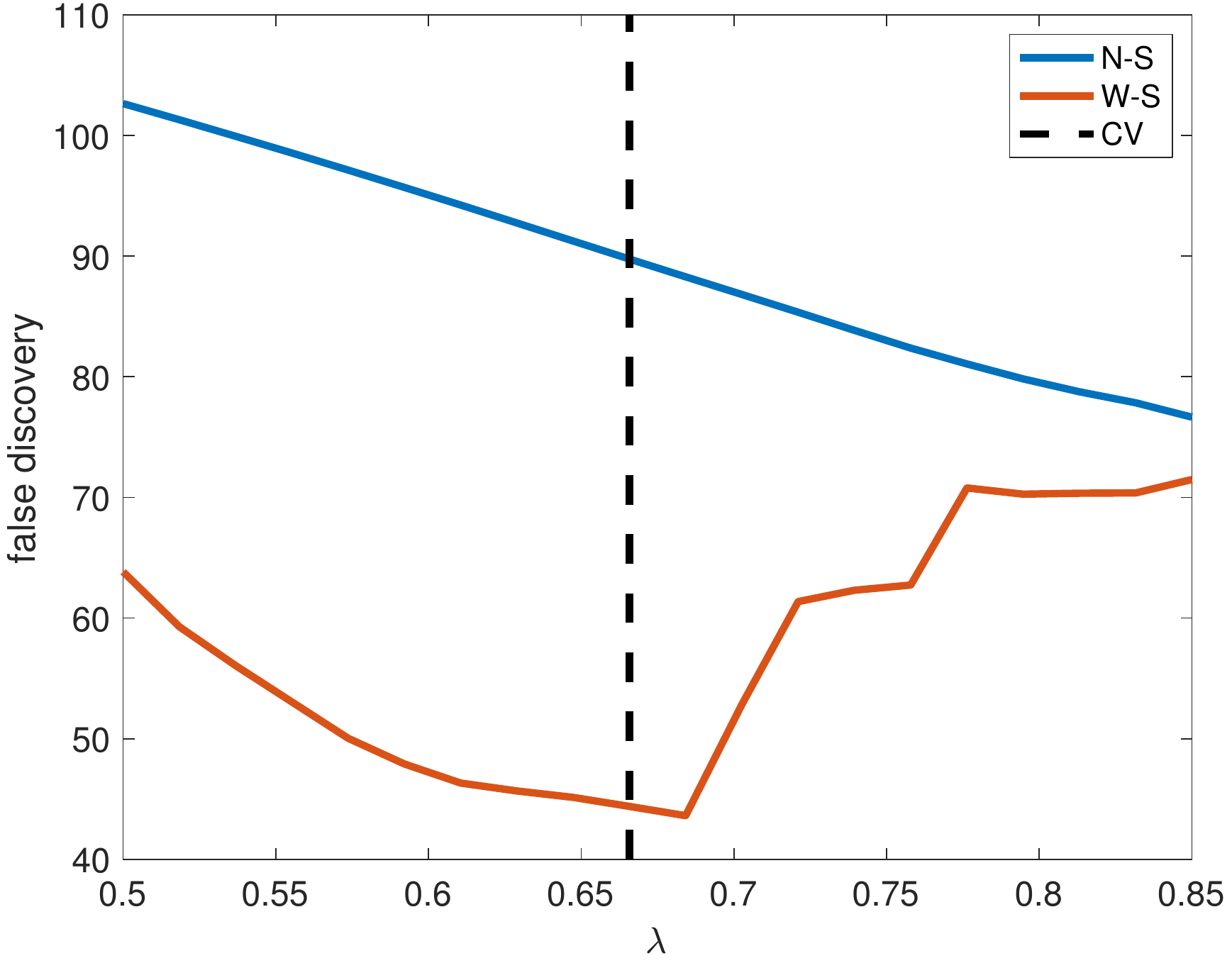}}
\caption{False discovery of subspace stability selection vs a non-subsampled approach with SNR $= 1.6,0.8$. Here, we choose a rank-$3$ approximation of the non-subsampled approach and $r_{\texttt{S3}} = 3$ in Algorithm $1$ of subspace stability selection.  The maximum possible amount of false discovery is $\text{dim}({T^\star}^\perp) = (70-10)^2 = 3600$. Furthermore,`N-S' denotes no subsampling and `W-S' denotes subspace stability selection.}
\label{fig:top_3_sing}
\end{figure}

{\par \emph{Illustration: subspace stability selection reduces false discovery}} -- Next, we demonstrate that subspace stability selection produces a tangent space which is different and usually of a higher quality (e.g. smaller expected false discovery) than the base estimator applied to the full dataset.  We choose the noise level so that SNR takes on one of the values in $\{1.5,2,2.5,3\}$.  On the one hand, we employ the procedure \eqref{eqn:nulcear_norm} on a subset of $2231$ observations (the training set) of the full set of $3186$ observations and the remaining subset of $955$ observations constitute the test set.  We use cross-validation to identify an optimal choice $\lambda^\star$ of the regularization parameter.  The estimate produced by \eqref{eqn:nulcear_norm} on the training set for this choice of $\lambda^\star$ is recorded as the output of the non-subsampled approach.  On the other hand, the estimator \eqref{eqn:nulcear_norm} with the choice $\lambda^\star$ is used in conjunction with $\alpha = 0.7$ to produce a subspace stability selection tangent space via Algorithm 1. For each of the four choices of SNR, we run $100$ experiments and average to find an empirical approximation to the expected false discovery \eqref{eqn:gfd}. Table~\ref{table:fd_illustration} compares the expected false discovery (with one sigma statistics) of the non-subsampled approach to that of the subspace stability selection procedure for the different problem settings.  Evidently, subspace stability selection yields a much smaller amount of false discovery compared to not employing subsampling.
\begin{center}
\begin{table}[htbp]
\scalebox{0.95}{

    \begin{tabular}{ | l | l | l | l | l | l | l|}
    \hline
    Method & SNR = 1.5 & SNR = 2 & SNR = 2.5 & SNR = 3\\\hline
    No subsampling         & 1274.6 $\pm$ 78.8& 1532.8 $\pm$ 68.5  & 1573.5$\pm$ 71.2  & 1417 $\pm$ 63.5   \\ \hline
    Subspace stability selection & 107.6 $\pm$ 11.5 & 89.7 $\pm$ 16.9 & 87.9 $\pm$ 18.7 & 87.9 $\pm$ 19.4  \\
\hline
\end{tabular}}
\caption{False discovery of subspace stability selection vs a non-subsampled approach on the stylized matrix completion problem. The maximum possible amount of false discovery is $\text{dim}({T^\star}^\perp) = (70-10)^2 = 3600$.}
\label{table:fd_illustration}
\end{table}
\end{center}
At this stage, it is natural to wonder whether the source of the improved false discovery control provided by subspace stability selection over not using subsampling is simply due to the non-subsampled approach providing estimates with a larger rank?  In particular, as an extreme hypothetical example, the zero-dimensional space is a stable tangent space and has zero expected false discovery, and more generally lower-rank tangent-space estimates are likely to have smaller expected false discovery.  Thus, is subsampling better primarily because it produces lower-rank estimates?  To address this point in our stylized setup, we consider a population $L^\star$ with associated incoherence parameter equal to $0.8$ \footnote{The incoherence of a matrix $M$ is $\max_{i} \max\{\|\Proj_{\text{col-space}(M)}(e_i)\|_2^2, \|\Proj_{\text{row-space}(M)}(e_i)\|_2^2\}$ where $e_i$ is the $i$'th standard basis vector, and it plays a prominent role in various analyses of the low-rank matrix completion problem \citep{matrix_completion}.}.  We sweep over the regularization parameter $\lambda$, and we compare the following two estimates: first, the estimate $\hat{L}$ obtained via \eqref{eqn:nulcear_norm} and then truncated to its first three singular values, and subsampled estimates obtained via Algorithm $1$ with $r_{\texttt{S3}}$ set to three.  The choice of three here is motivated by the fact that the population low-rank matrix $L^\star$ has three large components.  We perform this comparison for SNR $ = \{0.8,1.6\}$ and describe the results in the plots in Figure~\ref{fig:top_3_sing}.  In the high SNR regime, the performances of the subsampled and the non-subsampled approaches are similar.  However, in the low SNR regime, subspace stability selection yields a tangent space with far less false discovery across the entire range of regularization parameters.  Further, subspace stability selection provides a fundamentally different solution that cannot be reproduced simply by selecting the ``right'' regularization penalty in \eqref{eqn:nulcear_norm} applied to the entire dataset.

Similar behavior is also observed when the solution $\hat{L}$ is truncated at a different rank. As an example, with $\text{SNR = 0.8}$, we choose $\lambda$ via cross-validation and truncate $\hat{L}$ at rank $r = 1,2,\dots,5$ and compare its false discovery to the estimate produced by subspace stability selection with $r_\texttt{S3} = r$ (shown in Table 2).
\begin{center}
\begin{table}[h]
\scalebox{0.95}{
    \begin{tabular}{ | l | l | l | l | l | l | l|}
    \hline
    Method & rank = 1 & rank = 2 & rank = 3 & rank = 4 & rank = 5\\\hline
    No subsampling & 20.4 & 48.1 & 89.7& 146.7  & 218.8  \\ \hline
    Subspace stability selection & 12.4& 25.6 & 44.3 & 70.4 & 109 \\
\hline
\end{tabular}
}
\caption{False discovery of subspace stability selection vs a non-subsampled approach with SNR $= 0.8$ and rank of the estimate set to vary from $1$ to $5$. The maximum possible amount of false discovery is $\text{dim}({T^\star}^\perp) = 3600$.}
\end{table}
\end{center}
{\par\emph{Illustration: stability of tangent spaces to small changes in regularization parameter}}-- Finally, we note that in settings in which regularization is employed, the estimate can be extremely sensitive to the choice of regularization parameter.  For example, in nuclear-norm regularized formulations such as \eqref{eqn:nulcear_norm}, small changes to the parameter $\lambda$ can often lead to substantial changes in the optimal solution.  A virtue of subspace stability selection is that the estimates that it provides are generally very stable to small perturbations of $\lambda$.  To formalize this discussion, given two tangent spaces $T$ and $\tilde{T}$, we consider the quantity
$\mu(T,\tilde{T}) \triangleq 1-\frac{\text{trace}~(\mathcal{P}_{T}\mathcal{P}_{\tilde{T}})}{\max\{\text{dim}(T),\text{dim}(\tilde{T})\}}$
which measures the degree to which $T$ and $\tilde{T}$ are misaligned. If $T = \tilde{T}$, then $\mu(T,\tilde{T}) = 0$, and on the other hand, $T \subseteq \tilde{T}^\perp$ would yield $\mu(T,\tilde{T}) = 1$.  Hence, larger values of $\mu(T,\tilde{T})$ are indicative of greater deviations between $T$ and $\tilde{T}$. We use this metric to compare the stability of the non-subsampled approach with subspace stability selection.  In our stylized setup, we choose the noise level so that $SNR = 4$ and we select $\lambda = 0.03$ (based on cross-validation).  Letting $T$ be the tangent space of the estimator \eqref{eqn:nulcear_norm} with $\lambda = 0.03$ and $\tilde{T}$ with $\lambda = 0.05$, we find that $\mu(T, \tilde{T}) = 0.23$.  Setting $\alpha = 0.7$ with $B = 100$ complementary bags and computing the same metrics for the outputs of subspace stability selection, we find that $\mu(T, \tilde{T}) = 0.003$.  This contrast is observed for many other SNR levels.
\section{Experiments}
\label{section:experimental}
{\par}In this section, we demonstrate the utility of subspace stability selection in providing false discovery control both with synthetic and real data.  We consider the following types of low-rank estimation problems:
\begin{enumerate}
\item Low-rank linear measurements and matrix completion:
We consider noisy linear functions of a low-rank matrix $L^\star \in \mathbb{R}^{p_1 \times p_2}$ of the form $Y_i \approx \langle \mathcal{A}_i, L^\star \rangle, ~ i=1,\dots,n$ where each $\mathcal{A}_i \in \mathbb{R}^{p_1 \times p_2}$. In the linear measurement setting, $\mathcal{A}_i$ is an arbitrary sensing matrix, and in the matrix completion setting, $\mathcal{A}_i$ consists of zeros everywhere except a single entry which is equal to $1$. The matrix completion problem is similar to the one considered in the stylized demonstrations of Section~\ref{section:subspace_stability}.  One point of departure from that discussion in the present section is that in experiments where the dimensions $p_1, p_2$ are large, employing the nuclear norm regularized estimator \eqref{eqn:nulcear_norm} on each subsample is impractical.  Instead, we use on each subsample the following non-convex formulation:
\begin{eqnarray}
(\hat{U}~,~\hat{V}) = \underset{U \in \mathbb{R}^{p_1 \times{k}},V \in \mathbb{R}^{p_2 \times {k}}}{\mathrm{argmin}} && \sum_{i \in S} (Y_i - \langle \mathcal{A}_i, UV' \rangle )^2 + \lambda~(\|U\|_{F}^2 + \|V\|_{F}^2). \label{eqn:linear_measure}
\end{eqnarray}
where $\|U\|_{F}^2 + \|V\|_{F}^2$ is a surrogate for the nuclear norm penalty in \eqref{eqn:nulcear_norm}, $\lambda > 0$ is a regularization parameter, and $S \subset \{1,\dots,p_1\} \times \{1,\dots,p_2\}$ is the set of observed indices. By construction, $\hat{L} = \hat{U}\hat{V}'$ is constrained to have rank at most $k$, and this rank can be adjusted by appropriately tuning $\lambda$.  Fixing $U$ (resp. $V$) the above problem is convex in $V$ (resp. $U$), and thus a commonly employed approach in practice is alternating least-squares (ALS).

\item Factor analysis: We observe samples $\{Y^{(i)}\}_{i = 1}^n \subset \R^p$ of a random vector and we identify a factor model that best explains these observations, i.e., a model in which the coordinates of the observed vector are independent conditioned on a small number $k \ll p$ of latent variables.  In other words, our objective is to approximate the sample covariance of $\{Y^{(i)}\}_{i = 1}^n$ by a covariance matrix that is decomposable as the sum of a diagonal matrix and a positive-semidefinite low-rank matrix.  Using the Woodbury Inversion Lemma, we have that the precision matrix can be decomposed as a diagonal matrix minus a positive-semidefinite low-rank matrix.  The virtue of working with precision matrices is that the the log-likelihood function is concave with respect to this parametrization.  On each subsample, we use the following estimator \citep{Shapiro}:
\begin{eqnarray}
(\hat{D},\hat{L}) = \underset{L \in \mathbb{S}^{p},D \in \mathbb{S}^{p} }{\mathrm{argmin}} && -\log\det(D-L) + \text{trace}\left(\left(\tfrac{1}{|S|} \sum_{i \in S} Y^{(i)} {Y^{(i)}}' \right)(D-L) \right) + \lambda~\text{trace}(L).\label{eqn:factor_analysis}\\
\text{subject~to}  & &  D - L\succ 0,~ L \succeq 0,~ D\text{ is diagonal} \nonumber
\end{eqnarray}
Here $\text{trace}(\cdot)$ is the restriction of the nuclear norm to symmetric positive-semidefinite matrices.
\end{enumerate}

{
\subsection{{Synthetic Simulations}}
\vspace{0.1in}
We explore the role of the commutator in the false discovery bound of Theorem~\ref{thm:main} in a stylized matrix denoising problem. Specifically, we generate a population low-rank matrix $L^\star \in \mathbb{R}^{p \times p}$ with $p = 200$, with $\mathrm{rank}(L^\star)=6$, the nonzero singular values set to $\{120,100,80,30,20,10\}$, and the row and column spaces sampled uniformly from the Steifel manifold.  Once $L^\star$ is generated, we also choose a basis for the orthogonal complements of the row/column spaces of $L^\star$ and we let $U^\star{Q}{{V}^\star}'$ be the full SVD of $L^\star$, i.e., $U^\star,V^\star \in \mathbb{R}^{p \times p}$ are orthogonal matrices and $Q \in \mathbb{R}^{p \times p}$ is a diagonal matrix that is zero-padded.  We obtain $n$ noisy measurements of $L^\star$ of the form $Y_i = L^\star + \delta[{\gamma}U^\star{D}_i{V^\star}' + \epsilon_i]$ for $j = 1,2,\dots,n$, where $D_i$ is a diagonal matrix with i.i.d. standard Gaussian entries on the diagonal and $\epsilon_i \in \mathbb{R}^{p \times p}$ is a matrix with i.i.d. standard Gaussian entries.  The parameter $\delta > 0$ controls the signal-to-noise ratio and the parameter $\gamma > 0$ controls the commutator term inside Theorem~\ref{thm:main}. In particular, larger values of $\gamma$ leads to a smaller commutator term since the measurements $Y_i$ and $L^\star$ are all closer to being simultaneously diagonalizable. Geometrically, this corresponds to the principal angles between ${T^\star}^\perp$ and $\hat{T}(\mathcal{D}(n/2))$ concentrating around $0$ and $\pi/2$.  We vary $\gamma$ in the range $\{10,30\}$ and for each $\gamma$, we chose $\delta$ so that $\text{SNR} = 0.15$ (here $\text{SNR} = \mathbb{E}\left[{\|L^\star\|_2}/{\|\delta[\gamma{U}^\star{D}_iV^\star+\epsilon_i]\|_2}\right]$). We obtain $n = 2p$ measurements, and the estimator that we employ on a subsample computes best rank-$k$ approximation of the average over the data in the subsample (where $k$ is selected a-priori).  In our first illustration, the estimator computes rank-$6$ approximations.  We apply subspace stability selection with $\alpha \in [0.75,0.97]$ and $B = 100$ complementary bags, and we obtain an empirical approximation of the expected false discovery over $100$ trials. Since the population model is known, the quantities inside Theorem ~\ref{thm:main} are readily obtainable. We set the orthonormal basis elements $\{M_i\}_{i = 1}^{\mathrm{dim}({T^\star}^\perp)}$ needed to compute the term $F$ in \eqref{eqn:main2} to be $\{{U^\star}_{:,6+i}{V^\star}_{:,6+j}'\}_{i,j = 1}^{p-6}$. Figure \ref{fig:results_synthetic_factor}(a,b) compares the expected false discovery achieved by subspace stability selection with the bound of Theorem~\ref{thm:main}, the average number of discoveries of subspace stability selection (i.e., $\mathbb{E}[\mathrm{dim}(T)]$), and simply computing a rank-$6$ approximation of the entire data without any subsampling.  Figure \ref{fig:results_synthetic_factor}(c,d) shows a similar set of illustrations but with the estimator computing a rank-$10$ approximation.  A number of points are worth noting from these plots.  First, subspace stability selection performs far better than simply using computing low-rank approximations on the entire dataset; in particular, when the estimator computes rank-$6$ approximations and with $\gamma = 10$, subspace stability selection chooses a rank-$3$ model for $\alpha = 0.9$ and expected false discovery about $32.1$ while a rank-$6$ approximation on the entire dataset without subsampling yields an expected false discovery around $515.2$.  For comparison, the total amount of possible false discovery is $\mathrm{dim}({T^\star}^\perp) = 37636$.  Second, relative to the value of $\mathrm{dim}({T^\star}^\perp)$, the results provided by the theorem are very effective as they yield an expected false discovery bound between $300$ and $1100$ depending on the choice of $\alpha$.  Specifically, these bounds are also smaller than the average number of discoveries made by subspace stability selection as well as the expected false discovery of an estimator that operates on all the data with no subsampling.  As a final remark, we also note that smaller values of the commutator (larger choice of $\gamma$) leads to better bounds on the expected false discovery, as predicted by Theorem~\ref{thm:main}.

%Notice that subspace stability selection is substantially better than the full data approach or the theorem bound, as $L^\star$ has three strong components, and subspace stability selection teases those away from the noise and yields a small false discovery. We believe that bridging the gap between the achieved false discovery bound of subspace stability selection and Theorem 4 requires more careful bag dependent analysis, which is an interesting avenue for future research.  Figure \ref{fig:results_synthetic_factor}(c,d) demonstrates the Theorem 4 utility when $\lambda$ is conservatively chosen so that a rank-$10$ estimate is selected.   

\begin{figure}[thbp]
\centering
\subfigure[{$\gamma = 30~;~\text{rank sel.} = 6$ $\|[\Proj_{\hat{T}(\mathcal{D}(n/2))},\Proj_{{T^\star}^\perp}]\|_F \approx 52$}]{
\includegraphics[scale = 0.4]{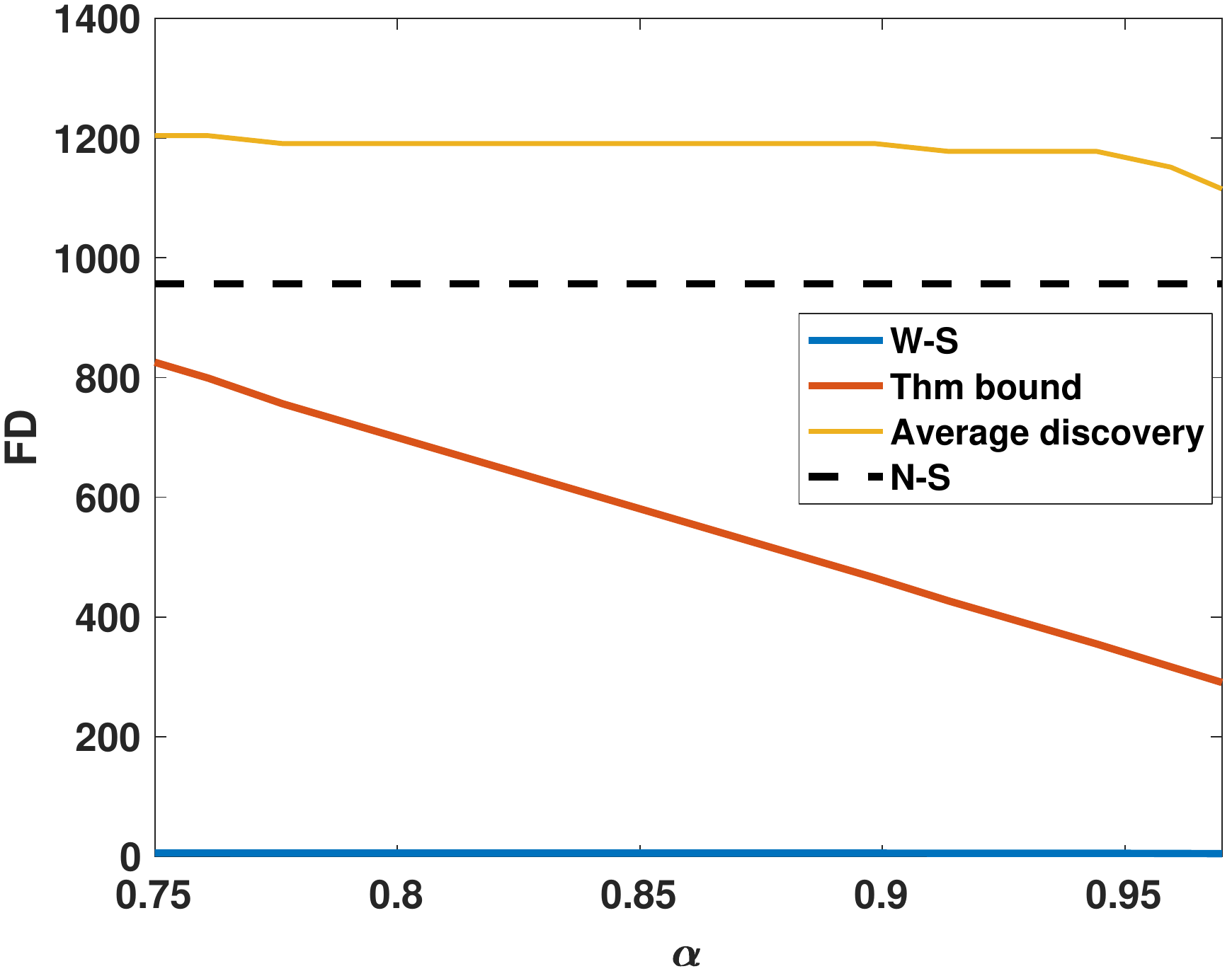}} \hspace{0.5cm}
\subfigure[{$\gamma = 10~;~\text{rank sel.}=6$ $\|[\Proj_{\hat{T}(\mathcal{D}(n/2))},\Proj_{{T^\star}^\perp}]\|_F \approx 226$}]{
\includegraphics[scale = 0.4]{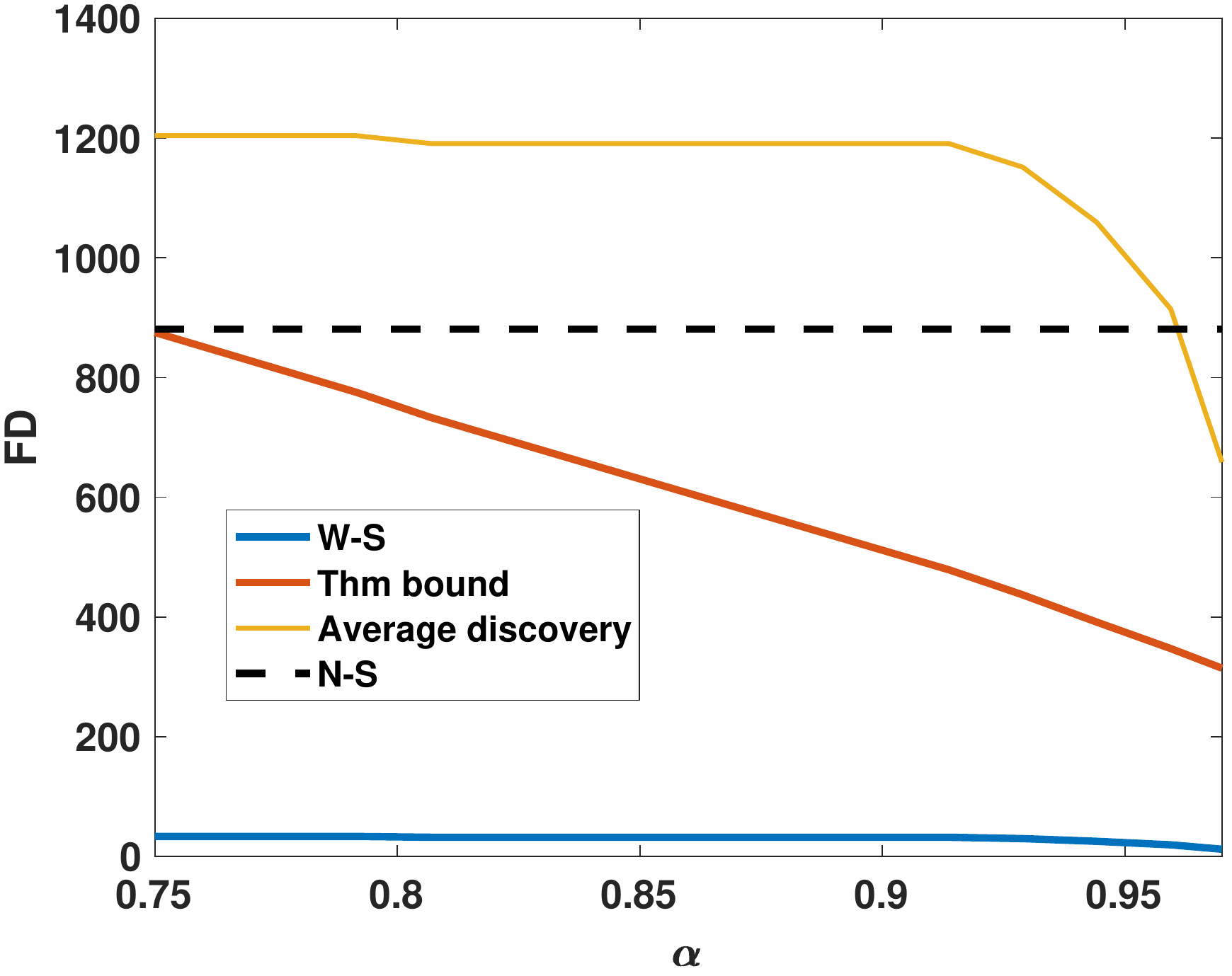}} \hspace{0.5cm}
\subfigure[{$\gamma = 30;\text{rank sel.} = 10$ $\|[\Proj_{\hat{T}(\mathcal{D}(n/2))},\Proj_{{T^\star}^\perp}]\|_F \approx 81$}]{
\includegraphics[scale = 0.4]{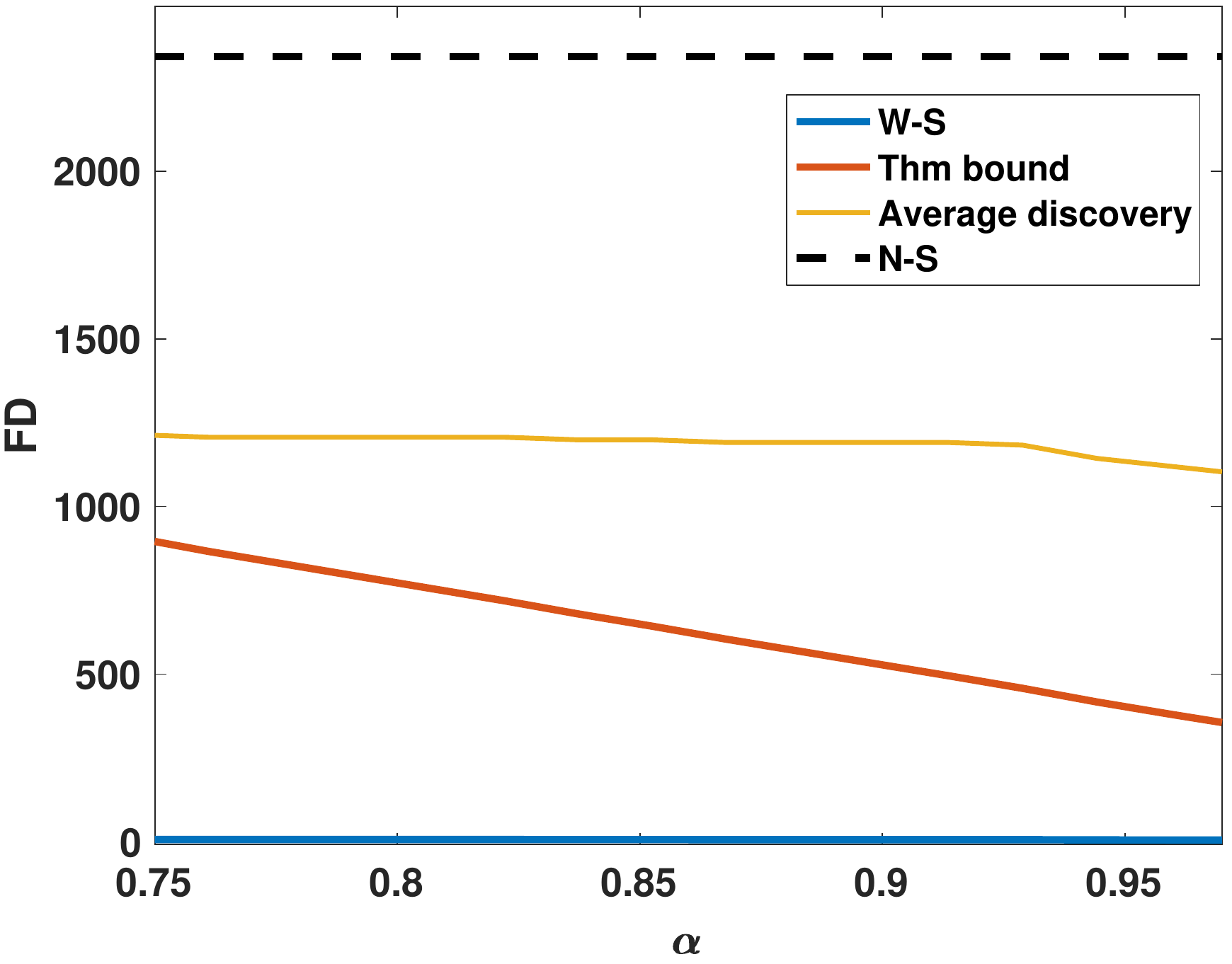}} \hspace{0.5cm}
\subfigure[{$\gamma = 10~;~\text{rank sel.} = 10$: $\|[\Proj_{\hat{T}(\mathcal{D}(n/2))},\Proj_{{T^\star}^\perp}]\|_F \approx 291$}]{
\includegraphics[scale = 0.4]{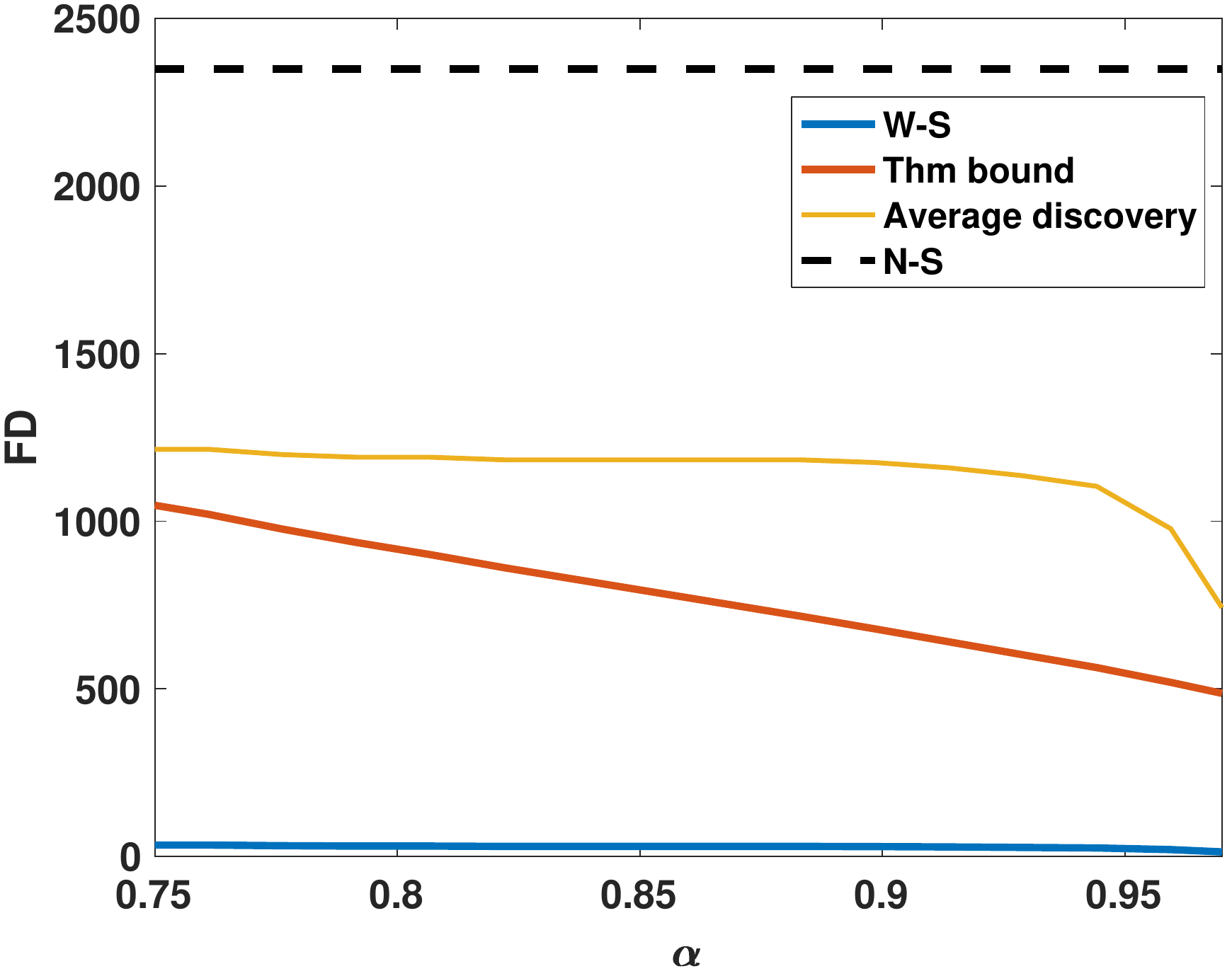}}
\caption{{False discovery of subspace stability selection as a function of $\alpha$ for matrix completion setting. The blue curve is false discovery obtained by subspace stability selection; the red curve is Theorem ~\ref{thm:main} bound; the yellow curve is average dimension of the selected tangent space; and the dotted line is false discovery from using entire data. Subspace stability selection has small but nonzero false discoveries. As an example, for $\gamma = 10$, rank selected $=6$, and $\alpha = 0.9$, subspace stability selection chooses typically a rank-3 model with $32.1$ false discoveries. Here $\mathrm{dim}({T^\star}^\perp) = 37636$.}}
\label{fig:results_synthetic_factor}
\end{figure}

{\par}Next, we explore the false discovery and power attributes of subspace stability selection in different noise and rank regimes. We consider the linear Gaussian measurement setting described earlier with $p = 60$, rank of $L^\star$ in the set $\{1,2,3,4\}$, the nonzero singular values set to $1$, and the row and column spaces sampled uniformly from the Steifel manifold. The measurements matrices $\{\mathcal{A}_i\}_{i=1}^{n}$ consist of iid entries drawn from $\mathcal{N}(0,1)$. We obtain noisy measurements $Y_i = \langle \mathcal{A}_i, L^\star + \epsilon \rangle, ~ i=1,\dots,n$ where $\epsilon \sim \mathcal{N}(0,\sigma^2)$. The observation noise level $\sigma^2$ is tuned so that SNR (here $\mathbb{E}[\langle \mathcal{A}_i, L^\star\rangle{/}\epsilon]$) lies in the set $\{1,2,3,4,5\}$. A fraction $n = 6p^2/10$ are used as training data for the estimator \eqref{eqn:linear_measure} with $\lambda$ chosen via holdout validation with a validation set of size $3p^2/20$ and the rank constraint $k$ set to $10$. With this choice of $\lambda$, we evaluate the expectation and standard deviations of false discovery and the power empirically over $100$ trials. As a point of comparison, we set $\alpha = 0.7$ with $B = 100$ complementary bags and compute the same metrics based on subspace stability selection. We repeat a similar experiment in the matrix completion setting where $L^\star \in \mathbb{R}^{p \times p}$ with $p = 100$, rank in the set $\{1,2,3,4\}$, row and column spaces chosen uniformly from Steifel manifold. We select a fraction $7/10$ of the total entries uniformly chosen at random as the observation set $\Omega$ so that $|\Omega| = 7p^2/10$. These observations are corrupted with Gaussian noise with variance selected so that SNR is in the range $\{0.5,0.875,1.25,1.625,2.00\}$. We use these observations as input to the estimator \eqref{eqn:linear_measure}, with $\lambda$ selected based on holdout validation on a $n_\text{test} = 7/20p^2$ validation set.

Figure ~\ref{fig:results_synthetic} compares the performance of the non-subsampled approach and subspace stability selection computed empirically over $100$ iterations for all the problem settings. For settings where either the false discovery standard deviation normalized by expected value or the power standard deviation normalized by expected value is greater than $0.01$, we plot the expected value with a cross and the one sigma around the mean with a rectangle. Several settings in Figure \ref{fig:results_synthetic} experience a significant loss in power using the subspace stability selection procedure. Those precisely correspond to models with high rank and low SNR regime where some components of the signal are overwhelmed by noise. To control false discoveries in these settings, subspace stability selection filters out some of the signal and as a result yields a small power.  
\begin{figure}[thbp]
\centering
\subfigure[Linear Measurements]{
\includegraphics[scale = 0.3]{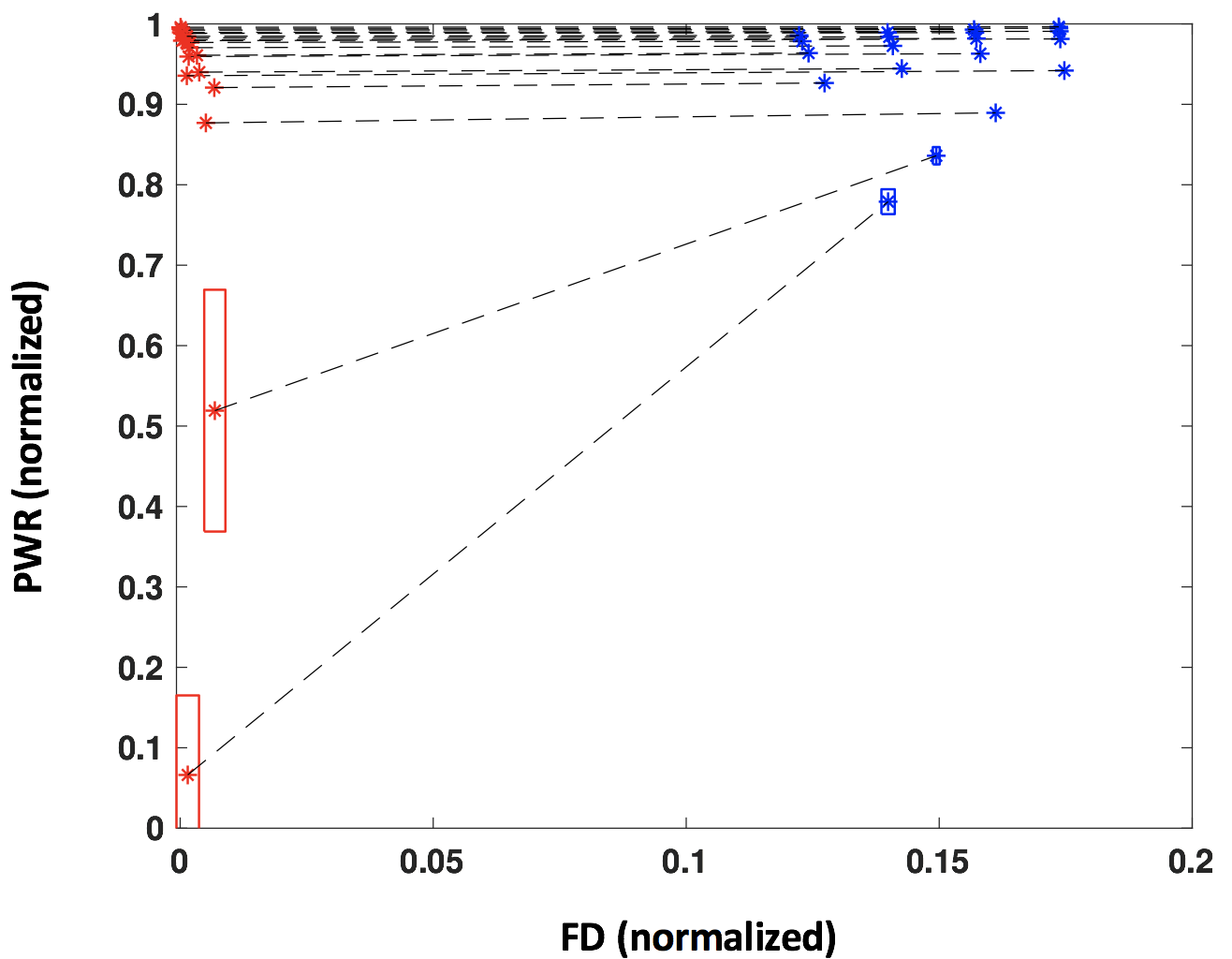}}
\subfigure[Matrix Completion]{
\includegraphics[scale = 0.3]{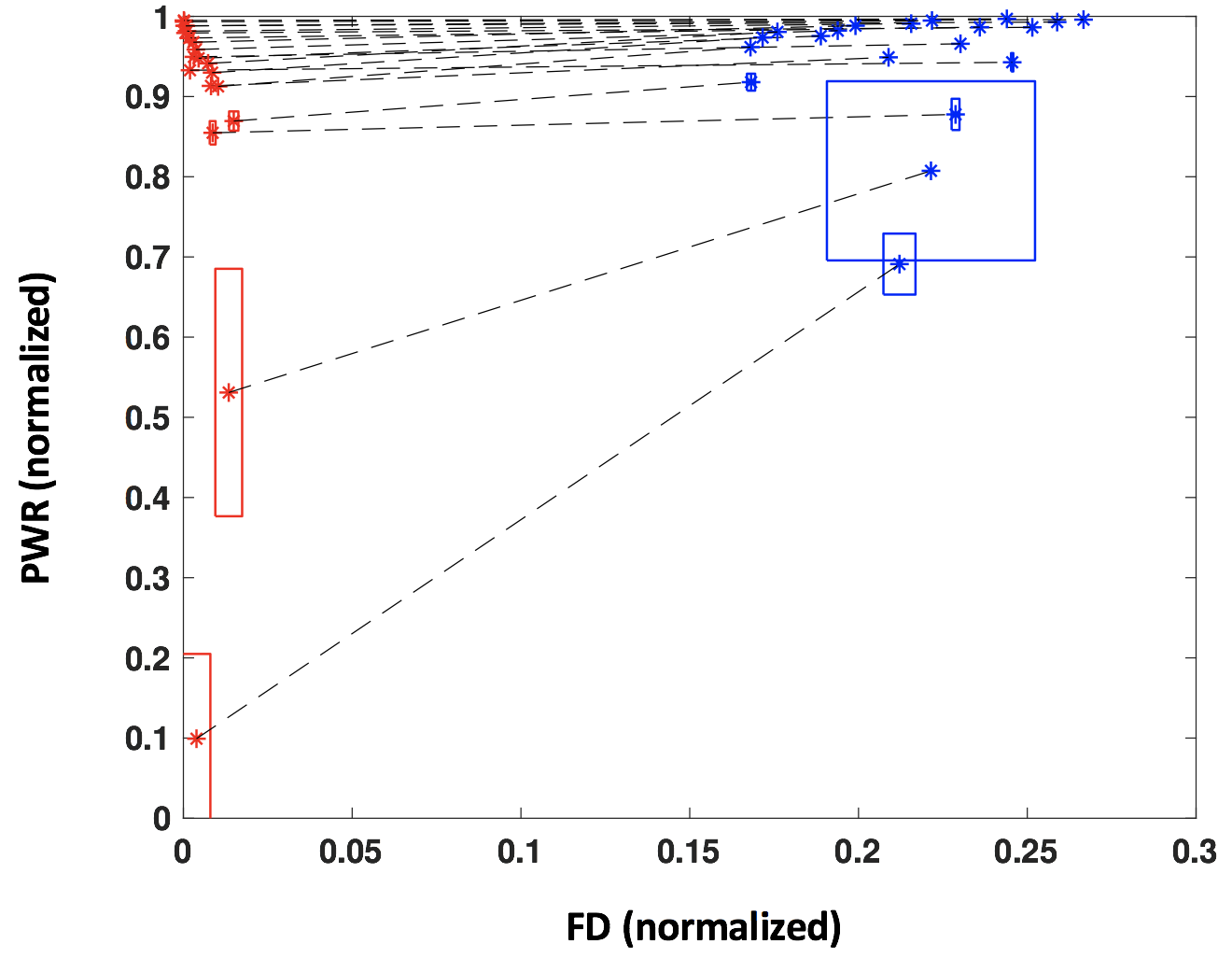}}
\caption{{False discovery vs power with (a) matrix completion and (b) linear measurements over 20 different problem instances (varying rank and noise level). Blue crosses corresponds to the performance of the non-subsampled approach and red crosses correspond to subspace stability selection with $\alpha = 0.7$. For the instances where standard deviation divided by mean is greater than $0.01$, we show one sigma rectangle around the mean. The lines connect dots corresponding to the same problem instance.  Both the false discovery and the power are normalized by dividing the expressions \eqref{eqn:gfd} and \eqref{eqn:gpw} by $\text{dim}({T^\star}^\perp)$ and $\text{dim}(T^\star)$, respectively.}}
\label{fig:results_synthetic}
\end{figure}
}
\subsection{Experimental Results on Real Datasets}
\subsubsection{Collaborative filtering}
\begin{figure}[thbp]
\centering
\subfigure[Amazon Video Games]{
%  % Requires \usepackage{graphicx}
 \includegraphics[scale = 0.40]{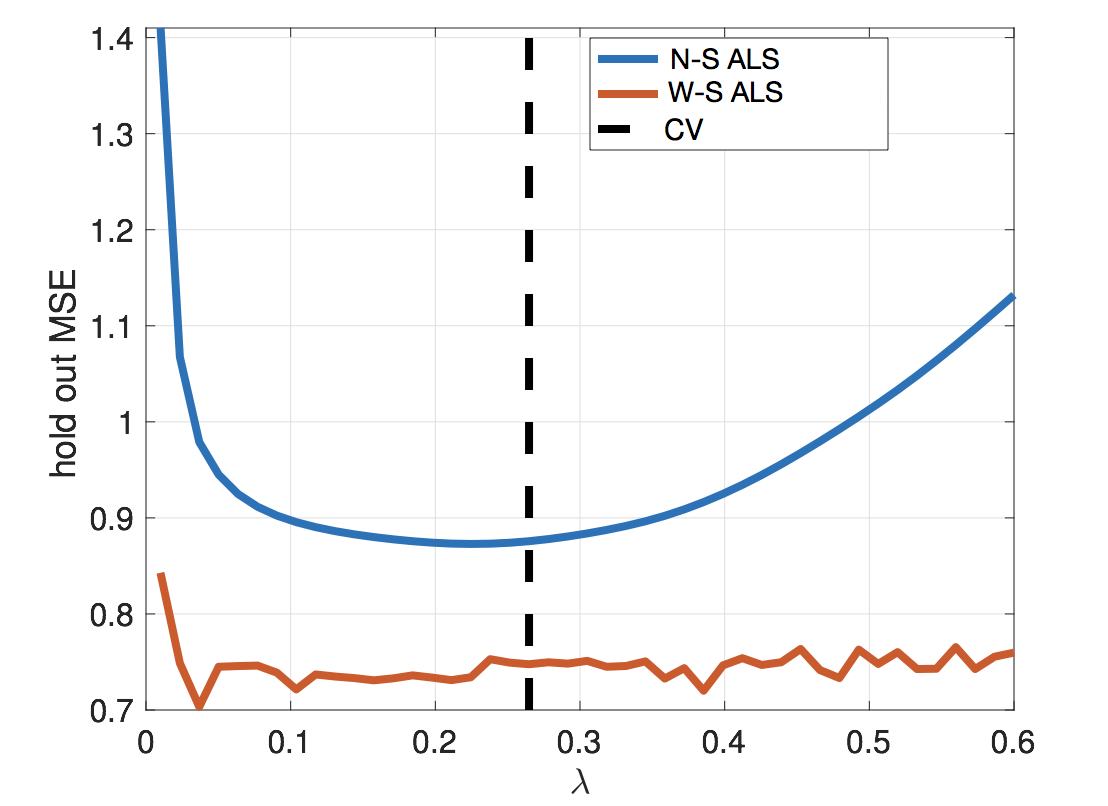}}
\subfigure[Amazon Book-Crossing]{
%  % Requires \usepackage{graphicx}
 \includegraphics[scale = 0.40]{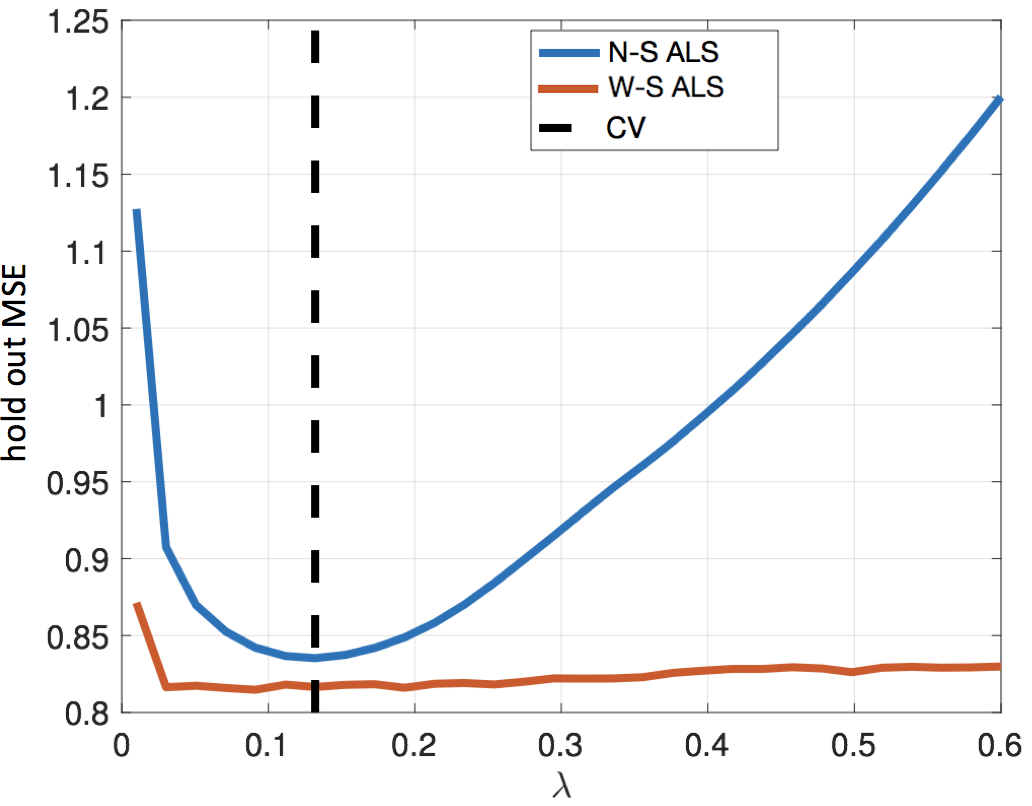}}
%\subfigure[MovieLens]{%  % Requires \usepackage{graphicx}
 %\includegraphics[scale = 0.25]{movielens.pdf}}
\caption{Collaborative filtering: MSE on holdout set of non-subsampled approach (denoted `N-S' and colored in blue) and subspace stability selection (denoted `W-S' and colored in red).  Dotted black line represents the cross-validated choice of $\lambda$ with the non-subsampled approach. }
\label{fig:recommend}
\end{figure}

\vspace{0.1in}
In collaborative filtering, one is presented with partially filled user-preference matrices in which rows are indexed by users and columns by items, with each entry specifying a user's preference for an item.  The objective is to infer the unobserved entries.  As discussed in Section~\ref{section:introduction}, such user-preference matrices are often well-approximated as low-rank, and therefore a popular approach to collaborative filtering is to frame it as a problem of low-rank matrix completion, and solve this problem based either on the convex relaxation \eqref{eqn:nulcear_norm} or the non-convex approach \eqref{eqn:linear_measure} via ALS.  We describe experimental results on two popular datasets in collaborative filtering: 1) the Amazon Book-Crossing dataset (obtained from \url{http://www2.informatik.uni-freiburg.de/~cziegler/BX/}) of which we consider a portion consisting of $p_1 = 1245$ users and $p_2 = 1054$ items with approximately $6\%$ of the ratings (integer values from 1 to 10) observed, and 2) the Amazon Video Games dataset (obtained from \url{http://jmcauley.ucsd.edu/data/amazon/}) of which we consider a portion consisting of $p_1 = 482$ users and $p_2 = 520$ items with approximately $3.5\%$ of the ratings (integer values from 1 to 5) observed.  In each case, we partition the dataset as follows: we set aside $85\%$ of the observations as a training set, $10\%$ of the observations as a holdout validation set, and the remaining $5\%$ as an evaluation set to assess the performance of our learned models.\\
{\indent}As these problems are relatively large in size, we employ ALS on the non-convex formulation \eqref{eqn:linear_measure} with $k = 80$ (the upper bound on the rank) and we apply the modification of Algorithm $1$ for subspace stability selection.  Finally, to obtain estimates of low-rank matrices (as this is the eventual object of interest in collaborative filtering) we use the formulation \eqref{eqn:tractable} given estimates of tangent spaces.  We set $\alpha = 0.7$ and $B=100$ complementary bags.  Figure~\ref{fig:recommend} illustrates the mean squared error of ALS and subspace stability selection on the holdout set for these two datasets for a range of values of the regularization parameter $\lambda$.  For both datasets, we observe that subspace stability selection yields models with better MSE on the holdout set over the entire range of regularization parameters.  On the Book-Crossings dataset, we further note that at the cross-validated $\lambda$, the rank of the estimate obtained from the non-subsampled approach is $80$ (i.e., the maximum allowable rank) with the first three singular values equal to $4329, 135.4, 63.1$. The MSE of this model on the evaluation set is equal to $0.83$. On the other hand, at the cross-validated $\lambda$ subspace stability selection yields a rank-$2$ model with an MSE of $0.81$ on the evaluation set.  Thus, we obtain a much simpler model with subspace stability selection that also offers better predictive performance.  Similarly, for the Amazon Video Games dataset, the rank of the estimate obtained from the non-subsampled approach is $39$ with the first five singular values equal to $1913.5, 49.4, 43.6, 28.4, 27.4$, with an MSE of $0.87$ on the evaluation set.  On the other hand, subspace stability selection yields a rank-$4$ solution with a much smaller MSE of $0.74$ on the evaluation set.  Finally, we observe for both datasets that subspace stability selection is much more stable across the range of regularization parameters.  Thus, subspace stability selection is far less sensitive to the particular choice of $\lambda$, which removes the need for fine-tuning $\lambda$.  
\subsubsection{Hyperspectral unmixing} 
\begin{figure}[thbp]
\centering
\subfigure{%  % Requires \usepackage{graphicx}
 \includegraphics[scale = 0.40]{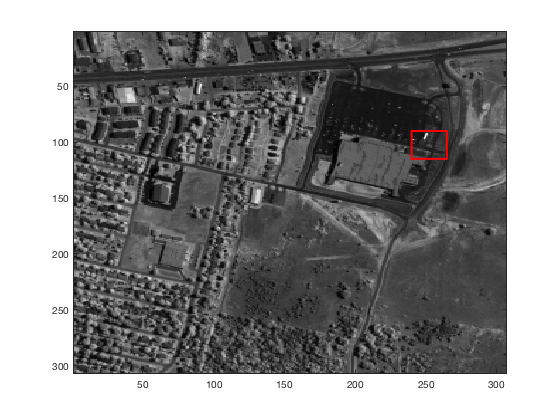}}
 \subfigure{%  % Requires \usepackage{graphicx}
 \includegraphics[scale = 0.40]{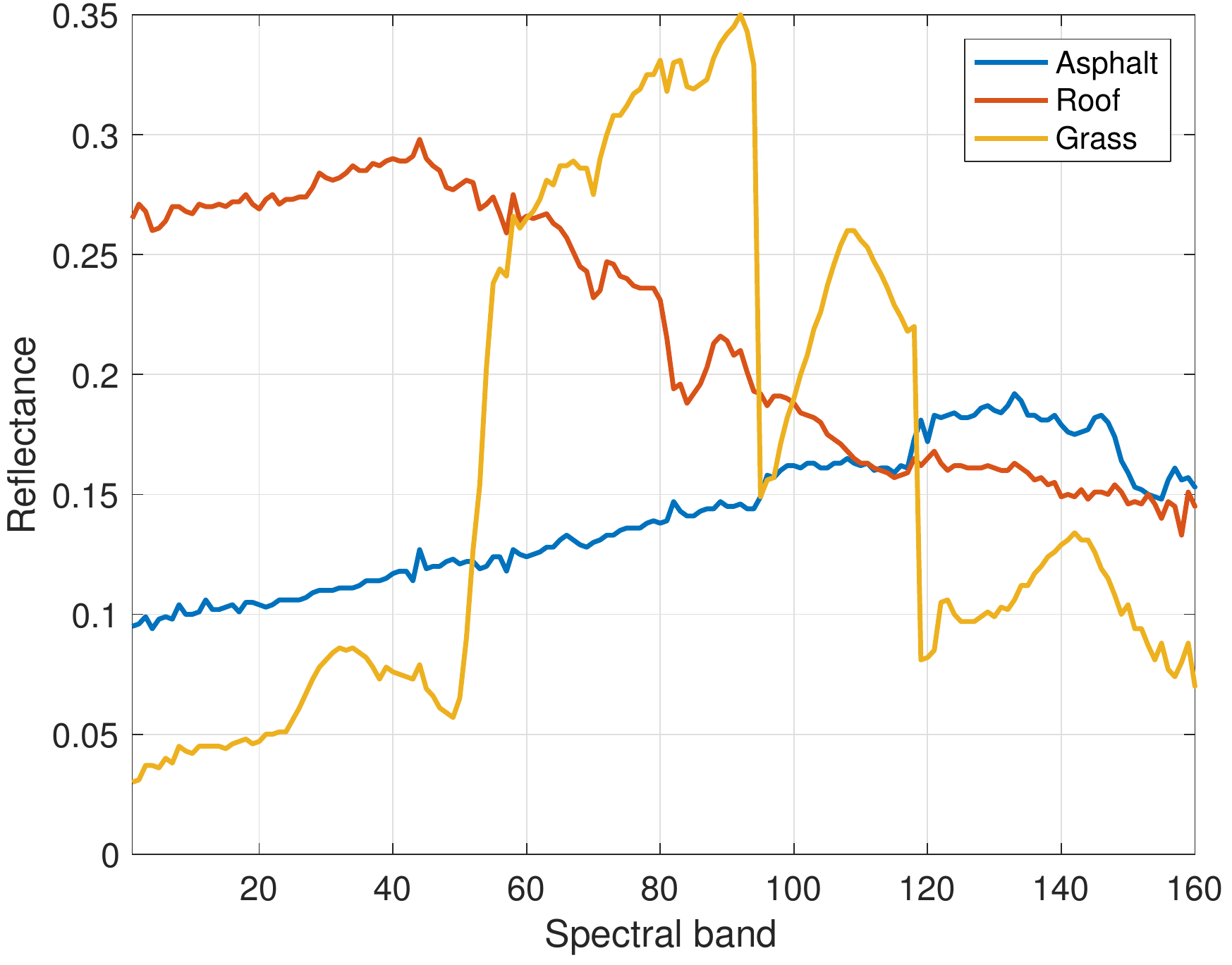}}
 \caption{Urban hyperspectral image (left) and spectra of three materials present in the image (right). The data and the population spectra are obtained from \url{http://www.escience.cn/people/feiyunZHU/Dataset_GT.html}.}
\label{fig:patch}
\end{figure}

\vspace{0.1in}
Here we give an illustration with real hyperspectral imaging data in which the underlying population parameters are known based on extensive prior experiments.  In this problem, we are given a hyperspectral image $Y \in \mathbb{R}^{p_1 \times p_2}$ consisting of $p_1$ frequency bands and $p_2$ pixels, where $Y_{i,j}$ is the reflectance of the $j$'th image pixel to the $i$'th frequency band. The spectral unmixing problem aims to find $W \in \mathbb{R}^{p_1 \times k}$ (called the endmember matrix) and $H\in \mathbb{R}^{k \times p_2}$ (called the abundance matrix) so that $Y \approx WH$, where $k \ll \min(p_1,p_2)$ is the number of endmembers \citep{Manolakis}.  Of particular interest is the $k$-dimensional column-space of $W$, which corresponds to the space spanned by the $k$ endmembers that are present in the image.  We discuss two natural hyperspectral unmixing problems that arise commonly in practice.  We focus on the Urban dataset (obtained from \url{http://www.escience.cn/people/feiyunZHU/Dataset_GT.html}), a hyperspectral image consisting of $307 \times 307$ pixels, each of which corresponds to a $2 \times 2 m^2$ area with $210$ wavelengths ranging from $400nm$ to $2500nm$. Following previous analyses of this dataset, we remove $48$ noisy channels to obtain $162$ wavelengths and select a $30 \times 25$ patch (equal to $750$ pixels) shown in Figure~\ref{fig:patch}(a). In the selected patch, there are a total of 3 endmembers (shown in Figure~\ref{fig:patch}(b)), with one strong signal and two weak signals.\\
In many settings, obtaining a complete hyperspectral image of a scene may be costly, and it is of interest to accurately reconstruct a hyperspectral image from partial observations.  This problem may be naturally formulated as one of low-rank matrix completion.  As with other application domains in which problems are reformulated as low-rank matrix completion, ALS applied to the non-convex formulation \eqref{eqn:linear_measure} is commonly employed.  To simulate such a hyperspectral unmixing problem, we randomly subsample $10\%$ of the hyperspectral data in the patch as training data.  We further select another $10\%$ of the remaining data as a holdout validation set.  We compare the amount of false discovery of a non-subsampled approach and subspace stability approach, with $k$ conservatively chosen to be equal to $20$ in the ALS procedure in each case.  Due to the scale of this problem being large, we use the modification of Algorithm 1 (with $\alpha = 0.7$ and $B = 100$ complementary bags) described in Section~\ref{section:subspace_stability} for subspace stability selection.  As the column space of the low-rank estimate is the principal object of interest for endmember detection, the quantities of interest for evaluating performance are based on \eqref{eqn:fd_colspace}: 
 $\overline{\text{FD}} = \mathbb{E}\left[\text{trace}\left(\mathcal{P}_{\text{col-space}(W^\star)^\perp}\mathcal{P}_{\text{col-space}(\hat{W})}\right)\right]$ and $\overline{\text{PW}} = \mathbb{E}\left[\text{trace}\left(\mathcal{P}_{\text{col-space}(W^\star)}\mathcal{P}_{\text{col-space}(\hat{W})}\right)\right]$. Here, the expectation is with respect to the randomness in the selection of the $10\%$ training data, $W^\star \in \mathbb{R}^{162 \times 3}$ is the matrix consisting of the spectra of the three endmemebers in Figure~\ref{fig:patch}(b)), and $\hat{W}$ is the estimated matrix.  We find a cross-validated choice of $\lambda = 1$ from one random selection of training data. With this $\lambda$ and over $100$ random trials in the selection of training data, no subsampling ALS produces on average rank-$20$ estimate with $\overline{\text{FD}} = 0.1 ~ \text{dim}(\text{col-space}({W^\star}^\perp))$ and $\overline{\text{PW}} = 0.97 ~ \text{dim}(\text{col-space}(W^\star))$.  In contrast, for the same $\lambda = 1$, subspace stability selection (operating on tangent spaces $T_n(\text{col-space}(\hat{W}))$ produces on average rank-$2.86$ with $\overline{\text{FD}} = 0.0007 ~ \text{dim}(\text{col-space}({W^\star}^\perp))$ and $\overline{\text{PW}} = 0.91 ~ \text{dim}(\text{col-space}(W^\star))$.  Furthermore, even if $\lambda$ is set large enough (for example, $\lambda = 29$) so that the non-subsampled ALS estimate has on average rank equal to $2.52$, the false discovery estimate is $\overline{\text{FD}} = 0.007 ~ \text{dim}({\text{col-space}(W^\star)}^\perp)$, which is still far larger than the amount of false discovery of subspace stability selection.

A different type of hyperspectral unmixing problem arises if the observations are corrupted by noise.  In particular, based on the decomposition $Y \approx WH$, the outer product $YY'$ is well approximated by a low-rank matrix.  Thus, another natural approach for endmember detection is to perform factor analysis by viewing each column of $Y$ (i.e., an entire collection of wavelengths corresponding to each pixel) as an observation and approximating the sample covariance of these observations as the sum of diagonal and low-rank matrices.  The row/column spaces of the low-rank component (which is symmetric, hence the row and column spaces are the same) serve as estimates of the subspace spanned by the endmembers.   We obtain $\{Y^{(i)}\}_{i=1}^{750} \subset \mathbb{R}^{162}$ spectral observations of the $750$ total pixels by applying white noise to the population parameters with the noise level chosen so that $\text{SNR} = 0.78$. We then set aside $80\%$ of the data as training data for the estimator  \eqref{eqn:factor_analysis}, which is solved using LogDetPPA solver \cite{Toh}. We set aside the remaining $20\%$ as a holdout validation set.  Employing the estimator \eqref{eqn:factor_analysis} without subsampling and with $\lambda$ chosen via cross-validation and expectations computed over $100$ yields false discovery $\text{FD} = 0.04 ~ \text{dim}({T^\star}^\perp)$ and power $\text{PW} = 0.48 ~ \text{dim}(T^\star)$.  (Here $T^\star$ represents the population tangent space.)  On the other hand, subspace stability selection with $\alpha = 0.7$ and $B = 100$ complementary bags yields a tangent space estimate with a false discovery and power $\text{FD} = 0.015 ~ \text{dim}({T^\star}^\perp)$ and $\text{PW} = 0.69 ~ \text{dim}(T^\star)$, respectively.  Evidently, subspace stability selection yields a substantial decrease in the amount of false discovery as well as an improvement in power.

%to identify an estimate $(\hat{L}, \hat{D})$. Since these quantities are in the precision domain, we apply the transformations $\tilde{L} =(\hat{D}-\hat{L})^{-1}-\hat{D}^{-1}$ and $\tilde{D} = \hat{D}^{-1}$, where $\tilde{L}$ is an estimate for $L^\star$ and $\tilde{D}$ is an estimate for $D^\star$. This yields a tangent space $\hat{T}$ where the normalized false discovery and normalized power are evaluated to be $\text{FD}_N = \frac{\mathbb{E}\left[\text{trace}\left(\mathcal{P}_{\hat{T}}\mathcal{P}_{{T^\star}^\perp}\right)\right]}{\text{dim}({{T}^\star}^\perp)} \approx 0.04$ and $\text{PWR}_N = \frac{\mathbb{E}\left[\text{trace}\left(\mathcal{P}_{{T}}\mathcal{P}_{{T^\star}}\right)\right]}{\text{dim}({T}^\star)} \approx 0.46$, respectively. 

%correspond to the span of the endmembers of the observations of factor modelinover $Y$, where the covariance of $Y$ is approximated by a sum of a low rank matrix $L^\star$ and a diagonal matrix $D^\star$ \citep{Manolakis}. The low rank component $L^\star$ identifies the influence of endmembers on the observed spectra; in particular the row and column spaces of $L^\star$ are spanned by the endmembers in the image. This gives rise to the tangent space $T^\star$, which is readily computable given the the population signals in Figure~\ref{fig:patch}(b). The diagonal component $D^\star$ models the noise in the observed spectra.
\section{Conclusions and Future Directions}
\label{section:conclusions}
{\par}In this paper, we describe a geometric framework for assessing false discoveries in low-rank estimation. The proposed framework has many appealing properties including that it is a natural generalization of false discovery in variable selection. We further describe the subspace stability selection algorithm to provide false discovery control in the low-rank setting.  This procedure is a generalization of the stability selection method of \cite{stability}.  The method is general and we demonstrate its utility with both synthetic and real datasets in a range of low-rank estimation tasks.
{\par}There are several interesting directions for further investigation that arise from our work.  First, within the context of Theorem~\ref{thm:main} on the expected false discovery of a stable tangent space produced by subspace stability selection, it would be useful to carry out a more refined bag-dependent analysis in the spirit of \cite{Samworth}.  Second, while Algorithm $1$ from Section~\ref{section:algorithm} outputs an estimate that does provide false discovery control, it is unclear whether this is the most powerful procedure possible.  In particular, it is of interest to obtain an optimal solution to the problem \eqref{eqn:optimal}, or to prove that Algorithm $1$ computes a near-optimal solution.  {Third, Algorithm 1 requires a user-specified $\alpha$ to produce an estimate that provides a false discovery bound as stated in Theorem~\ref{thm:main}. In exploratory settings, one may wish to examine the data first, and choose $\alpha$ to obtain a desired amount discovery while still retaining some false discovery guarantees.  This viewpoint, considered by \cite{Solari}, reverses the traditional role of the analyst and the inference procedure.  Building on their perspective, it would be of interest to develop false discovery bounds for subspace stability selection that remain valid despite post-hoc selection of $\alpha$}.  Fourth, a significant topic of contemporary interest in variable selection -- especially when there are a large number of possible predictors -- is to control for the false discovery rate.  In Section~ \ref{section:false_discovery}
 we gave a formulation of false discovery rate in the low-rank setting, and it is natural to seek procedures that provide false discovery rate control in settings with high-dimensional matrices.  One obstacle that arises with this effort is that every proof of false discovery rate control of a variable selection method (of which we are aware) relies strongly on the simultaneous diagonalizability of the projection matrices associated with the population tangent space and the estimated tangent space (when translated to the geometric viewpoint of our paper).  Finally, the geometric framework developed in this paper for assessing false discovery is potentially relevant beyond the specific setting of low-rank estimation.  For example, our setup extends naturally to latent-variable graphical model selection \cite{Chand2012} as well as low-rank tensor estimation \cite{kolda}, both of which are settings in which the underlying geometry is similar to that of low-rank estimation.  More broadly, the perspective presented here may be useful in addressing many other structured estimation problems.

\section*{Acknowledgements}
The authors were supported in part by National Science Foundation grant CCF-1350590, Air Force Office of Scientific Research grant FA9550-16-1-0210, Sloan Fellowship, and the Resnick Fellowship.

\appendix
\section{Appendix}

\subsection{Proof of Theorem 4 (main paper)}
\vspace{0.1in}

We first prove the basis-dependent bound.  For each $\ell = 1,\dots,B$ and for each $i=1,\dots,\mathrm{dim}({T^\star}^\perp)$ we have that
\begin{equation} \label{eq:masterdecomposition}
\begin{aligned}
    \mathrm{trace}(\Proj_T \Proj_{\mathrm{span}(M_i)}) ~ = ~ & \mathrm{trace}(\Proj_{\hat{T}(\mathcal{D}_\ell)} \Proj_T \Proj_{\hat{T}(\mathcal{D}_\ell)} \Proj_{\mathrm{span}(M_i)}) + \mathrm{trace}(\Proj_{{\hat{T}(\mathcal{D}_\ell)}^\perp} \Proj_T \Proj_{{\hat{T}(\mathcal{D}_\ell)}^\perp} \Proj_{\mathrm{span}(M_i)}) \\ & + \mathrm{trace}(\Proj_{\hat{T}(\mathcal{D}_\ell)} \Proj_T \Proj_{{\hat{T}(\mathcal{D}_\ell)}^\perp} \Proj_{\mathrm{span}(M_i)}) + \mathrm{trace}(\Proj_{\hat{T}{(\mathcal{D}_\ell)}^\perp} \Proj_T \Proj_{\hat{T}(\mathcal{D}_\ell)} \Proj_{\mathrm{span}(M_i)}).
\end{aligned}
\end{equation}
The last two terms may be simplified as follows:
\begin{equation*}
\begin{aligned}
    & \mathrm{trace}(\Proj_{\hat{T}(\mathcal{D}_\ell)} \Proj_T \Proj_{{\hat{T}(\mathcal{D}_\ell)}^\perp} \Proj_{\mathrm{span}(M_i)}) + \mathrm{trace}(\Proj_{\hat{T}{(\mathcal{D}_\ell)}^\perp} \Proj_T \Proj_{\hat{T}(\mathcal{D}_\ell)} \Proj_{\mathrm{span}(M_i)}) \\ & = ~ \mathrm{trace}(\Proj_{\hat{T}(\mathcal{D}_\ell)} (\Proj_T \Proj_{{\hat{T}(\mathcal{D}_\ell)}^\perp} - \Proj_{{\hat{T}(\mathcal{D}_\ell)}^\perp} \Proj_T) \Proj_{\mathrm{span}(M_i)}) \\ & ~~~~ +  \mathrm{trace}((\Proj_{\hat{T}{(\mathcal{D}_\ell)}^\perp} \Proj_T - \Proj_T \Proj_{\hat{T}{(\mathcal{D}_\ell)}^\perp}) \Proj_{\hat{T}(\mathcal{D}_\ell)} \Proj_{\mathrm{span}(M_i)}) \\
    & = ~ \mathrm{trace}([\Proj_T, \Proj_{\hat{T}{(\mathcal{D}_\ell)}^\perp}] \times [\Proj_{\mathrm{span}(M_i)}, \Proj_{\hat{T}(\mathcal{D}_\ell)}]).
\end{aligned}
\end{equation*}
The first equality follows by noting that $\Proj_{\hat{T}{(\mathcal{D}_\ell)}^\perp} \Proj_{\hat{T}{(\mathcal{D}_\ell)}} = \Proj_{\hat{T}{(\mathcal{D}_\ell)}} \Proj_{\hat{T}{(\mathcal{D}_\ell)}^\perp} = 0$ for each $\ell = 1,\dots,B$.  The second equality follows from the definition of the commutator and the cyclicity of trace.  We label the various terms of \eqref{eq:masterdecomposition} combined with the above simplification in terms of commutators as follows for each $\ell=1,\dots,B$ and $i=1,\dots,\mathrm{dim}({T^\star}^\perp)$:
\begin{equation} \label{eq:fgh}
\begin{aligned}
    f_{\ell,i} & = \mathrm{trace}(\Proj_{\hat{T}(\mathcal{D}_\ell)} \Proj_T \Proj_{\hat{T}(\mathcal{D}_\ell)} \Proj_{\mathrm{span}(M_i)}) \\ g_{\ell,i} & = \mathrm{trace}(\Proj_{{\hat{T}(\mathcal{D}_\ell)}^\perp} \Proj_T \Proj_{{\hat{T}(\mathcal{D}_\ell)}^\perp} \Proj_{\mathrm{span}(M_i)}) \\ h_{\ell,i} & = \mathrm{trace}([\Proj_T, \Proj_{\hat{T}{(\mathcal{D}_\ell)}^\perp}] \times [\Proj_{\mathrm{span}(M_i)}, \Proj_{\hat{T}(\mathcal{D}_\ell)}]).
\end{aligned}
\end{equation}
Therefore, we have for each $\ell=1,\dots,B$ and $i=1,\dots,\mathrm{dim}({T^\star}^\perp)$ that:
\begin{equation*}
    \mathrm{trace}(\Proj_T \Proj_{\mathrm{span}(M_i)}) = f_{\ell,i} + g_{\ell,i} + h_{\ell,i}.
\end{equation*}
Fix a pair of complementary bags indexed by $\{2j-1,2j\}$ for some $j \in \{1,\dots,\tfrac{B}{2}\}$.  For this pair, we have that:
\begin{equation}
\begin{aligned}
\mathrm{trace}(\Proj_T \Proj_{\mathrm{span}(M_i)}) & =  \min\{f_{2j-1,i}+g_{2j-1,i}+h_{2j-1,i}, f_{2j,i}+g_{2j,i}+h_{2j,i}\} \\ & \leq \min\{f_{2j-1,i}+g_{2j-1,i}, f_{2j,i}+g_{2j,i}\} + \max\{h_{2j-1,i}, h_{2j,i}\} \\ & \leq \min\{f_{2j-1,i}, f_{2j,i}\} + g_{2j-1,i} + g_{2j,i} + \max\{h_{2j-1,i}, h_{2j,i}\}. 
\end{aligned}
\label{bound:inter}
\end{equation}
The first equality holds because the two terms in the minimum are equal.  The first inequality holds because $\min\{u_0+v_0, u_1+v_1\} \leq \min\{u_0,u_1\} + \max\{v_0,v_1\}$ if $u_0+v_0 = u_1+v_1$ (here $u_k = f_{2j-k,i}+g_{2j-k,i}$ and $v_k=h_{2j-k,i}$ for $k=0,1$).  The second inequality holds because $\min\{u_0+v_0,u_1+v_1\} \leq \min\{u_0,u_1\} + v_0 + v_1$ for $v_0,v_1 \geq 0$ (here $u_k = f_{2j-k,i}$ and $v_k=g_{2j-k,i}$ for $k=0,1$). The bound \eqref{bound:inter} holds for all $j = 1,2,\dots,B/2$ and for each $i=1,\dots,\mathrm{dim}({T^\star}^\perp)$.  We can thus minimize the upper bounds as follows:
\begin{equation*}
\begin{aligned}
\mathrm{trace}(\Proj_T \Proj_{\mathrm{span}(M_i)}) &\leq \min_{j =1,2,\dots,B/2}\min\{f_{2j-1,i}, f_{2j,i}\} + g_{2j-1,i} + g_{2j,i} + \max\{h_{2j-1,i}, h_{2j,i}\}  \\
&\leq \frac{2}{B}\sum_{j = 1}^{B/2}\min\{f_{2j-1,i}, f_{2j,i}\} + g_{2j-1,i} + g_{2j,i} + \max\{h_{2j-1,i}, h_{2j,i}\}, 
\end{aligned}
\end{equation*}
where the second inequality follows from the fact that the minimum over a collection of numbers is bounded above by their average.  Since $\mathrm{trace}(\Proj_T\Proj_{{T^\star}^\perp}) = \sum_{i = 1}^{\text{dim}({T^\star}^\perp)} \mathrm{trace}(\Proj_T\Proj_{\text{span}(M_i)})$, we have the following bound after taking expectations:
\begin{equation*}
\begin{aligned}
\mathbb{E}\left[\mathrm{trace}(\Proj_T \Proj_{{T^\star}^\perp})\right] & \leq \underbrace{\mathbb{E}\left[\sum_{i = 1}^{\text{dim}({T^\star}^\perp)}\frac{2}{B}\sum_{j = 1}^{B/2}\min\{f_{2j-1,i}, f_{2j,i}\}\right]}_{\text{Term 1}} + \underbrace{\mathbb{E}\left[\sum_{i = 1}^{\text{dim}({T^\star}^\perp)}\frac{2}{B}\sum_{j = 1}^{B/2}(g_{2j-1,i}+g_{2j,i})\right]}_{\text{Term 2}} \\ & ~~~ + \underbrace{\mathbb{E}\left[\sum_{i = 1}^{\text{dim}({T^\star}^\perp)}\frac{2}{B}\sum_{j = 1}^{B/2}\max\{h_{2j-1,i}, h_{2j,i}\}\right]}_{\text{Term 3}}.
\end{aligned}
\end{equation*}
We focus on bounding each term separately.  First, considering Term 1, we have for each $\ell=1,\dots,B$ and each $i=1,\dots,\mathrm{dim}({T^\star}^\perp)$ that:
\begin{equation*}
\begin{aligned}
f_{\ell,i}  &= \mathrm{trace}(\Proj_{\hat{T}(\mathcal{D}_{\ell})} \Proj_T \Proj_{{\hat{T}(\mathcal{D}_{\ell})}} \Proj_{\mathrm{span}(M_i)}) \\
&= \mathrm{trace}(\Proj_T \Proj_{\hat{T}(\mathcal{D}_{\ell})} \Proj_{\mathrm{span}(M_i)} \Proj_{\hat{T}(\mathcal{D}_{\ell})} \Proj_T) \\ &= \|\Proj_T \Proj_{\hat{T}(\mathcal{D}_{\ell})}(M_i)\|_F^2 \\ &\leq \|\Proj_{\hat{T}(\mathcal{D}_{\ell})}(M_i)\|_F^2.
\end{aligned}
\end{equation*}
Here the second equality follows from the idempotence of projection operators and the cyclicity of trace; the third equality by the definition of the Frobenius norm; and the inequality from the property that projection reduces the Frobenius norm of a matrix.  With this relation, we bound Term 1 as follows:
\begin{equation*}
\begin{aligned}
\text{Term 1} &\leq \mathbb{E}\left[\sum_{i = 1}^{\text{dim}({T^\star}^\perp)}\frac{2}{B}\sum_{j = 1}^{B/2}\min\{\|\Proj_{\hat{T}(\mathcal{D}_{2j-1})}(M_i)\|_F^2,\|\Proj_{\hat{T}(\mathcal{D}_{2j})}(M_i)\|_F^2\}\right] \\
&\leq \mathbb{E}\left[\sum_{i = 1}^{\text{dim}({T^\star}^\perp)}\frac{2}{B}\sum_{j= 1}^{B/2}\|\Proj_{\hat{T}(\mathcal{D}_{2j-1})}(M_i)\|_F\|\Proj_{\hat{T}(\mathcal{D}_{2j})}(M_i)\|_F\right]\\
&= \sum_{i = 1}^{\text{dim}({T^\star}^\perp)}[\mathbb{E}\|\Proj_{\hat{T}(\mathcal{D}(n/2))}(M_i)\|_F]^2.
\end{aligned}
\end{equation*}
Here the second inequality follows from the property that minimum of two positive quantities is bounded above by the product of their square roots, and the equality follows from $\hat{T}(\mathcal{D}_{2j-1})$ and $\hat{T}(\mathcal{D}_{2j})$ being independent, and $\hat{T}(\mathcal{D}_\ell)$ being identically distributed for all $\ell = 1,2,\dots,\ell$. Turning next to Term 2, we have that:
\begin{equation*}
    \begin{aligned}
        \text{Term 2} &= 2 ~ \mathbb{E}\left[\frac{1}{B}\sum_{\ell = 1}^{B}\sum_{i = 1}^{\text{dim}({T^\star}^\perp)}\mathrm{trace}(\Proj_{\hat{T}(\mathcal{D}_\ell)^\perp}\Proj_T\Proj_{\hat{T}(\mathcal{D}_\ell)^\perp}\Proj_{\text{span}(M_i)})\right]\\
        &= 2 ~ \mathbb{E}\left[\frac{1}{B}\sum_{\ell = 1}^{B}\mathrm{trace}(\Proj_{\hat{T}(\mathcal{D}_\ell)^\perp}\Proj_T\Proj_{\hat{T}(\mathcal{D}_\ell)^\perp}\Proj_{{T^\star}^\perp})\right]\\
        &\leq 2 ~ \mathbb{E}\left[\frac{1}{B}\sum_{\ell = 1}^{B}\mathrm{trace}(\Proj_{\hat{T}(\mathcal{D}_\ell)^\perp}\Proj_T\Proj_{\hat{T}(\mathcal{D}_\ell)^\perp})\right] \\
        &= 2 ~ \mathbb{E}\left[\mathrm{trace}(\Proj_T(\mathcal{I}-\Proj_\texttt{avg})\Proj_T)\right] \\
        &\leq 2(1-\alpha)\text{dim}(T).
    \end{aligned}
\end{equation*}
Here the second equality follows from $\sum_{i=1}^{\mathrm{dim}({{T^\star}^\perp})} \Proj_{\text{span}(M_i)} = \Proj_{{T^\star}^\perp}$; the first inequality follows from the inequality $\mathrm{trace}(AB) \leq \mathrm{trace}(A)\|B\|_2$ for symmetric and positive-semidefinite $A$; the third equality from the definition of $\Proj_\texttt{avg}$, the idempotence of projection operators, and the cyclicity of trace; and the last inequality from the choice of $T$.  Term 3 is simply taken as is.  This concludes the basis-dependent bound.

Next we consider the basis-independent bound.  We begin with a decomposition analogous to that of \eqref{eq:masterdecomposition} along with the subsequent simplification in terms of commutators for each $\ell = 1,\dots,B$:
\begin{equation}
\begin{aligned}
\mathrm{trace}\left(\Proj_{T}\Proj_
{{T^\star}^\perp}\right) & = 
\mathrm{trace}\left(\Proj_{\hat{T}(\mathcal{D}_{\ell})}\Proj_{T}\Proj_{\hat{T}(\mathcal{D}_{\ell})}\Proj_
{{T^\star}^\perp}\right)+ \mathrm{trace}\left(\Proj_{\hat{T}(\mathcal{D}_{\ell})^\perp}\Proj_{T}\Proj_{\hat{T}(\mathcal{D}_{\ell})^\perp}\Proj_
{{T^\star}^\perp}\right)
 \\ & ~~~ + \mathrm{trace}\left(\left[\Proj_T, \Proj_{\hat{T}(\mathcal{D}_{\ell})^\perp} \right] \times \left[\Proj_
{{T^\star}^\perp},\Proj_{\hat{T}(\mathcal{D}_{\ell})}\right]\right).
\end{aligned}
\end{equation}
The remainder of the proof proceeds in an analogous fashion.

\subsection{Proof of Proposition 5 (main paper)}
\vspace{0.05in}
We use the terminology of subsection A.1 above.  We prove a bound on $\kappa_{\text{bag}}(\alpha)$ in both the basis-dependent and basis-independent settings based on the following observation for each $j=1,\dots,B/2$:
\begin{equation*}
    \max\left\{\sum_{i=1}^{\mathrm{dim}({T^\star}^\perp)} h_{2j-1,i}, \sum_{i=1}^{\mathrm{dim}({T^\star}^\perp)} h_{2j,i} \right\} \leq \sum_{i=1}^{\mathrm{dim}({T^\star}^\perp)} \max\left\{h_{2j-1,i}, h_{2j,i} \right\}.
\end{equation*}
Taking expectations on both sides and averaging over the collection of complementary pairs of bags indexed by $j=1,\dots,B/2$, the left-hand-side corresponds to the basis-independent version of $\kappa_{\text{bag}}(\alpha)$ while the right-hand-side corresponds to the basis-dependent version of $\kappa_{\text{bag}}(\alpha)$.  Consequently, it suffices to just bound the right-hand-side.  Consider the following sets for each $j=1,\dots,B/2$ and each $i=1,\dots,\mathrm{dim}({T^\star}^\perp)$: 
\begin{eqnarray*}
\mathcal{S}_{j}^{1} &=& \{i ~|~ h_{2j-1,i} = \max\{h_{2j-1},h_{2j,i}\}\}\\
\mathcal{S}_{j}^{0} &=& \{i ~|~ h_{2j,i} = \max\{h_{2j-1},h_{2j,i}\}\}.
\end{eqnarray*}
If there are some $i$ such that $h_{2j-1} = h_{2j,i}$, then the corresponding $i$ should be assigned (arbitrarily) to one of $\mathcal{S}_{j}^{0}$ or $\mathcal{S}_{j}^{1}$, exclusively, so that the sets $\mathcal{S}_{j}^{0}, \mathcal{S}_{j}^{1}$ partition $\{1,\dots,\mathrm{dim}({T^\star}^\perp)\}$.  With this notation, $\kappa_\text{bag}(\alpha)$ (basis-dependent or basis-independent) may be bounded as:
\begin{equation}
\kappa_\text{bag}(\alpha) \leq \mathbb{E}\Bigg[\frac{2}{B} \sum_{j=1}^{B/2}\left\{\sum_{i\in \mathcal{S}_j^1}h_{2j-1,i}+\sum_{i\in \mathcal{S}_j^0}h_{2j,i}\right\}\Bigg].
\label{eqn:kappa_bnd}
\end{equation}
We first bound the term $\sum_{i\in \mathcal{S}_j^0}h_{2j,i}$ as follows:
\begin{equation*}
\begin{aligned}
\sum_{i\in \mathcal{S}_j^0}h_{2j,i} &= \mathrm{trace}\left([\Proj_T, \Proj_{\hat{T}{(\mathcal{D}_{2j})}^\perp}] \times \left[\sum_{i\in \mathcal{S}_j^0}\Proj_{\mathrm{span}(M_i)} , \Proj_{\hat{T}(\mathcal{D}_{2j})} \right] \right)\\
&\leq  \|[\Proj_T, \Proj_{\hat{T}{(\mathcal{D}_{2j})}^\perp}]\|_\star \left\|\left[\sum_{i\in \mathcal{S}_j^0}\Proj_{\mathrm{span}(M_i)} , \Proj_{\hat{T}(\mathcal{D}_{2j})} \right]\right\|_2 \\
&\leq \frac{1}{2}\|[\Proj_T, \Proj_{\hat{T}{(\mathcal{D}_{2j})}^\perp}]\|_\star \\ &\leq \| \Proj_T \Proj_{\hat{T}{(\mathcal{D}_{2j})}^\perp} \|_\star \\ &\leq \| \Proj_T \Proj_{\hat{T}{(\mathcal{D}_{2j})}^\perp} \|_F \sqrt{\mathrm{dim}(T)}.
\end{aligned}
\end{equation*}
Here the first inequality holds because of the tracial H\"older inequality; the second inequality holds because the spectral norm of the commutator between two projection matrices is bounded above by $\frac{1}{2}$; the third inequality follows from the triangle inequality; and the final inequality follows from $\|A\|_\star \leq \|A\|_F\sqrt{\text{rank}(A)}$.  We can similarly bound $\sum_{i \in \mathcal{S}_j^1}h_{2j-1,i}$. Applying this to \eqref{eqn:kappa_bnd}, we obtain:
\begin{equation*}
    \begin{aligned}
    \kappa_\text{bag}(\alpha) &\leq \mathbb{E}\left[\frac{2}{B}\sum_{j = 1}^{B/2} (\|\Proj_T \Proj_{\hat{T}{(\mathcal{D}_{2j-1})}^\perp} \|_F + \| \Proj_T \Proj_{\hat{T}{(\mathcal{D}_{2j})}^\perp} \|_F) \sqrt{\mathrm{dim}(T)}  \right] \\ &= 2~\mathbb{E}\left[ \left(\frac{1}{B}\sum_{\ell=1}^{B} \|\Proj_T \Proj_{\hat{T}{(\mathcal{D}_{\ell})}^\perp} \|_F \right) \sqrt{\mathrm{dim}(T)}  \right] \\
    &\leq 2~\mathbb{E}\left[\left(\sqrt{\frac{1}{B} \sum_{\ell = 1}^{B} \|\Proj_T \Proj_{\hat{T}{(\mathcal{D}_{\ell})}^\perp} \|_F^2} \right) \sqrt{\text{dim}(T)}  \right] \\
    &\leq 2~\mathbb{E}\left[\left(\sqrt{\frac{1}{B} \sum_{\ell = 1}^{B} \mathrm{trace}(\Proj_T \Proj_{\hat{T}{(\mathcal{D}_{\ell})}^\perp} \Proj_T)} \right) \sqrt{\text{dim}(T)}  \right] \\
    &= 2~\mathbb{E}\left[\sqrt{\mathrm{trace}(\Proj_T(\mathcal{I}-\Proj_\texttt{avg})\Proj_T)}\sqrt{\text{dim}(T)}  \right] \\
    &\leq 2\sqrt{1-\alpha} ~ \mathbb{E}[\text{dim}(T)].
    \end{aligned}
\end{equation*}
Here the second inequality follows from concavity of the square root function; and the final two steps follow from the definition of $\Proj_\texttt{avg}$ and the fact that $T$ is a stable tangent space.

Next we conclude that $\mathbb{E}[\mathrm{dim}(T)] \leq \frac{q}{\alpha}$ via the following sequence of inequalities:
\begin{equation*}
    \mathbb{E}[\mathrm{dim}(T)] \leq \tfrac{1}{\alpha} \mathbb{E}[\sigma_{\text{min}}(\Proj_T \Proj_\texttt{avg} \Proj_T)] \leq \tfrac{1}{\alpha} \mathbb{E}[\mathrm{trace}(\Proj_T \Proj_\texttt{avg} \Proj_T)] \leq \tfrac{1}{\alpha} \mathbb{E}[\mathrm{trace}( \Proj_\texttt{avg})] = \tfrac{q}{\alpha}.
\end{equation*}

\subsection{Proof of Bound in Remark 4 (main paper)}
\label{section:sparseproof}
\vspace{0.05in}
We employ the notation of $f,g,h$ as presented in \eqref{eq:fgh}.  Considering the decomposition \eqref{eq:masterdecomposition} and noting that the projection operators $\Proj_{\hat{T}(\mathcal{D}_\ell)}, \Proj_{\mathrm{span}(M_i)}, \Proj_{T}$ all commute with each other in variable selection, we have that $h_{2j-1,i} = h_{2j,i} = 0$ for each $j=1,\dots,B/2$ and each $i=1,\dots,\mathrm{dim}({T^\star}^\perp)$. Hence, for each $j = 1,2,\dots,B/2$ and each $i=1,\dots,\mathrm{dim}({T^\star}^\perp)$:
\begin{equation}
\begin{aligned}
\mathrm{trace}\left(\Proj_{T}\ProjM\right) &=  f_{2j-1,i}+g_{2j-1,i} \\
\mathrm{trace}\left(\Proj_{T}\ProjM\right) &=  f_{2j,i}+g_{2j,i}.
\end{aligned}
\label{eqn:variable}
\end{equation}
Furthermore,
\begin{eqnarray*}
g_{2j-1,i} &=& \mathrm{trace}\left(\Proj_{\hat{T}(\mathcal{D}_{2j-1})^\perp}\Proj_{T}\Proj_{\hat{T}(\mathcal{D}_{2j-1})^\perp}\ProjM\right)\\
&{=}&\mathrm{trace}\left(\Proj_{T \cap \text{span}(M_i)}\Proj_{\hat{T}(\mathcal{D}_{2j-1})^\perp}\right).
\end{eqnarray*}
The second equality holds from commutativity of the projection operators in variable selection. Noticing that $\mathrm{trace}\left(\Proj_{T}\ProjM\right) = \mathrm{trace}\left(\Proj_{T \cap \text{span}(M_i)}\right)$, we move $g_{2j-1,i}$ and $g_{2j,i}$ to the left side and conclude the relation :
\begin{eqnarray}
\mathrm{trace}\left(\Proj_{T \cap \text{span}(M_i)}\Proj_{\hat{T}(\mathcal{D}_{2j-1})}\right) =  f_{2j-1}~~;~~
\mathrm{trace}\left(\Proj_{T \cap \text{span}(M_i)}\Proj_{\hat{T}(\mathcal{D}_{2j})}\right) =  f_{2j}.
\label{eqn:interm_var}
\end{eqnarray}
Notice $\Proj_{T \cap \text{span}(M_i)}$ is a diagonal matrix with with all zeros except potentially one nonzero in the diagonal. Hence,  $\mathrm{trace}(\Proj_{T \cap \text{span}(M_i)})= \{0,1\}$ and is equal to $0$ only if $\Proj_{T \cap \text{span}(M_i)}$ is an identically zero matrix. Thus, an equivalent reformulation of \eqref{eqn:interm_var} is:
\begin{equation}
\begin{aligned}
\mathrm{trace}\left(\Proj_{T \cap \text{span}(M_i)}\right)\mathrm{trace}\left(\Proj_{T \cap \text{span}(M_i)}\Proj_{\hat{T}(\mathcal{D}_{2j-1})}\right) &=&  f_{2j-1}\\
\mathrm{trace}\left(\Proj_{T \cap \text{span}(M_i)}\right)\mathrm{trace}\left(\Proj_{T \cap \text{span}(M_i)}\Proj_{\hat{T}(\mathcal{D}_{2j})}\right) &=&  f_{2j}.
\end{aligned}
\label{eqn:inter_var2}
\end{equation}
Taking the minimum over complementary bags  yield:
\begin{eqnarray*}
\min\{f_{2j-1},f_{2j}\} &=& \mathrm{trace}\left(\Proj_{T \cap \text{span}(M_i)}\right) \min_k\mathrm{trace}\left(\Proj_{T \cap \text{span}(M_i)}\Proj_{\hat{T}(\mathcal{D}_{2j-k})}\right)\\
&\geq&\mathrm{trace}\left(\Proj_{T \cap \text{span}(M_i)}\right)\prod_k\mathrm{trace}\left(\Proj_{T \cap \text{span}(M_i)}\Proj_{\hat{T}(\mathcal{D}_{2j-k})}\right)\\
&\geq& \mathrm{trace}\left(\Proj_{T \cap \text{span}(M_i)}\right)\left\{\sum_{k}\mathrm{trace}\left(\Proj_{T \cap \text{span}(M_i)}\Proj_{\hat{T}(\mathcal{D}_{2j-k})}\right)-1\right\}.
\end{eqnarray*}
Here the first inequality follows from the fact that $\min\{a,b\} \geq ab$ for $a,b \in [0,1]$. The second inequality follows from $ab \geq a+b-1$ for $a,b \in [0,1]$. We then bound $\frac{2}{B}\sum_{j = 1}^{B/2}\min\{f_{2j-1},f_{2j}\}$
\begin{eqnarray}
\frac{2}{B}\sum_{j = 1}^{B/2}\min\{f_{2j-1},f_{2j}\} &\geq& \mathrm{trace}\left(\Proj_{T \cap \text{span}(M_i)}\right) [2~\mathrm{trace}\left(\Proj_{T \cap \text{span}(M_i)}\Proj_{\texttt{avg}}\right)-1].\nonumber
\end{eqnarray}
Suppose $\Proj_{T \cap \text{span}(M_i)}$ is not zero-dimensional. Then, as $T \in \mathcal{T}_\alpha$, we find that:
\begin{eqnarray}
\mathrm{trace}\left(\Proj_{T \cap \text{span}(M_i)}\right) \leq \frac{1}{2\alpha-1}\frac{2}{B}\sum_{j = 1}^{B/2}\min\{f_{2j-1},f_{2j}\}.
\label{eqn:inter_var}
\end{eqnarray}
If $\Proj_{T \cap \text{span}(M_i)}$ is zero-dimensional, the bound \eqref{eqn:inter_var} continues to hold as $f_{2j-1},f_{2j}$ are non-negative quantities. Via the inequality $\mathrm{trace}(AB) \leq \mathrm{trace}(A)\|B\|_2$ for positive-semidefinite $A$, we have that $f_{2j-1,i} = \mathrm{trace}(\Proj_{\hat{T}(\mathcal{D}_{2j-1})}\allowbreak \Proj_T \Proj_{\hat{T}(\mathcal{D}_{2j-1})} \Proj_{\mathrm{span}(M_i)}) \leq \mathrm{trace}(\Proj_{\hat{T}(\mathcal{D}_{2j-1})} \Proj_{\mathrm{span}(M_i)})$. We substitute this into  \eqref{eqn:inter_var} to find:
\begin{eqnarray*}
\mathbb{E}[\mathrm{trace}(\Proj_T\Proj_{{T^\star}^\perp})] &=& \sum_{i = 1}^{\text{dim}({T^\star}^\perp)}\mathbb{E}\left[\mathrm{trace}\left(\Proj_{T \cap \text{span}(M_i)}\right)\right]\\
&\leq& \sum_{i = 1}^{\text{dim}({T^\star}^\perp)}\frac{\mathbb{E}\left[\frac{2}{B}\sum_{j = 1}^{B/2}\min\{\mathrm{trace}( \Proj_{\hat{T}(\mathcal{D}_{2j-1})} \Proj_{\mathrm{span}(M_i)}),\mathrm{trace}( \Proj_{\hat{T}(\mathcal{D}_{2j})} \Proj_{\mathrm{span}(M_i)})\}\right]}{2\alpha-1}\\
&=& \sum_{i = 1}^{\text{dim}({T^\star}^\perp)}\frac{\mathbb{E}\left[\frac{2}{B}\sum_{j = 1}^{B/2}\mathrm{trace}( \Proj_{\hat{T}(\mathcal{D}_{2j-1})} \Proj_{\mathrm{span}(M_i)})\mathrm{trace}( \Proj_{\hat{T}(\mathcal{D}_{2j})} \Proj_{\mathrm{span}(M_i)})\right]}{2\alpha-1}\\
&=& \sum_{i = 1}^{\text{dim}({T^\star}^\perp)}\frac{\mathbb{E}[\mathrm{trace}( \Proj_{\hat{T}(\mathcal{D}(n/2))} \Proj_{\mathrm{span}(M_i)})]^2}{2\alpha-1}.
\end{eqnarray*}
Here the second equality follows from $\mathrm{trace}( \Proj_{\hat{T}(\mathcal{D}_{2j-1})} \Proj_{\mathrm{span}(M_i)}) \in \{0,1\}$ and $\mathrm{trace}( \Proj_{\hat{T}(\mathcal{D}_{2j})} \Proj_{\mathrm{span}(M_i)}) \in \{0,1\}$; the final equality holds from $\hat{T}(\mathcal{D}_{2j-1})$ and $\hat{T}(\mathcal{D}_{2j})$ being independent and that $\hat{T}(\mathcal{D}_\ell)$ is identically distributed for all $\ell = 1,2,\dots,B$.

\subsection{When are Assumptions 1 and 2 in (3.6) Satisfied?}
\label{section:assumption}
\vspace{0.1in}
Are there reasonable estimators and models in the low-rank setting that satisfy Assumptions 1 and 2 in (3.7) (main paper)?  This section aims to address this question.

Assumption 1 is rather benign.  Specifically, fix any $k \leq \min\{p_1,p_2\}$. Let $U \in \mathbb{R}^{p_1 \times k}$ and $V \in \mathbb{R}^{p_2 \times k}$ be drawn respectively from a Haar measure on the Stiefel Manifold. Then it is straightforward to check that the tangent space $\hat{T} = T(\mathrm{span}(U),\mathrm{span}(V))$ satisfies the following condition:
\begin{eqnarray*}
\frac{\mathbb{E}\left[\text{trace}\left(\Proj_{{T^\star}^\perp}\Proj_{\hat{T}}\right)\right]}{\text{dim}({T^\star}^\perp)} = \frac{\mathbb{E}\left[\text{trace}\left(\Proj_{{T^\star}}\Proj_{\hat{T}}\right)\right]}{\text{dim}({T^\star})}.
\end{eqnarray*}
In other words, the case of equality in Assumption 1 is satisfied if the row and column space estimates are drawn uniformly at random as above, and Assumption 1 merely requires that the low-rank estimator under consideration is better than such a procedure which makes no use of any observations.

Assumption 2 is more stringent, although it is fulfilled in some natural classes of models / estimators. In particular, this assumption is satisfied when the estimator as well as the data generation process are both invariant under orthogonal conjugation.  Consider for example:
\begin{itemize}
    \item Linear regression with Gaussian functionals: consider the linear matrix regression setting where we obtain $n$ linear measurements of $L^\star$ in the form $y_i =  \langle \mathcal{A}_i,L^\star\rangle + \epsilon_i$, with each $\mathcal{A}_i \in \mathbb{R}^{p_1 \times p_2}$ consisting of i.i.d. standard Gaussian entries (and the $\mathcal{A}_i$'s being independent of each other) and $\epsilon \in \R^n$ being a standard Gaussian vector.  Consider estimators of the form:
    \begin{eqnarray*}
    \hat{L} = \argmin_L & \sum_{i=1}^n (y_i - \langle \mathcal{A}_i, L \rangle )^2 + \lambda~\mathcal{R}(L), 
    \label{eqn:genclass_Linear}
    \end{eqnarray*}
which includes as special cases a convex approach with $\mathcal{R}(L) = \|L\|_\star$ as well as a non-convex approach (solved via alternating least squares)  with $\mathcal{R}(L) = \|U\|_F^2 + \|V\|_F^2$ with $L = UV'$ corresponding to the estimator (4.2) (main paper).

\item Matrix denoising: suppose we are given $n$ observations of $L^\star$ of the form $Y_i = L^\star +\epsilon_i$. Here $\epsilon_i$ is a random matrix with i.i.d. standard Gaussian entries (and the $\epsilon_i$'s are independent of each other). Consider any spectral estimator (such as soft thresholding or hard thresholding of the singular values) applied to $\bar{Y} = \frac{1}{n} \sum_{i = 1}^{n} Y_i$ to estimate $L^\star$. 
\end{itemize}
In Section \ref{section:subspace_theory}, we provide a PCA model and a corresponding estimator that would satisfy a version of Assumption 2 suitable for subspace estimation problems. 

\subsection{Proof of Proposition 6 (main paper)}
\label{section:gammaproof}
\vspace{0.1in}
Recall that $F = \sum_{i = 1}^{\mathrm{dim}({T^\star}^\perp)} \mathbb{E}[\|\Proj_{\hat{T}(\mathcal{D}(n/2))}(M_i)\|_F]^2$ in the basis-dependent bound. Consider a collection of rank-1 basis elements $\{M_i\}_{i = 1}^{\text{dim}({T^\star}^\perp)}$. By Assumption 2, $\mathbb{E}[\|\Proj_{\hat{T}(\mathcal{D}(n/2))}(M_i)\|_F]^2 = \mathbb{E}[\|\Proj_{\hat{T}(\mathcal{D}(n/2))}(M)\|_F]^2$ for any fixed rank-1 matrix $M \in {T^\star}^\perp$ with $\|M\|_F = 1$. Letting $\delta_1 =\mathbb{E}[\|\Proj_{\hat{T}(\mathcal{D}(n/2))}(M)\|_F]$, we thus have $F = \text{dim}({T^\star}^\perp)\delta_1^2$. Define the quantity $\delta_2 = \mathbb{E}[\|\Proj_{\hat{T}(\mathcal{D}(n/2))}(M)\|_F^2]$. Then, $F$ can be bounded in terms of $\delta_2$ and $|\delta_1-\delta_2|$ as follows:
\begin{equation}
\begin{aligned}
F &=\text{dim}({T^\star}^\perp) \delta_1^2\\
&= \text{dim}({T^\star}^\perp)\{\delta_2^2 + (\delta_1-\delta_2)^2+2\delta_2(\delta_1-\delta_2)\}\\
&\leq \text{dim}({T^\star}^\perp)\{\delta_2^2 + (\delta_1-\delta_2)^2+2\delta_2|\delta_1-\delta_2|\}.
\end{aligned}
\label{bound:delta}
\end{equation}
We focus on bounding $\delta_2$ and $|\delta_1-\delta_2|$. To bound $\delta_2$, note that:
\begin{eqnarray*}
\mathbb{E}\left[\text{trace}\left(\Proj_{\hat{T}(\mathcal{D}(n/2))}\Proj_{{T^\star}^\perp}\right) \right] + \mathbb{E}\left[\text{trace}\left(\Proj_{\hat{T}(\mathcal{D}(n/2))}\Proj_{{T^\star}}\right) \right] = \mathbb{E}[\text{dim}(\hat{T}(\mathcal{D}({n/2})))] =q. 
\end{eqnarray*}
Employing ``better than random guessing" Assumption 1, we find that:
\begin{eqnarray*}
\mathbb{E}\left[\text{trace}\left(\Proj_{\hat{T}(\mathcal{D}({n/2}))}\Proj_{{T^\star}^\perp}\right) \right]\left(1+\frac{\text{dim}(T^\star)}{\text{dim}({T^\star}^\perp)}\right) \leq q.
\end{eqnarray*}
Since $\text{dim}(T^\star)+\text{dim}({T^\star}^\perp) = p_1p_2$, we find
\begin{eqnarray}
\mathbb{E}\left[\text{trace}\left(\Proj_{\hat{T}(\mathcal{D}({n/2}))}\Proj_{{T^\star}^\perp}\right) \right] \leq \frac{q}{p_1p_2}\text{dim}({T^\star}^\perp).
\label{eqn:betterthan}
\end{eqnarray}
We now express the left-hand side of \eqref{eqn:betterthan} in terms of $\delta_2$. By  Assumption 2, we have that \\ $\mathbb{E}\left[\text{trace}\left(\Proj_{\hat{T}(\mathcal{D}({n/2}))}\Proj_{{T^\star}^\perp}\right) \right] = \text{dim}({T^\star}^\perp)\delta_2$. Combining this with \eqref{eqn:betterthan}, we obtain the bound:
\begin{eqnarray}
\delta_2\leq \frac{q}{p_1p_2}.
\label{eqn:0}
\end{eqnarray}

Now we focus on bounding $|\delta_1-\delta_2|$. We proceed by bounding $\delta_1-\delta_2$ and $\delta_2-\delta_1$ by the same quantity. In particular, we show that $\delta_1-\delta_2 \leq \kappa_\text{indiv}$ where $\kappa_\text{indiv} =\mathbb{E}\left\|\left[\Proj_{\hat{T}(\mathcal{D}(n/2))},\Proj_{\mathrm{span}(M)}\right]\right\|_F$:
\begin{equation*}
\begin{aligned}
\delta_1-\delta_2 &=\mathbb{E}\left[\|\Proj_{\hat{T}(\mathcal{D}({n/2}))}(M)\|_F\right]-\mathbb{E}\left[\|\Proj_{\hat{T}(\mathcal{D}({n/2}))}(M)\|_F^2\right] \nonumber\\
&{\stackrel{(a)}=} \mathbb{E}\left[\left\|\Proj_{\hat{T}(\mathcal{D}({n/2}))}\Proj_{\mathrm{span}(M)}\right\|_F\right] - \mathbb{E}\left[\left\|\Proj_{\hat{T}(\mathcal{D}({n/2}))}\Proj_{\mathrm{span}(M)}\right\|_F^2\right] \nonumber \\
&{\stackrel{(b)}=} \mathbb{E}\left[\left\|\Proj_{\hat{T}(\mathcal{D}({n/2}))}\Proj_{\mathrm{span}(M)}\right\|_F\right] - \mathbb{E}\left[\left\|\Proj_{\hat{T}(\mathcal{D}({n/2}))}\Proj_{\mathrm{span}(M)}\Proj_{\hat{T}(\mathcal{D}({n/2}))}\right\|_F\right] \nonumber\\
&{\stackrel{(c)}=} \mathbb{E}\left[\|\Proj_{\hat{T}(\mathcal{D}({n/2}))}\Proj_{\mathrm{span}(M)}]\|_F\right]\nonumber\\& ~~ -\mathbb{E}\Bigg[\Big\|\Proj_{\mathrm{span}(M)}\Proj_{\hat{T}(\mathcal{D}({n/2}))} +\left[\Proj_{\hat{T}(\mathcal{D}({n/2}))},\Proj_{\mathrm{span}(M)}\right]\Proj_{\hat{T}(\mathcal{D}(n/2))} \Big\|_F\Bigg] \nonumber \\
&{\stackrel{(d)}\leq} \mathbb{E}\left\|\left[\Proj_{\hat{T}(\mathcal{D}(n/2))},\Proj_{\mathrm{span}(M)}\right]\Proj_{\hat{T}(\mathcal{D}(n/2))}\right\|_F {\stackrel{(e)}\leq}\mathbb{E}\left\|\left[\Proj_{\hat{T}(\mathcal{D}(n/2))},\Proj_{\mathrm{span}(M)}\right]\right\|_F = \kappa_\text{indiv}.
\label{eqn:2}
\end{aligned}
\end{equation*}
Here ${\stackrel{(a)}=}$ follows from the property that $\mathbb{E}\left[\allowbreak\|\Proj_{\hat{T}(\mathcal{D}({n/2}))}(M)\|_F\right] = \mathbb{E}\left[\|\Proj_{\hat{T}(\mathcal{D}({n/2}))}\Proj_{\mathrm{span}(M)}\|_F\right]$ and that \\$\mathbb{E}\left[\|\Proj_{\hat{T}(\mathcal{D}({n/2}))}\allowbreak(M)\|_F^2\right] = \mathbb{E}\left[\|\Proj_{\hat{T}(\mathcal{D}({n/2}))}\Proj_{\mathrm{span}(M)}\|_F^2\right]$; 
${\stackrel{(b)}=}$ follows from noting that $\Proj_{\mathrm{span}(M)}$ has rank-1 by construction so that $\left\|\Proj_{\hat{T}(\mathcal{D}({n/2}))}\allowbreak\Proj_{\mathrm{span}(M)}\allowbreak\Proj_{\hat{T}(\mathcal{D}({n/2}))}\right\|_F = \mathrm{trace}\left(\Proj_{\hat{T}(\mathcal{D}({n/2}))}\Proj_{\mathrm{span}(M)}\Proj_{\hat{T}(\mathcal{D}({n/2}))}\right) = \left\|\Proj_{\hat{T}(\mathcal{D}({n/2}))}\Proj_{\mathrm{span}(M)}\right\|_F^2$; ${\stackrel{(c)}=}$ follows from the definition of a commutator; ${\stackrel{(d)} \leq }$ follows from reverse triangle inequality; and ${\stackrel{(e)} \leq }$ follows from the following reasoning for matrices $A,B$ where $B$ is a projection matrix: $\|AB\|_F = \sqrt{\|AB\|_F^2} = \sqrt{\mathrm{trace}(A'AB)} \leq \sqrt{\mathrm{trace}(A'A)\|B\|_2} \leq \sqrt{\mathrm{trace}(A'A)} = \|A\|_F$.  \\

Similar logic shows that $\delta_2-\delta_1 \leq \kappa_\text{indiv}$ which leads to the conclusion that $|\delta_1-\delta_2| \leq \kappa_\text{indiv}$. Plugging in the bounds for $\delta_2$ and $|\delta_1-\delta_2|$ into \eqref{bound:delta}, we find that:
\begin{eqnarray*}
F \leq \frac{q^2}{p_1p_2}+\text{dim}({T^\star}^\perp)\kappa_\text{indiv}^2 + 2q\kappa_\text{indiv},
\end{eqnarray*}
as desired.

\subsection{Goodness of the Data-Driven Heuristic in Remark 5}
\label{section:heuristic}
\vspace{0.1in}
Recall that we chose ${M} = uv'$, where $u,v$ are selected to be the smallest singular vectors associated with $\Proj_\texttt{avg}^\mathcal{C}$ and $\Proj_\texttt{avg}^\mathcal{R}$, respectively. Notice that $\Proj_{{T^\star}^\perp}(M)$ will be of rank less than or equal to $1$ since $\Proj_{{T^\star}^\perp}(M) = \Proj_{{\mathcal{C}^\star}^\perp}M\Proj_{{\mathcal{R}^\star}^\perp}$. Hence, the cosine of the largest principal angle between ${T^\star}^\perp$ and $M$, given by $\|\Proj_{{T^\star}^\perp}\Proj_{\mathrm{span}(M)}\|_F$, will be achieved between the direction spanned by $M$ and a rank-1 direction in ${T^\star}^\perp$. As such, we next prove that if the estimator has good power, $\|\Proj_{{T^\star}^\perp}\Proj_{\mathrm{span}(M)}\|_F$ will be close to $1$. 
\begin{lemma} Let $\tau := \mathbb{E}[\min\{\sigma_\mathrm{min}(\Proj_{{\mathcal{C}^\star}}\Proj_{{\hat{\mathcal{C}}(\mathcal{D}(n/2))}}\Proj_{{\mathcal{C}}^\star}),\sigma_\mathrm{min}(\Proj_{{\mathcal{R}^\star}}\Proj_{{\hat{\mathcal{R}}(\mathcal{D}(n/2))}}\Proj_{{\mathcal{R}^\star}})\}]$ and\\ $\delta := \mathbb{E}[\max\{\sigma_\mathrm{min}(\Proj_\texttt{avg}^\mathcal{C}),\sigma_\mathrm{min}(\Proj_\texttt{avg}^\mathcal{R})\}]$. Then, the expected cosine of the principal angle between the data-driven $M$ and ${T^\star}^\perp$ is lower-bounded by:
$$\mathbb{E}\left[\|\Proj_{{T^\star}^\perp}\Proj_{\mathrm{span}(M)}\|_F^2\right] \geq 2\tau-1-2(\delta+\sqrt{\delta}).$$
\end{lemma}
Evidently, when the estimator has good power, i.e. $\tau$ is close to $1$, and the expected smallest singular values $\delta$ is close to $0$, the data-driven approach produces an $M$ that is close to ${T^\star}^\perp$. We next prove this lemma.
\begin{proof}
Notice that:
\begin{eqnarray}
\mathbb{E}\left[\mathrm{trace}\left(\Proj_{{T^\star}^\perp}\Proj_{\mathrm{span}(M)}\right)\right] &=& \mathbb{E}\left[\mathrm{trace}\left(\Proj_{{\mathcal{C}^\star}^\perp}\Proj_{\mathrm{span}(u)}\right)\mathrm{trace}\left(\Proj_{{\mathcal{R}^\star}^\perp}\Proj_{\mathrm{span}(v)}\right)\right] \nonumber\\
&\geq& \mathbb{E}\left[\mathrm{trace}\left(\Proj_{{\mathcal{C}^\star}^\perp}\Proj_{\mathrm{span}(u)}\right)\right]+\mathbb{E}\left[\mathrm{trace}\left(\Proj_{{\mathcal{R}^\star}^\perp}\Proj_{\mathrm{span}(v)}\right)\right]-1\nonumber\\
&=& 1-\mathbb{E}\left[\mathrm{trace}\left(\Proj_{{\mathcal{C}^\star}}\Proj_{\mathrm{span}(u)}\right)\right]-\mathbb{E}\left[\mathrm{trace}\left(\Proj_{{\mathcal{R}}^\star}\Proj_{\mathrm{span}(v)}\right)\right],
\label{eqn:decomp_good}
\end{eqnarray}
where the first equality is due to the property that $\Proj_{{T^\star}^\perp} = {\mathcal{C}^\star}^\perp \otimes {\mathcal{R}^\star}^\perp$ and the inequality is due to the property $ab \geq a+b-1$ for $a,b \in [0,1]$. This decomposition implies that upper bounds for $\mathbb{E}\left[\mathrm{trace}\left(\Proj_{{\mathcal{C}^\star}}\Proj_{\mathrm{span}(u)}\right)\right]$ and $\mathbb{E}\left[\mathrm{trace}\left(\Proj_{{\mathcal{R}^\star}}\Proj_{\mathrm{span}(v)}\right)\right]$ yield a lower-bound for  $\mathbb{E}\left[\mathrm{trace}\left(\Proj_{{T^\star}^\perp}\Proj_{\mathrm{span}(M)}\right)\right]$. Proceeding with upper-bounding $\mathbb{E}\left[\mathrm{trace}\left(\Proj_{{\mathcal{C}^\star}}\Proj_{\mathrm{span}(u)}\right)\right]$, we consider the following decomposition:
\begin{eqnarray*}
\mathrm{trace}\left(\Proj_{{\mathcal{C}^\star}}\Proj_{\mathrm{span}(u)}\right) &=& \mathrm{trace}\left(\Proj_{{\mathcal{\hat{C}}(\mathcal{D}_\ell)}^\perp}\Proj_{{\mathcal{C}^\star}}\Proj_{{\mathcal{\hat{C}}(\mathcal{D}_\ell)}^\perp}\Proj_{\mathrm{span}(u)}\right)+\mathrm{trace}\left(\Proj_{{\mathcal{\hat{C}}(\mathcal{D}_\ell)}}\Proj_{{\mathcal{C}^\star}}\Proj_{{\mathcal{\hat{C}}(\mathcal{D}_\ell)}}\Proj_{\mathrm{span}(u)}\right)\\
&+& \mathrm{trace}\left(\Proj_{{\mathcal{\hat{C}}(\mathcal{D}_\ell)}^\perp}\Proj_{{\mathcal{C}^\star}}\Proj_{{\mathcal{\hat{C}}(\mathcal{D}_\ell)}}\Proj_{\mathrm{span}(u)}\right)+\mathrm{trace}\left(\Proj_{{\mathcal{\hat{C}}(\mathcal{D}_\ell)}}\Proj_{{\mathcal{C}^\star}}\Proj_{{\mathcal{\hat{C}}(\mathcal{D}_\ell)}^\perp}\Proj_{\mathrm{span}(u)}\right) \\
&\leq & \left\|\Proj_{{\mathcal{\hat{C}}(\mathcal{D}_\ell)}^\perp}\Proj_{{\mathcal{C}^\star}}\Proj_{{\mathcal{\hat{C}}(\mathcal{D}_\ell)}^\perp}\right\|_2 + \mathrm{trace}\left(\Proj_{{\mathcal{\hat{C}}(\mathcal{D}_\ell)}}\Proj_{\mathrm{span}(u)}\Proj_{{\mathcal{\hat{C}}(\mathcal{D}_\ell)}}\right)+\left\|\Proj_{{\mathcal{\hat{C}}(\mathcal{D}_\ell)}}\Proj_{\mathrm{span}(u)}\right\|_\star,
\end{eqnarray*}
where the inequality is due to $\mathrm{trace}(AB) \leq \mathrm{trace}(A)\|B\|_2$ for $A \succeq 0$, $\mathrm{trace}(AB) \leq \|A\|_\star\|B\|_2$, the idempotence of projection operators and that $\|[\Proj_{T_1},\Proj_{T_2}]\|_2 \leq\frac{1}{2}$ for any two subspaces $T_1$ and $T_2$. Since the choice of $\ell$ was arbitrary, we minimize over the entire collection:
\begin{eqnarray*}
\mathbb{E}\left[\mathrm{trace}\left(\Proj_{{\mathcal{C}^\star}}\Proj_{\mathrm{span}(u)}\right)\right] &\leq& \mathbb{E}\Bigg[\min_{\ell = 1,2,\dots,B} \left\|\Proj_{{\mathcal{\hat{C}}(\mathcal{D}_\ell)}^\perp}\Proj_{{\mathcal{C}^\star}}\Proj_{{\mathcal{\hat{C}}(\mathcal{D}_\ell)}^\perp}\right\|_2 + \mathrm{trace}\left(\Proj_{{\mathcal{\hat{C}}(\mathcal{D}_\ell)}}\Proj_{\mathrm{span}(u)}\Proj_{{\mathcal{\hat{C}}(\mathcal{D}_\ell)}}\right)\\&+&
\left\|\Proj_{{\mathcal{\hat{C}}(\mathcal{D}_\ell)}}\Proj_{\mathrm{span}(u)}\right\|_\star\Bigg] \\
&{\stackrel{(a)}\leq}& \frac{1}{B}\sum_{\ell = 1}^B \mathbb{E}\left[\left\|\Proj_{{\mathcal{\hat{C}}(\mathcal{D}_\ell)}^\perp}\Proj_{{\mathcal{C}^\star}}\Proj_{{\mathcal{\hat{C}}(\mathcal{D}_\ell)}^\perp}\right\|_2\right] + \mathbb{E}\left[\mathrm{trace}\left(\Proj_{{\mathcal{\hat{C}}(\mathcal{D}_\ell)}}\Proj_{\mathrm{span}(u)}\Proj_{{\mathcal{\hat{C}}(\mathcal{D}_\ell)}}\right)\right]\\&+&
\mathbb{E}\left[\left\|\Proj_{{\mathcal{\hat{C}}(\mathcal{D}_\ell)}}\Proj_{\mathrm{span}(u)}\right\|_\star\right] \\
&{\stackrel{(b)}\leq}& \frac{1}{B}\sum_{\ell = 1}^B \mathbb{E}\left[\left\|\Proj_{{\mathcal{\hat{C}}(\mathcal{D}_\ell)}^\perp}\Proj_{{\mathcal{C}^\star}}\Proj_{{\mathcal{\hat{C}}(\mathcal{D}_\ell)}^\perp}\right\|_2\right] + \mathbb{E}\left[\mathrm{trace}\left(\Proj_{\mathrm{span}(u)}\Proj_\texttt{avg}^\mathcal{C}\Proj_{\mathrm{span}(u)}\right)\right]\\&+& \mathbb{E}\left[\sqrt{\mathrm{trace}\left(\Proj_{\mathrm{span}(u)}\Proj_\texttt{avg}^\mathcal{C}\Proj_{\mathrm{span}(u)}\right)}\right] \\
&{\stackrel{(c)}\leq}& \frac{1}{B}\sum_{\ell = 1}^B \mathbb{E}\left[\left\|\Proj_{{\mathcal{\hat{C}}(\mathcal{D}_\ell)}^\perp}\Proj_{{\mathcal{C}^\star}}\Proj_{{\mathcal{\hat{C}}(\mathcal{D}_\ell)}^\perp}\right\|_2\right] +\delta+\sqrt{\delta} \\
&{\stackrel{(d)}=}& 1-\tau +\delta+\sqrt{\delta}.
\end{eqnarray*}
Here ${\stackrel{(a)}\leq}$ follows from the fact that minimum over a collection is bounded by their average; ${\stackrel{(b)}\leq}$ follows from $\|A\|_\star \leq \|A\|_F \text{rank}(A)$ and the concavity of square root function; and ${\stackrel{(c)}\leq}$ follows from the fact that $u$ is selected to be the smallest singular vector of $\Proj_\texttt{avg}^\mathcal{C}$, concavity of square function and Jensen's inequality, and the definition of $\delta$, and ${\stackrel{(d)}=}$ follows from the fact that $\|\Proj_{T_1^\perp}\Proj_{T_2}\Proj_{T_1^\perp}\|_2 = 1-\sigma_\mathrm{min}(\Proj_{T_2}\Proj_{T_1}\Proj_{T_2})$ and that $\hat{\mathcal{C}}(\mathcal{D}_\ell)$ is identically distributed for all $\ell$. Repeating the same steps for the row-space and combining with \eqref{eqn:decomp_good} gives the desired result. 
\end{proof}

\subsection{Sensitivity of Subspace Stability Selection to $\alpha$}
\label{section:alpha_sensitivity}
\vspace{0.1in}
The tuning parameter $\alpha \in [0,1]$ plays an important role in how much discovery is made by subspace stability selection. In our experience, the output of subspace stability selection (which selects a stable tangent space) is rather robust to $\alpha$ in moderate to high SNR settings. As a result, in all our experiments, we select $\alpha$ to equal $0.70$.

\begin{figure}[thbp]
\centering
\subfigure[FD normalized: rank = 1]{
%  % Requires \usepackage{graphicx}
 \includegraphics[scale = 0.28]{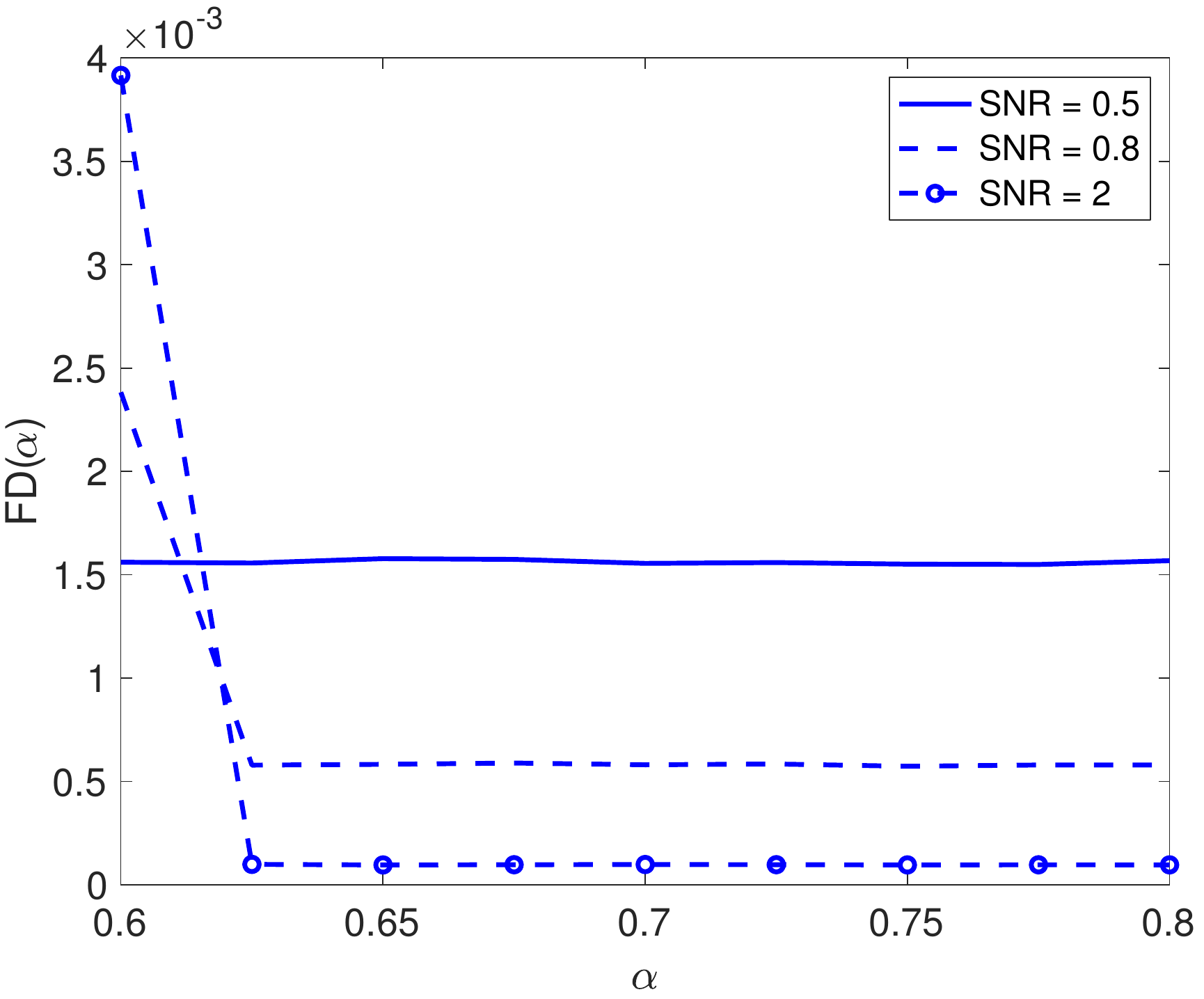}}
\subfigure[normalized PWR : rank = 1]{
%  % Requires \usepackage{graphicx}
 \includegraphics[scale = 0.28]{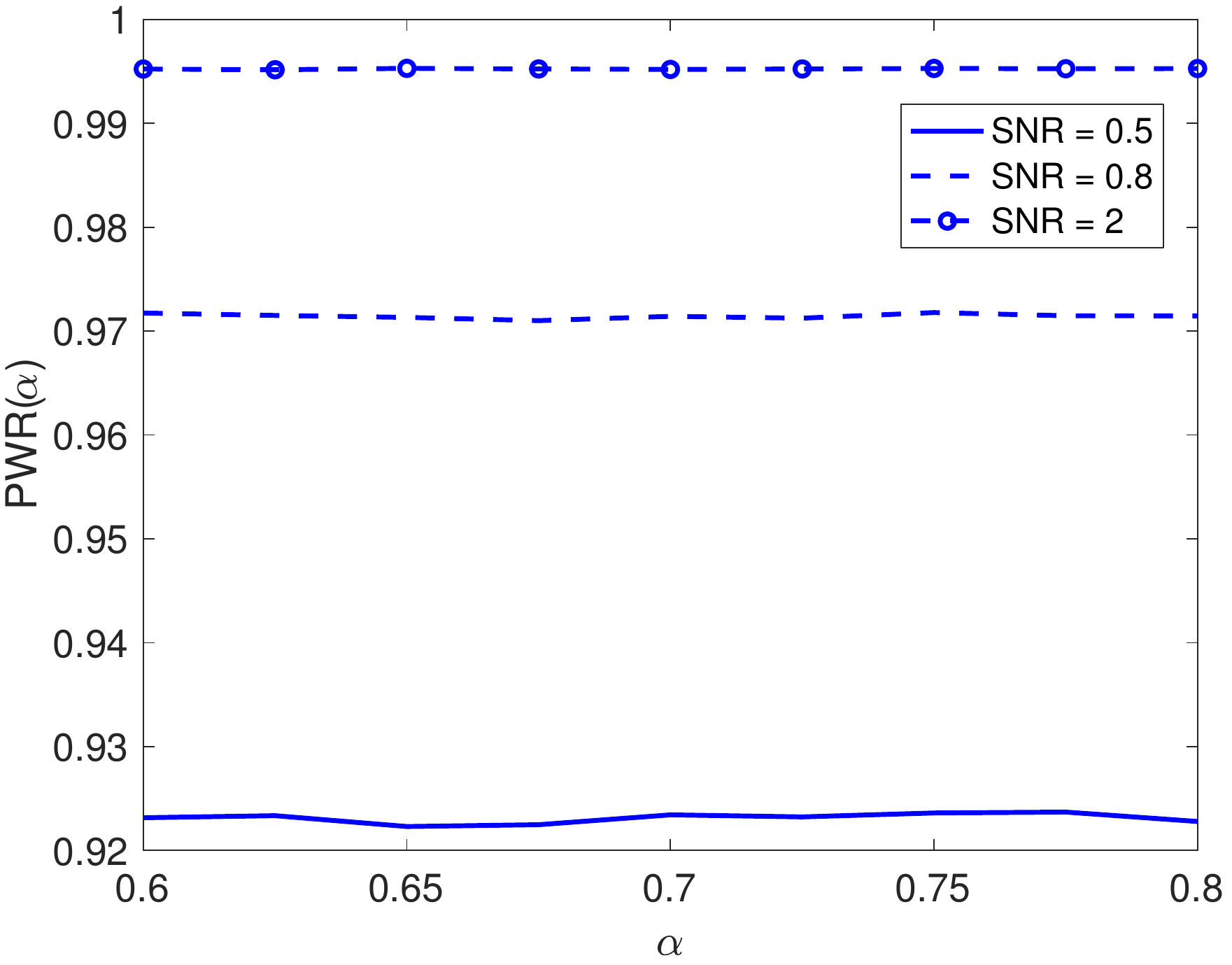}}
\subfigure[normalized FD: rank = 3]{
%  % Requires \usepackage{graphicx}
 \includegraphics[scale = 0.28]{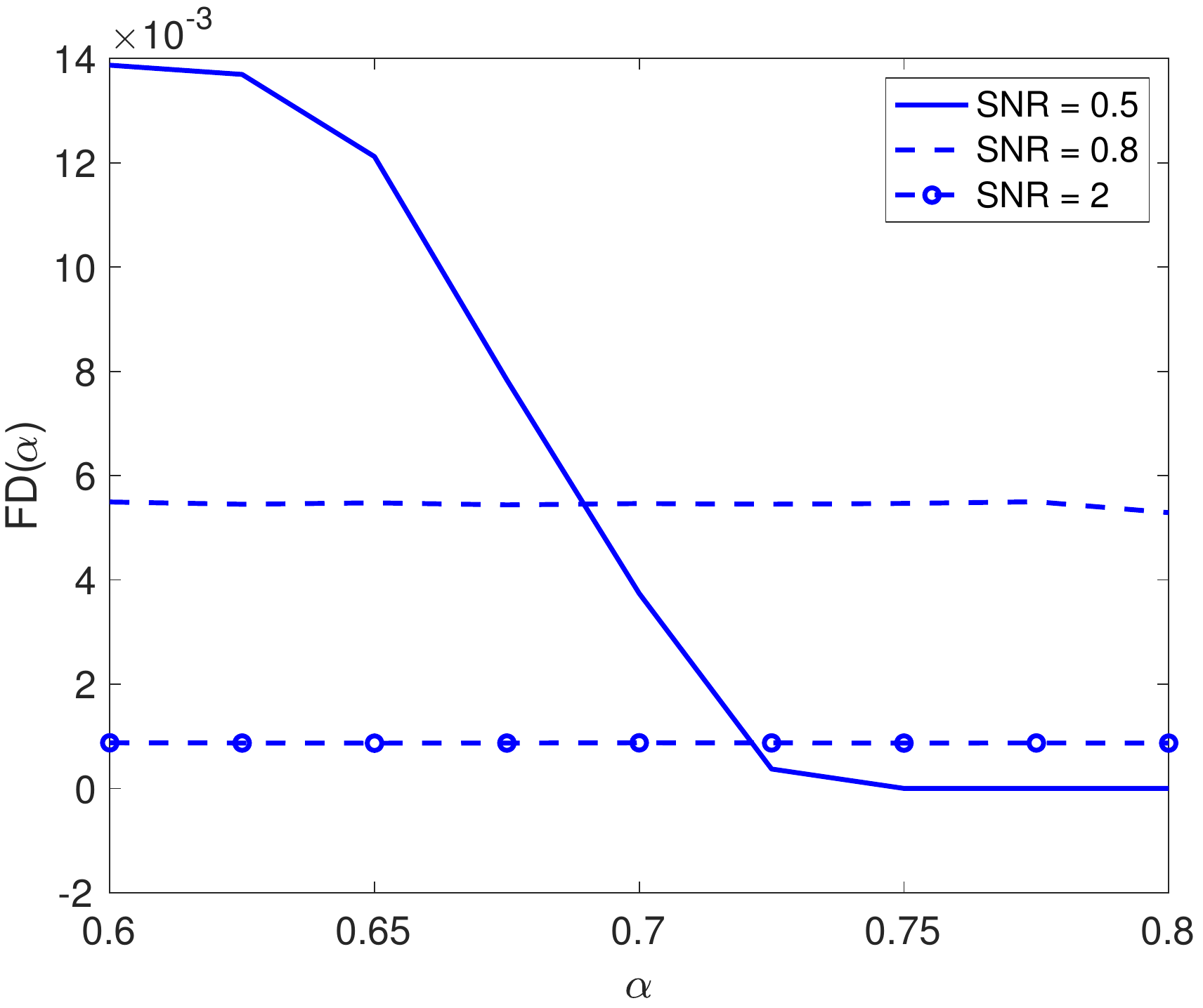}}
\subfigure[normalized PWR: rank = 3]{
%  % Requires \usepackage{graphicx}
 \includegraphics[scale = 0.28]{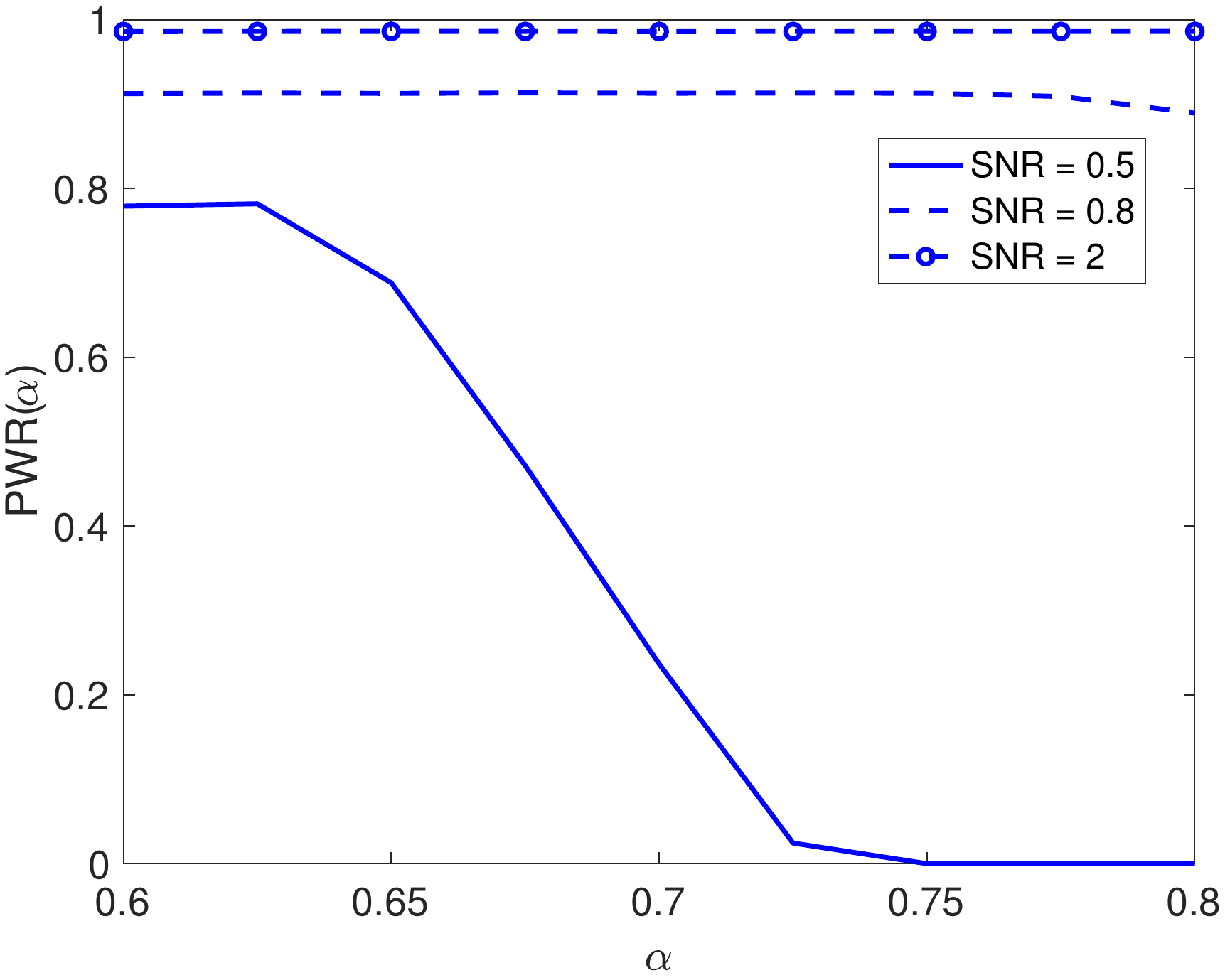}}
\subfigure[normalized FD: rank = 5]{
%  % Requires \usepackage{graphicx}
 \includegraphics[scale = 0.28]{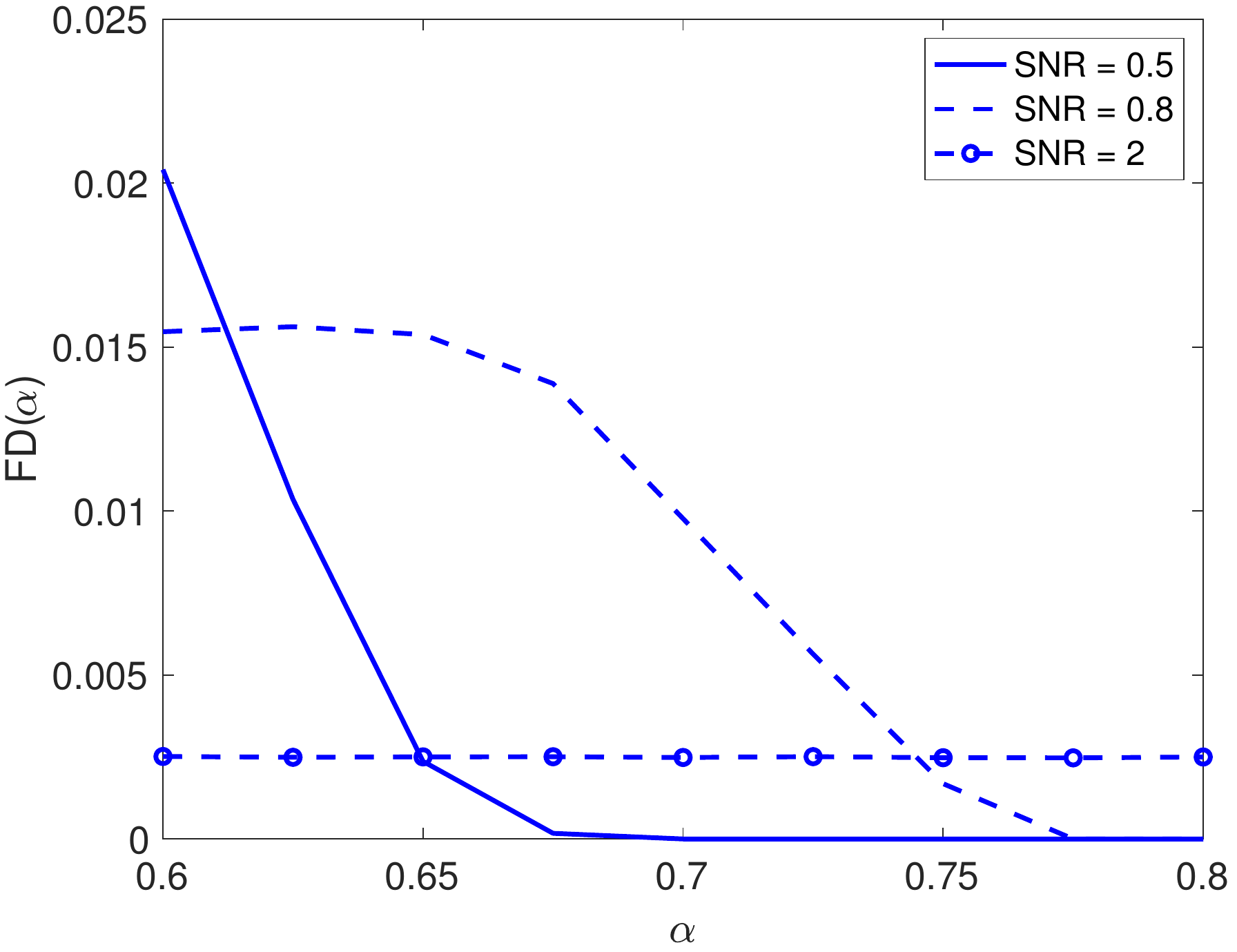}}
\subfigure[normalized PWR: rank = 5]{
%  % Requires \usepackage{graphicx}
 \includegraphics[scale = 0.28]{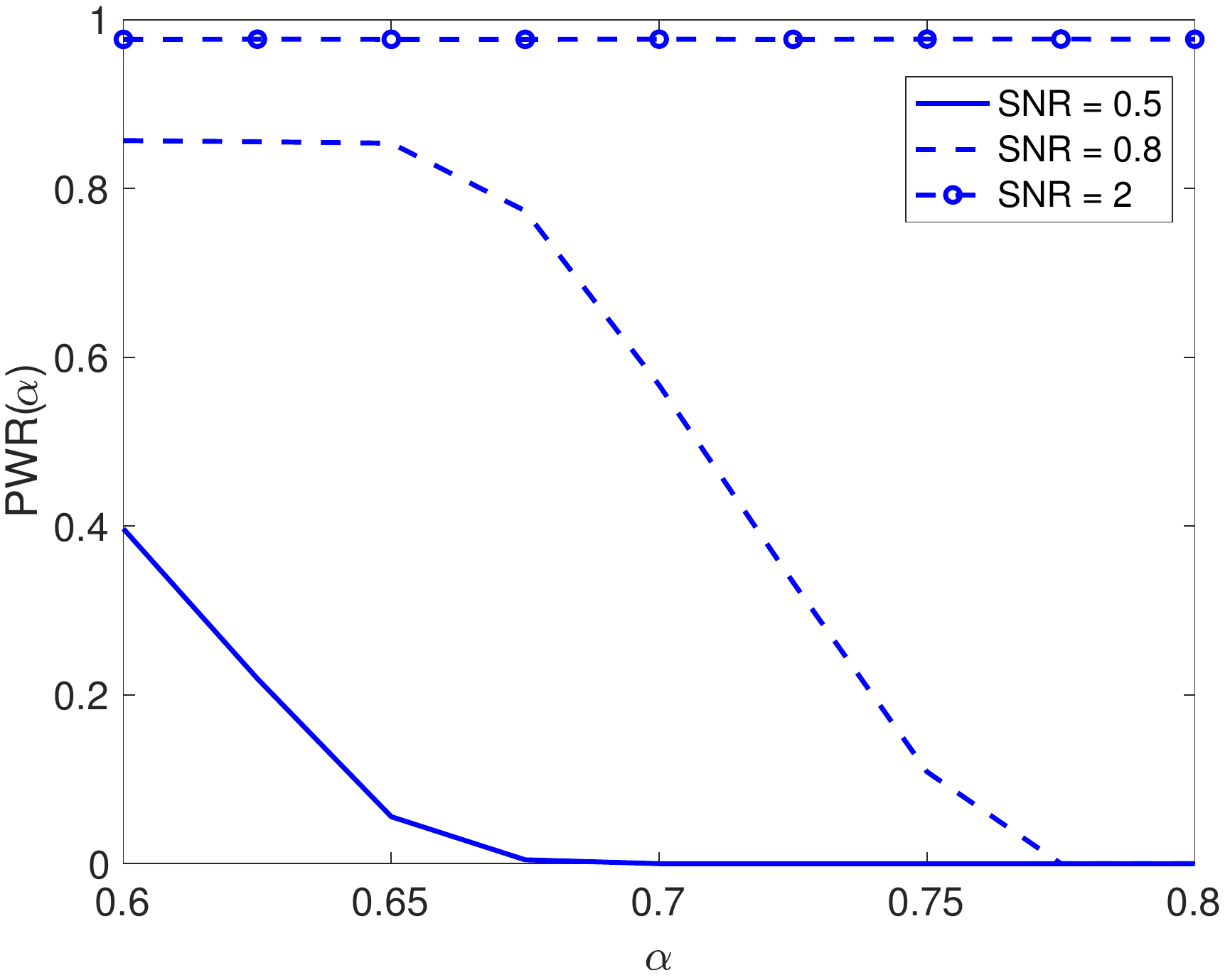}}
%\subfigure[MovieLens]{%  % Requires \usepackage{graphicx}
 %\includegraphics[scale = 0.25]{movielens.pdf}}
\caption{Variation in false discovery $\mathbb{E}\left[\mathrm{trace}\left(\Proj_{T_\texttt{S3}(\alpha)}\Proj_{{T^\star}^\perp}\right)\right]/\mathrm{dim}({T^\star}^\perp)$ and power $\mathbb{E}\left[\mathrm{trace}\left(\Proj_{T_\texttt{S3}(\alpha)}\Proj_{{T^\star}}\right)\right]/\mathrm{dim}(T^\star)$ as a function of $\alpha$ for different SNR and rank regimes.}
\label{fig:alphasta}
\end{figure}

To more systematically explore the sensitivity of the subspace stability selection algorithm to the choice of $\alpha$, we consider the following matrix completion setup where $L^\star \in \mathbb{R}^{p \times p}$ with $p = 100$, rank of $L^\star$ in the set $\{1,3,5\}$, and row/column spaces chosen uniformly at random from the Steifel manifold. We select a fraction $7/10$ of the total entries uniformly at random as the observation set $\Omega$ so that $|\Omega| = 7p^2/10$. These observations are corrupted with Gaussian noise with variance selected so that the SNR is one of the values $\{0.5,0.8,2\}$, for a total number of nine problem instances (three different noise levels and three different ranks). We use these observations as input to the estimator (4.1) (main paper), with $\lambda$ selected based on holdout validation on a $n_\text{test} = 7/20p^2$ validation set. We fix $B = 100$ and vary the choice of in Algorithm 1 (main paper) over the values in the set $\alpha_\mathrm{set} = \{0.6, 0.625, 0.65, 0.675, 0.7, 0.725, 0.75, 0.775, 0.8\}$. For each $\alpha$, we obtain an associated stable tangent space $T_{\texttt{S3}(\alpha)}$. Figure ~\ref{fig:alphasta} demonstrates the variation in the normalized false discovery $\mathbb{E}\left[\mathrm{trace}\left(\Proj_{T_\texttt{S3}(\alpha)}\Proj_{{T^\star}^\perp}\right)\right]/\mathrm{dim}({T^\star}^\perp)$ and normalized power $\mathbb{E}\left[\mathrm{trace}\left(\Proj_{T_\texttt{S3}(\alpha)}\Proj_{{T^\star}}\right)\right]/\mathrm{dim}(T^\star)$ as a function of $\alpha$. We notice that for $\text{SNR} = 2$, both the false discovery and power are very stable with respect to $\alpha$ for all ranks. Even for a lower value of $\text{SNR}= 0.8$, the normalized false discovery and power remain stable to changes in $\alpha$ for small ranks, but are less stable for larger ranks. In summary, this experiment indicates that the subspace stability selection algorithm tends to be robust to perturbations of $\alpha$ for moderate-to-high SNR regimes and small ranks.  

We note that the choice of $\alpha$ can also be guided by our theoretical results. In particular, in cases where the signal strength is strong so that the commutator terms are small (see the theoretical statements in Section 3.2 (main paper)), we recommend selecting a large $\alpha$ to maximize power while controlling for false discoveries. \\

\subsection{Proof of Proposition 7 (main paper)}
\label{section:largescaleproof}
\vspace{0.1in}
Let $T$ be a tangent space produced by the modified algorithm with associated column and row spaces $(\mathcal{C},\mathcal{R})$.  We proceed by obtaining an upper bound on $\|\mathcal{P}_T(\mathcal{I}-\mathcal{P}_{\texttt{avg}})\mathcal{P}_T\|_2$, which gives a lower bound on $\sigma_{\text{min}}(\mathcal{P}_T \mathcal{P}_{\texttt{avg}} \mathcal{P}_T)$:
\begin{eqnarray*}
\|\mathcal{P}_T(\mathcal{I}-\mathcal{P}_{\texttt{avg}})\mathcal{P}_T\|_2
&=&
\max_{M \in T, \|M\|_{F} = 1} \frac{1}{B}\text{trace}\Big(\sum_{\ell = 1}^{B} M'\mathcal{P}_{\hat{T}(\mathcal{D}_{\ell})^\perp}(M)\Big) \\
&{\stackrel{(a)}=}&
\max_{M \in T, \|M\|_{F} = 1} \frac{1}{B}\sum_{\ell = 1}^{B} \|\mathcal{P}_{\hat{\mathcal{C}}(\mathcal{D}_{\ell})^\perp}M\mathcal{P}_{\hat{\mathcal{R}}(\mathcal{D}_{\ell})^\perp}\|_F^2 \\
&{\stackrel{(b)}\leq}& 
\max_{M \in T, \|M\|_{F} = 1} \frac{2}{B}\sum_{\ell = 1}^{B}\|\mathcal{P}_{\hat{\mathcal{C}}(\mathcal{D}_{\ell})^\perp}\Proj_{\mathcal{C}}M\mathcal{P}_{\hat{\mathcal{R}}(\mathcal{D}_{\ell})^\perp}\|_F^2\\
&+& \frac{2}{B}\sum_{\ell = 1}^{B} \|\mathcal{P}_{\hat{\mathcal{C}}(\mathcal{D}_{\ell})^\perp}\Proj_{\mathcal{C}^\perp}M\Proj_{\mathcal{R}}\mathcal{P}_{\hat{\mathcal{R}}(\mathcal{D}_{\ell})^\perp}\|_F^2 \\
&{\stackrel{(c)}\leq}&
\max_{M \in T, \|M\|_{F} = 1} \frac{2}{B}\sum_{\ell = 1}^{B} \|\mathcal{P}_{\hat{\mathcal{C}}(\mathcal{D}_{\ell})^\perp}\Proj_{\mathcal{C}}M\|_F^2 +  \frac{2}{B}\sum_{\ell = 1}^{B}\|\mathcal{P}_{\hat{\mathcal{R}}(\mathcal{D}_{\ell})^\perp}\Proj_{\mathcal{R}}M'\|_F^2\\
&=&
\max_{M \in T, \|M\|_{F} = 1} 2~\mathrm{trace}(\Proj_\mathcal{C}(\mathcal{I}-\Proj_\texttt{avg})\Proj_\mathcal{C}MM') + 2~\mathrm{trace}(\Proj_\mathcal{R}(\mathcal{I}-\Proj_\texttt{avg})\Proj_\mathcal{R}M'M)\\
&\leq&
2~\|\Proj_\mathcal{C}(\mathcal{I}-\Proj_\texttt{avg})\Proj_\mathcal{C}\|_2+2~\|\Proj_\mathcal{R}(\mathcal{I}-\Proj_\texttt{avg})\Proj_\mathcal{R}\|_2 \leq 4(1-\alpha).
\end{eqnarray*}
Here $(a)$ follows from the cyclicity of the trace functional and the idempotence of projection maps; $(b)$ from the fact that $M \in  T$ implies that $M = \mathcal{P}_{\mathcal{C}} M + \mathcal{P}_{\mathcal{C}^\perp} M \mathcal{P}_{\mathcal{R}}$ and the elementary inequality $(a+b)^2\leq 2a^2+2b^2$; and $(c)$ from the property $\|A \Proj\|_{F} \leq \|A\|_{F}$ for any projection matrix $\Proj$.

\subsection{Tangent Spaces for Column-Space Estimation}
\label{section:subspace_tangent}
\vspace{0.1in}
In certain domains such as hyperspectral imaging, one only requires estimates of the column-space of a low-rank matrix, and we seek an appropriate tangent space that represents the discoveries in this context.

We begin by considering the tangent space with respect to the determinantal variety $\mathcal{V}(r) \subset \mathbb{R}^{p_1 \times p_2}$ at a rank-$r$ matrix $L = UV' \in \mathbb{R}^{p_1 \times p_2}$ with column/row spaces $(\mathcal{C},\mathcal{R})$.  To compute this space, we consider differences of the form $(U+\Delta_1)(V+\Delta_2)' - UV' = \Delta_1 V' + U \Delta_2' + \Delta_1 \Delta_2' \approx \Delta_1 V' + U \Delta_2'$ for $\Delta_1 \in \mathbb{R}^{p_1 \times r}, \Delta_2 \in \mathbb{R}^{p_2 \times r}$ small.  However, such elements involve attributes of the neighborhood of $L$ that do not concern the estimation of an accurate column space, and therefore we must \emph{quotient} out the irrelevant directions.  Specifically, the directions consisting of column-space components in $\mathcal{C}^\perp$ are not relevant to the accurate estimation of $\mathcal{C}$.  The matrices in $\mathcal{V}(r)$ that lie in a neighborhood around $L$ with deviations in the column-space purely in directions in $\mathcal{C}^\perp$ are given by $L + \Proj_{\mathcal{C}^\perp} \Delta$ for $\Delta \in \mathbb{R}^{p_1 \times p_2}$.  Therefore, we consider the following \emph{equivalence class} associated to each rank-$r$ matrix $L \in \mathcal{V}(r)$:
\begin{eqnarray}
[L] = \{L + \Proj_{\mathcal{C}^\perp} \Delta \Proj_{\mathcal{R}} ~|~ \Delta \in \mathbb{R}^{p_1 \times p_2}\}.
\label{eqn:equiv}
\end{eqnarray}

The tangent space at $L$ with respect to the \emph{quotient manifold} $\mathcal{V}(r) \backslash [L]$ then signifies the discoveries of interest for column-space estimation.  The tangent spaces at $L$ with respect to the equivalence class $[L]$ and with respect to the quotient manifold $\mathcal{V}(r) \backslash [L]$ form complementary subspaces of the tangent space at $L$ with respect to $\mathcal{V}(r)$, and these are known respectively as the \emph{vertical space} and the \emph{horizontal space}.  One can check that the vertical space is given by $T_\mathrm{vertical} = \{\Proj_{\mathcal{C}^\perp}\Delta\Proj_{\mathcal{R}} ~|~ \Delta \in \mathbb{R}^{p_1 \times p_2}\}$ while the horizontal space is given by $T_\mathrm{horizontal} = \{\Proj_{\mathcal{C}}\Delta ~|~ \Delta \in \mathbb{R}^{p_1 \times p_2}\}$ so that $T(\hat{\mathcal{C}},\hat{\mathcal{R}}) = T_\mathrm{vertical} \oplus T_\mathrm{horizontal}$. Our tangent space of interest is thus the subspace $T_\mathrm{horizontal}$, which is solely a function of the column space $\mathcal{C}$.\\

Observing that $\Proj_{T_\text{horizontal}} = \Proj_{\mathcal{C}} \otimes \mathcal{I}$ and $\Proj_{T_\text{horizontal}^\perp} = \Proj_{\mathcal{C}^\perp} \otimes \mathcal{I}$, the expected false discovery, power, and false discovery rate in the context of column-space estimation associated to an estimator $\hat{C}$ are defined as:
\begin{eqnarray}
\begin{aligned}
\mathrm{FD} &=& \mathbb{E}\left[\mathrm{trace}\left(\Proj_{\hat{\mathcal{C}}}\Proj_{{\mathcal{C}^\star}^\perp} \right)\right] \\
\mathrm{PW} &=& \mathbb{E}\left[\mathrm{trace}\left(\Proj_{\hat{\mathcal{C}}}\Proj_{{\mathcal{C}^\star}} \right)\right] \\
\mathrm{FDR} &=& \mathbb{E}\left[\frac{\mathrm{trace}\left(\Proj_{\hat{\mathcal{C}}}\Proj_{{\mathcal{C}^\star}^\perp} \right)}{\mathrm{dim}(\hat{\mathcal{C}})}\right].
\end{aligned}
\label{eqn:horizontal_fd}
\end{eqnarray}

\subsection{False Discovery Guarantees for Column-Space Estimation}
\label{section:subspace_theory}
\vspace{0.1in}
In this section, we provide false discovery control guarantees of subspace stability selection for column-space estimation problems.  Suppose there exists a population column-space $\mathcal{C}^\star \in \mathbb{R}^{p_1}$, and we are given i.i.d observations from a model parameterized by $\mathcal{C}^\star$. Let $\hat{\mathcal{C}}$ be a subspace estimator that operates on samples drawn from the model parameterized by $\mathcal{C}^\star$. Let $\mathcal{D}(n)$ denote a dataset consisting of $n$ i.i.d observations from these models; we assume $n$ is even and that we are given $B$ subsamples $\{\mathcal{D}_{\ell}\}_{i = 1}^{B}$ via complementary partitions of $\mathcal{D}(n)$. 

We omit the proof of each of these statements as their proof is similar in spirit to those from the main paper. 
\begin{theorem}
[False Discovery Control of Subspace Stability Selection]~Consider the setup described above.  Let $\hat{C}(\mathcal{D}_\ell)$ denote the subspace estimates obtained from each of the subsamples, and let $\mathcal{P}^\mathcal{C}_{\texttt{avg}}$ denote the associated average projection operator computed via (3.2) (main paper).  Fix any $\alpha \in (0,1)$ and let $\mathcal{C}$ denote any selection of an element of the associated set $\mathcal{T}_\alpha$ of stable tangent spaces.  Then for any fixed collection of orthonormal basis elements $\{M_i\}_{i = 1}^{\mathrm{dim}({\mathcal{C}^\star}^\perp)}$ of ${\mathcal{C}^\star}^\perp$
\begin{eqnarray}
\mathbb{E}\left[\mathrm{trace}\left(\Proj_{\mathcal{C}}\Proj_{{\mathcal{C}^\star}^\perp}\right)\right] \leq F+\kappa_\text{bag}(\alpha)+ {2({1-\alpha})}\mathbb{E}[\mathrm{dim}(\mathcal{C})]\label{eqn:main3}.
\end{eqnarray}
For basis-dependent bound, $F \sum_{i = 1}^{\mathrm{dim}({\mathcal{C}^\star}^\perp)} \mathbb{E}\left[\left\|\Proj_{\hat{\mathcal{C}}(\mathcal{D}(n/2))}(M_i)\right\|_F\right]^2$ and $\kappa_{\text{bag}}(\alpha) =\sum_{i = 1}^{\text{dim}({\mathcal{C}^\star}^\perp)}\tfrac{2}{B} \sum_{j=1}^{B/2}\allowbreak\mathbb{E}[ \max_{k\in\{0,1\}}\allowbreak\mathrm{trace}([\Proj_{\mathcal{C}}, \Proj_{\hat{\mathcal{C}}{(\mathcal{D}_{2j-k})}^\perp}] \times [\Proj_{\mathrm{span}(M_i)}, \Proj_{\hat{\mathcal{C}}(\mathcal{D}_{2j-k})}])]$, whereas for a basis-independent bound, 
$F \leq \mathbb{E}[\mathrm{trace}(\allowbreak\Proj_{\hat{\mathcal{C}}(\mathcal{D}(n/2))}\Proj_{{\mathcal{C}^\star}^\perp})^{1/2}]^2$ and  $\kappa_{\text{bag}}(\alpha) =\tfrac{2}{B} \sum_{j=1}^{B/2}\mathbb{E}[ \max_{k\in\{0,1\}}\mathrm{trace}([\Proj_{\mathcal{C}}, \Proj_{\hat{\mathcal{C}}{(\mathcal{D}_{2j-k})}^\perp}] \times [\Proj_{{\mathcal{C}^\star}^\perp}, \Proj_{\hat{\mathcal{C}}(\mathcal{D}_{2j-k})}])]$.
\label{thm:main2}
Here the expectation is with respect to randomness in the observations.  The set $\mathcal{D}(n/2)$ denotes a collection of $n/2$ i.i.d. observations drawn from the model parametrized by $\mathcal{C}^\star$.
\end{theorem}
The next proposition provides an upper bound for $\kappa_\text{bag}(\alpha)$ and also provides a bag independent bound:
\begin{proposition} [Bounding $\kappa_\text{bag}$ and a Bag Independent Result] Consider the setup of Theorem~\ref{thm:main2}. Then the following bound holds for both the basis-independent and basis-dependent $\kappa_\text{bag}(\alpha)$: $\kappa_\text{bag}(\alpha) \leq 2\sqrt{1-\alpha}\mathbb{E}[\text{dim}(\mathcal{C})]$. Furthermore, letting the average number of discoveries from $n/2$ observations be denoted by $q := \mathbb{E}[\mathrm{dim}(\hat{\mathcal{C}}(\mathcal{D}(n/2)))]$, we also have that $\mathbb{E}[\text{dim}(\mathcal{C})] \leq \frac{q}{\alpha}$.  Thus, we obtain the following false discovery bound for any $B \geq 2$: 
\begin{eqnarray}
\mathbb{E}\left[\mathrm{trace}\left(\Proj_{\mathcal{C}}\Proj_{{\mathcal{C}^\star}^\perp}\right)\right] &\leq& F+\frac{2q}{\alpha}({1-\alpha}+\sqrt{1-\alpha}). \label{eqn:bag_dependent}
\end{eqnarray}
\end{proposition}

Finally, we obtained a refined bound under ``better than random guessing" and exchangeability assumptions: 
\begin{eqnarray}
\begin{gathered}
\text{Assumption 3:}~~  \frac{\mathbb{E}\left[\text{trace}\left(\Proj_{{\mathcal{C}^\star}^\perp}\Proj_{\hat{\mathcal{C}}(\mathcal{D}({n/2}))}\right)\right]}{\text{dim}({\mathcal{C}^\star}^\perp)} \leq \frac{\mathbb{E}\left[\text{trace}\left(\Proj_{{\mathcal{C}^\star}}\Proj_{\hat{\mathcal{C}}(\mathcal{D}({n/2}))}\right)\right]}{\text{dim}({\mathcal{C}^\star})} \\
\text{Assumption 4:} ~~\text{The distribution of } \|\Proj_{\hat{\mathcal{C}}(\mathcal{D}(n/2))}(M)\|_F \text{ is the same for all } M \in {\mathcal{C}^\star}^\perp, \|M\|_F = 1.
\end{gathered}
\label{eqn:exchangeable_column}
\end{eqnarray} 
The idea behind these two assumptions are similar to Assumptions 1 and 2 in (3.7) (main paper). In particular, a similar argument as with Assumption 1 demonstrates that Assumption 3 is very benign. Assumption 4 is satisfied for data generation processes and estimators that are both invariant under orthogonal conjugation. In particular, consider the PCA model $y = \mathcal{B}^\star{z} + \epsilon$ for $\mathcal{B}^\star \in \mathbb{R}^{p_1 \times k}$ and $\epsilon$ is a Gaussian vector with independent and identically distributed coordinates. Consider the PCA-estimator that finds top components of the empirical covariance of $y$ from observations. Then the estimator satisfies Assumption 4 in \eqref{eqn:exchangeable_column}.\\[0.1in]
\begin{proposition} [Refined False Discovery Bound] Consider the setup in Theorem~\ref{thm:main2}. Suppose additionally that Assumptions 3 and 4 in \eqref{eqn:exchangeable_column} are satisfied. Let the average number of discoveries from $n/2$ observations be denoted by $q := \mathbb{E}[\mathrm{dim}(\hat{\mathcal{C}}(\mathcal{D}(n/2)))]$. Then, for any fixed $M \in {\mathcal{C}^\star}^\perp$ with $\|M\|_2 = 1$, the expected false discovery of a stable column-space $\mathcal{C}$ is bounded by:
\begin{eqnarray}
\mathbb{E}\left[\mathrm{trace}\left(\Proj_{\mathcal{C}}\Proj_{{\mathcal{C}^\star}^\perp}\right)\right] \leq \frac{q^2}{p_1}+ f\left(\kappa_\text{indiv}\right) +\frac{2q}{\alpha}(1-\alpha+\sqrt{1-\alpha}),
\label{eqn:gammaboundcpl}
\end{eqnarray}
where $\kappa_\text{indiv} := \mathbb{E}\left[\|[\Proj_{\mathrm{span}(M)},\Proj_{{\mathcal{C}^\star}^\perp}]\|_F\right]$ and $f(\kappa_\text{indiv}) = p_1\kappa_\text{indiv}^2+2q\kappa_\text{indiv}$.
\label{prop:columnspace}
\end{proposition}

\end{document}